\documentclass[structabstract]{aa}  
\usepackage{graphicx}
\usepackage[varg]{txfonts}
\usepackage{multirow} 
\usepackage[FIGTOPCAP, center, nooneline]{subfigure}
\usepackage{longtable}
\usepackage[]{natbib}

\newcommand{\HII}{H\,{\sc ii}}
\newcommand{\HI}{H\,{\sc i}}
\newcommand{\CII}{C\,{\sc ii}}

\bibpunct{(}{)}{;}{a}{}{,}

\begin{document}
   \title{The chemistry and spatial distribution of small hydrocarbons \\ in UV-irradiated molecular clouds: the Orion Bar PDR\thanks{Based on observations obtained with the IRAM 30m telescope. IRAM is supported by INSU/CNRS (France), MPG (Germany), and IGN (Spain).}}

     \author{S. Cuadrado\inst{\ref{inst1},\ref{inst2}}\and J. R. Goicoechea\inst{\ref{inst1},\ref{inst2}}\and P. Pilleri\inst{\ref{inst3},\ref{inst4}}\and J. Cernicharo\inst{\ref{inst1},\ref{inst2}}\and A. Fuente\inst{\ref{inst5}}\and C. Joblin\inst{\ref{inst3},\ref{inst4}}}

   \institute{
   Grupo de Astrof\'{\i}sica Molecular. Instituto de Ciencia de Materiales de Madrid (ICMM, CSIC). Sor Juana Ines de la Cruz 3, 28049 Cantoblanco, Madrid, Spain.  \email{s.cuadrado@icmm.csic.es}\label{inst1}    
   \and Centro de Astrobiolog\'{\i}a (CSIC-INTA), Carretera de Ajalvir km 4, 28850 Torrej\'on de Ardoz, Madrid, Spain.
 \label{inst2}
   \and Universit\'e de Toulouse, UPS-OMP, IRAP, Toulouse, France. \label{inst3}
   \and CNRS, IRAP, 9 Av. colonel Roche, BP 44346, 31028, Toulouse Cedex 4, France. \label{inst4}
   \and Observatorio Astron\'omico Nacional, Apdo. 112, 28803 Alcal\'a de Henares, Madrid, Spain. \label{inst5}}

\date{Received <date> / Accepted <date>}

\abstract{Carbon chemistry plays a pivotal role in the interstellar medium (ISM) but even the synthesis of the simplest hydrocarbons and how they relate to polycyclic aromatic hydrocarbons (PAHs) and grains is not well understood.} 
{We study the spatial distribution and chemistry of small hydrocarbons in the Orion Bar photodissociation region (PDR), a prototypical environment in which to investigate molecular gas irradiated by strong UV fields.} 
{We used the IRAM 30m telescope to carry out a millimetre line survey towards the Orion Bar edge, 
complemented with $\sim$2'$\times$2' maps of the C$_{2}$H and \mbox{$c$-C$_{3}$H$_{2}$} emission.
We analyse the excitation of the detected hydrocarbons
and constrain the physical conditions of the emitting regions with non-LTE radiative transfer models. We compare the inferred column 
densities with updated gas-phase photochemical models including $^{13}$CCH and C$^{13}$CH isotopomer fractionation.} 
 {Approximately 40$\%$ of the lines in the survey 
arise from hydrocarbons (C$_{2}$H, C$_{4}$H, $c$-C$_{3}$H$_{2}$, \mbox{$c$-C$_{3}$H}, C$^{13}$CH, $^{13}$CCH, 
\mbox{$l$-C$_{3}$H}, and \mbox{$l$-H$_2$C$_{3}$} in decreasing order of abundance). 
We detect new lines from \mbox{$l$-C$_{3}$H$^{+}$}  and  improve
its rotational spectroscopic constants. Anions or deuterated hydrocarbons are not detected, but we 
provide accurate upper limit abundances: \mbox{[C$_{2}$D]/[C$_{2}$H]<0.2$\%$}, 
\mbox{[C$_{2}$H$^{-}$]/[C$_{2}$H]<0.007$\%$}, and \mbox{[C$_{4}$H$^{-}$]/[C$_{4}$H]<0.05$\%$}. }
{ Our models can reasonably match the observed column densities of most hydrocarbons (within factors of $<$3). 
Since the observed spatial distribution of the C$_{2}$H and $c$-C$_{3}$H$_{2}$ emission is similar but does not follow the PAH emission, we conclude that, in high UV-flux PDRs, photodestruction of PAHs is not a necessary requirement  to explain the observed abundances  of the smallest hydrocarbons.
Instead, gas-phase endothermic reactions (or with barriers) between C$^+$, radicals, and H$_{2}$  enhance the formation \mbox{of simple} hydrocarbons.
Observations and models suggest that the [C$_{2}$H]/[$c$-C$_{3}$H$_{2}$]  ratio ($\sim$32 at the PDR edge) decreases 
with \mbox{the UV field} attenuation. The observed low cyclic-to-linear C$_3$H column density ratio ($\leq$3)
 is consistent with a high electron abundance ($x_e$) PDR environment. 
In fact, the poorly constrained $x_e$ gradient 
influences much of the hydrocarbon chemistry in the more UV-shielded gas. The inferred hot rotational temperatures for C$_{4}$H and $l$-C$_3$H$^+$  also suggest that radiative IR
 pumping affects their excitation.
We propose that reactions of C$_{2}$H isotopologues with $^{13}$C$^+$ and H atoms can explain  the 
observed \mbox{[C$^{13}$CH]/[$^{13}$CCH]=1.4$\pm$0.1 fractionation level.}}

 

   
   
   


\titlerunning{The chemistry and spatial distribution of small hydrocarbons in UV irradiated molecular clouds.}
 \authorrunning{S. Cuadrado, et al.}
 \maketitle
 
%

\section{Introduction}

Bright photodissociation regions (PDRs) are the transition layers between the ionised gas directly irradiated by strong UV fields (e.g. from  massive OB stars) and the cold neutral gas shielded from radiation \citep[e.g.][]{Tielens_1985a}. 
Photodissociation regions are found in many astrophysical environments and spatial scales, from the nuclei of starburst galaxies \citep[e.g.][]{Fuente_2008} to the illuminated surfaces of protoplanetary disks \citep[e.g.][]{Agundez_2008a}. All of them show a characteristic chemistry that can be understood in terms of an active UV photochemistry. The closest and brightest example of such PDR is the so-called Orion Bar, at the interface between the Orion Molecular Cloud (OMC) and the \HII\,region illuminated by the Trapezium  stars. The Orion Bar is a prototypical  high-UV flux, hot PDR, with a far-UV radiation field (\mbox{FUV, 6.0 eV < h$\nu$ < 13.6 eV}) of a few $10^{4}\,$ times the mean interstellar field \citep{Marconi_1998}. Because of its proximity (414$\pm$7 pc to the Orion Nebula cluster, \citealt{Menten_2007}), the Orion Bar offers the opportunity to determine the chemical content, spatial stratification of different species, and chemical formation routes in UV illuminated gas.

The transition from  ionised to neutral gas in the Orion Bar has been extensively mapped 
in various atomic and molecular tracers
\citep[see e.g.][]{Tielens_1993, Hogerheijde_1995, vanderWerf_1996,Walmsley_2000, Ossenkopf_2013}. 
The detailed analysis of these observations suggested an inhomogeneous density distribution. 
The most commonly accepted scenario is that an extended gas component, with mean gas densities
of 10$^{4-5}$\,cm$^{-3}$, causes the chemical stratification seen perpendicular to the dissociation 
front as the FUV field is attenuated. In this context, the low energy transitions of different
molecules, including CO, would arise from this extended  
interclump medium \citep{Hogerheijde_1995, Jansen_1995, Simon_1997, Wiel_2009, Habart_2010, van_der_Tak_2013}.
 In addition, another component of higher density clumps was 
invoked to fit the observed high-$J$ CO, CO$^+$, and other high density and temperature tracers 
\citep[][]{Burton_1990, Parmar_1991, Stoerzer_1995, YoungOwl_2000, Batrla_2003}. Owing to its small filling factor,
 this clumpy structure would allow FUV radiation to permeate the region. 
Although this scenario is still controversial, recent 3D models of the Orion Bar structure are compatible with this morphology \citep{Andree-Labsch_2014}.

Depending on the FUV field strength and on the gas density, different processes contribute to the gas heating. Photoelectrons from
polycyclic aromatic hydrocarbons (PAHs) and 
grains heat the interclump gas from \mbox{$\sim$85 K} \citep{Hogerheijde_1995} to
\mbox{$\gtrsim$500 K} at the dissociation front \citep{Allers_2005,vanderWerf_2013}. In addition, the temperature at the surface 
($A_{\rm V}\negthickspace<$1) of dense clumps can go above \mbox{1000 K} because collisional deexcitation of vibrationally excited H$_{2}$ dominates the gas heating \citep{Burton_1990}. The presence of both hot gas and FUV-pumped, vibrationally excited H$_{2}$ (observed in the near-IR at $\sim$2.1 $\mu$m) triggers a distinctive PDR chemistry where highly endothermic reactions and reactions with large activation barriers can proceed quickly \citep{Agundez_2010}. Examples of this peculiar chemistry are the reactions of H$_{2}$ with C$^{+}$, O, and S$^{+}$ (all very abundant in PDR edges) that allow the formation of CH$^{+}$, OH, and SH$^{+}$
and represent the first steps of the  PDR chemistry \citep[see recent detections by][]{Habart_2010,Goicoechea_2011,Nagy_2013,Muller_2014}. 
In addition to these simple hydrides, the abundance of carbon bearing radicals (CN, C$_{2}$H, \mbox{$c$-C$_{3}$H$_{2}$}) 
was found to increase close to the dissociation front \citep[e.g.][]{Jansen_1995, Fuente_1996}.

The formation and chemical behaviour of hydrocarbon molecules in the interstellar medium (ISM) are long standing problems in astrochemistry. 
Since the early 1970s it has been evident that hydrocarbons, the simplest organic molecules, are ubiquitous in the 
ISM \citep[e.g.][]{Tucker_1974, Thaddeus_1985a, Thaddeus_1985b, Yamamoto_1987a}. These molecular species have peculiar chemical structures (e.g. very rigid compounds such as polyynes or cumulenes) and are quite reactive and polar because of the presence of unpaired electrons on the carbon atoms (e.g. carbenes). Owing to their abundance and ubiquity, these molecules often have a bright rotational spectrum which makes them easily detected in the ISM \citep[e.g.][]{Guelin_1978, Cernicharo_1991a}. \citet{Cernicharo_1984} were the first to show that carbon chain radicals and cyanopolyynes were present in several cold cores in the Taurus region.  
The C$_{2}$H and \mbox{$c$-C$_{3}$H$_{2}$} molecules, among the most abundant hydrocarbons, are detected in very different environments, ranging from diffuse clouds \citep[e.g.][]{Lucas_2000} to cold dark clouds \citep[e.g.][]{Fosse_2001}. They have even been detected towards extragalactic sources \citep{Fuente_2005, Meier_2005, Aladro_2011, Meier_2012}. The longest hydrocarbon chain radicals found in the ISM so far are C$_{6}$H, C$_{7}$H, and C$_{8}$H \citep{Cernicharo_1987, Guelin_1987, Guelin_1997, Cernicharo_1996}. 
Despite many studies on their abundance and formation routes in different environments  \citep[e.g.][]{Sakai_2010,Liszt_2012,Pilleri_2013}, the synthesis 
of hydrocarbons in the ISM is still poorly understood.
Observations towards diffuse interstellar clouds and low-FUV flux PDRs suggest that the inferred abundances are significantly higher than current pure gas-phase model predictions \citep[e.g.][]{Fosse_2000, Fuente_2003}. The good spatial correlation between the hydrocarbon emission and the PAH emission towards the Horsehead PDR led \citet{Pety_2005} to suggest that the photo-erosion of PAHs and small carbonaceous grains may dominate the formation of small hydrocarbon molecules. In more strongly irradiated environments like the Orion Bar (>300 times higher FUV radiation fluxes than the Horsehead) the situation may not necessarily be the same. In particular, the molecular gas attains much higher temperatures and new gas-phase formation routes, endothermic reactions, and reactions with activation energy barriers, become efficient.

In the context of investigating the chemistry in hot molecular gas irradiated by strong FUV radiation fields, we have 
performed a complete millimetre line survey towards the Orion Bar PDR. This line survey has allowed us to unveil the molecular content,  accurately  determine the abundances of the detected species, and constrain their formation mechanisms. Approximately 40$\%$ of the detected lines arise from small hydrocarbon molecules. In this paper, we report all hydrocarbon lines detected and investigate the spatial distribution of C$_{2}$H and \mbox{$c$-C$_{3}$H$_{2}$} throughout the region. The paper is organised as follows. In \mbox{Sect. 2} we describe the line survey and the mapping  observations. In \mbox{Sect. 3} we report the observational features of the detected hydrocarbons, while in \mbox{Sect. 4} we present the C$_{2}$H and $c$-C$_{3}$H$_{2}$ integrated line-intensity maps at \mbox{3 mm} and \mbox{1 mm}. The data analysis is explained in \mbox{Sect. 5} and the PDR chemical models of hydrocarbons are reported in \mbox{Sect. 6}. In \mbox{Sect. 7} we discuss the results, and finally in \mbox{Sect. 8} we summarise the main conclusions. The results of the whole survey will be reported in a subsequent paper. 

\begin{table}
\centering 
\caption{Observed frequency ranges and telescope parameters.}
\label{Table_efficiences}     
\begin{tabular}{c c c c c @{\vrule height 10pt depth 5pt width 0pt}} 
\hline\hline      
\bf Rec.\tablefootmark{a}	&	\bf Obs. Freq.\tablefootmark{b}	& \bf Backend &	\bf{$\mathrm{\eta_{_{MB}}}$\tablefootmark{c}}	&	\bf HPBW\tablefootmark{d} \\
	&	[GHz]	&		&		&	[arcsec]		\\
\hline   
E0	& \ \ 80.0-117.0 &	FFTS	&	0.87-0.82	&	30.8-21.0		\\  \hline   
E1	& 128.0-175.6 	&	WILMA	&	0.80-0.74	&	19.1-14.0		\\  \hline
E2 	& 202.0-275.0 	&	FFTS	&	0.70-0.56	&	12.2-8.9		\\  \hline
\multirow{2}{*}[-0.1cm]{E3}  & 275.0-304.5 	&	FFTS	&	0.56-0.50	&	8.9-8.1  \\  
   	& 327.8-359.0	&	FFTS	&	0.46-0.40   &   7.5-6.9	 \\  
\hline
\hline 
\end{tabular}
\tablefoot{
\tablefoottext{a}{Emir receiver.} 
\tablefoottext{b}{Observed frequency range.}
\tablefoottext{c}{Antenna efficiencies.}
\tablefoottext{d}{The half power beam width can be well fitted by \mbox{HPBW[arcsec]$\approx$2460/Frequency[GHz]}.}}
\end{table}

	\section{Observations and data reduction}

	\subsection{Line survey}

In 2009 we started a shallow line survey with the aim to investigate the chemistry of the Orion Bar PDR. The observations were conducted with IRAM 30m telescope at Pico Veleta (Sierra Nevada, Spain), at the position 
\mbox{$\mathrm{\alpha_{2000}=05^{h}\,35^{m}\,20.8^{s}\,}$}, \mbox{$\mathrm{\delta_{2000}=-\,05^{\circ}25'17.0''}$}, 
corresponding to the dissociation front of the Orion Bar, close to what \citet{Stoerzer_1995} call the "CO$^{+}$ peak''.

We began the survey with the EMIR receivers (E0 and E1) and WILMA (wideband line multiple autocorrelator) backend, the only broadband backend 
at that time. In April 2012, with the implementation of the new FFTS (fast fourier transform spectrometer) backends, we began a higher spectral resolution line survey and since then we have observed with the E0, E2, and E3 receivers at \mbox{200 kHz} spectral resolution covering a total of \mbox{217 GHz} along 3, 2, 1, and \mbox{0.8 mm} bands with both backends. The receivers were configured in dual sideband (2SB) for bands E0, E2, and E3 (covering 16 GHz of instantaneous bandwidth per polarization), and in single sideband (SSB) for E1 (8~GHz per polarization). The observing procedure was position switching (PSW) with the reference position located at an offset \mbox{(-600",0")} to avoid the extended molecular emission from the OMC complex. The telescope pointing and focus were checked every two hours through azimuth-elevation cross scans on the nearby continuum source (the 0420-014 quasar).
Atmospheric opacity was corrected by calibrating the data using the ATM code (Cernicharo 1985, IRAM internal report; \citealt{Pardo_2001}). The antenna temperature, $T^{*}_{_{\rm A}}$, was converted to the main beam temperature, $T_{_{\rm MB}}$, through the \mbox{$T_{_{\rm MB}}$=$T^{*}_{_{\rm A}}/ \eta_{_{\rm MB}}$} relation, where $\mathrm{\eta_{_{MB}}}$ is the antenna efficiency, which is defined as the ratio between main beam efficiency, $\mathrm{B_{eff}}$, and forward efficiency, $\mathrm{F_{eff}}$. All intensities in tables and figures are in main beam temperature. A local standard of rest (LSR) of \mbox{10.7 km s$^{-1}$} has been assumed in the line survey target position in the Orion Bar dissociation front. Table~\ref{Table_efficiences} shows an overview of the frequency ranges observed with each backend, as well as the variation in the telescope efficiencies, $\mathrm{\eta_{_{MB}}}$, and the half power beam width (HPBW) across the covered frequency range. 

\begin{figure}[t]
\centering
\includegraphics[scale=0.47,angle=0]{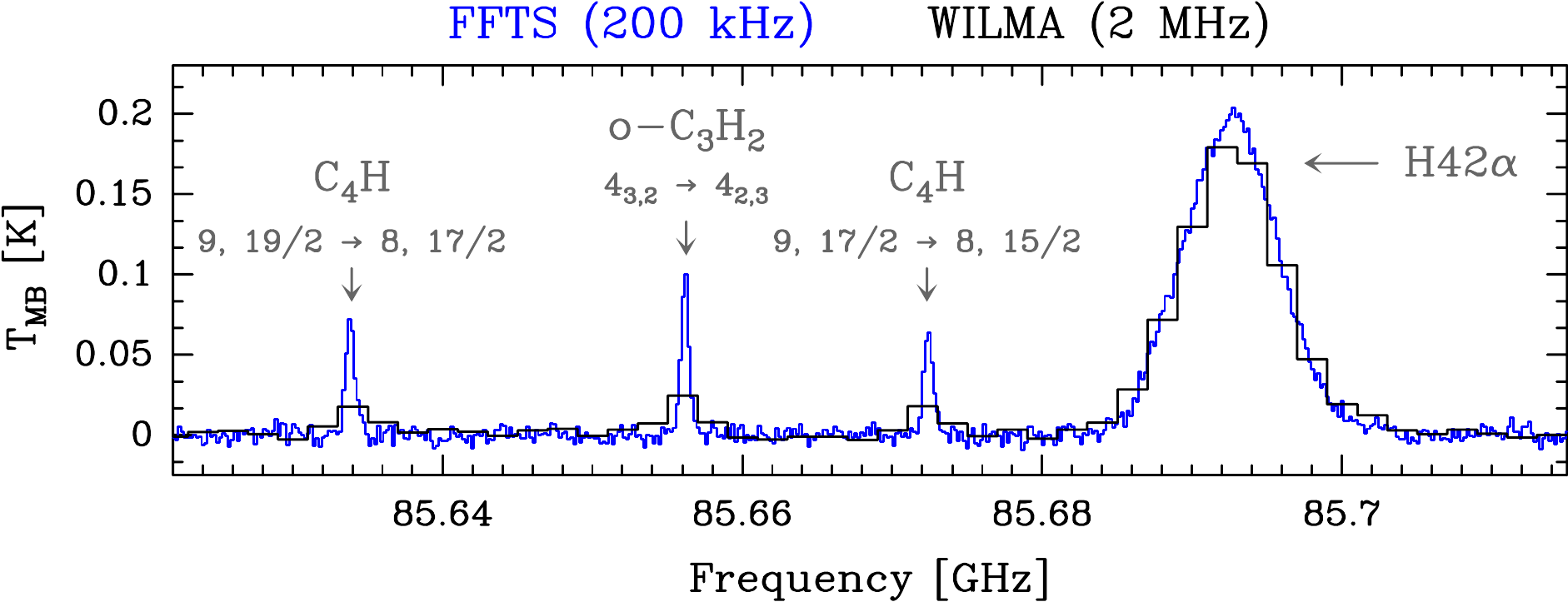}
\caption{Zoom to the \mbox{3 mm} window where both FFTS (\mbox{200 kHz} spectral resolution) and WILMA spectra (\mbox{2 MHz} spectral
 resolution) are shown. Both data have similar rms(channel) $\times$ channel width [K km s$^{-1}$] noise values, but molecular
  line profiles are only resolved with the FFTS. The broad hydrogen recombination lines are resolved at both
   resolutions.}\label{fig:FFTS_vs_WILMA}
\end{figure}

The intrinsic molecular line widths towards the Orion Bar PDR are typically \mbox{$\sim$2 km s$^{-1}$}, so the spectral resolution of the backend has to be high enough to resolve the line profiles. The spectral resolution of WILMA backend does not allow the narrow molecular line profiles of the PDR to be resolved, and therefore it only gives information about the integrated line intensity. Most of the hydrocarbon molecular lines were observed with the FFTS backends which does allow the molecular lines to be resolved, thus providing line profile information. Hydrogen recombination lines from ionised gas in the adjacent \HII\,region are resolved with both backends because they are intrinsically broad (Fig.~\ref{fig:FFTS_vs_WILMA}).

The data were reduced using the CLASS software of the GILDAS package\footnote{http://www.iram.fr/IRAMFR/GILDAS/}. 
A polynomial  baseline of low order (typically second or third order) was subtracted from each \mbox{$\sim$200 MHz} wide spectrum after all scans were added. The rms noise of our observations obtained by integration during \mbox{$\sim$4 h} ranges between 4 mK and 20 mK. Despite the good attenuation of the image band signal \mbox{($>$10 dB)}, bright lines arising from the image band had to be eliminated in the data processing. The image band lines were identified following the procedure developed by \citet{Tercero_2010} according to which each setting was repeated at a slightly shifted frequency \mbox{($\sim$50 MHz)}. The change of the frequency for the lines coming from the image side band when shifting the observing sky frequency allows all possible contaminating lines of the image band to be identified and removed.

\begin{table}
\centering 
\caption{Dipole moments ($\mu$), electronic ground state (E.G.S.), and number of hydrocarbon detected lines in this work.}
\label{Table_detected_hydrocarbon}     
\begin{tabular}{c c c c c @{\vrule height 10pt depth 5pt width 0pt}} 
\hline\hline      
{\bf Molecule}    &    \bf  $\mathrm{\bf \mu}$     & {\bf E.G.S.}  &    \bf Detected & {\bf Ref.}        \\
    &    [Debye]    &            &    \bf Lines    &    \\
\hline   
C$_{2}$H                 &   0.77      & $^{2}\Sigma^{+}$        & 25   & 1    \\  
$^{13}$CCH             &     0.77     & $^{2}\Sigma^{+}$       & 7     & 2  \\ 
C$^{13}$CH             &     0.77     & $^{2}\Sigma^{+}$       & 8     & 2  \\  
$l$-C$_{3}$H$^{+}$   &    3.00     & $^{1}\Sigma^{+}$        & 9     & 3  \\ 
$l$-C$_{3}$H            &    3.55     & $^{2}\Pi^{+}$              & 25    & 1  \\  
$c$-C$_{3}$H           &    2.40     & $^{2}\mathrm{B}_{2}$  & 24    & 4  \\  
$c$-C$_{3}$H$_{2}$  &    3.43     & $^{1}\mathrm{A}_{1}$  & 50    & 5    \\  
$l$-H$_{2}$C$_{3}$   &    4.10     & $^{1}\mathrm{A}_{1}$  & 18    & 6    \\  
C$_{4}$H                 &    0.87     & $^{2}\Sigma^{+}$        & 40    & 1    \\  
\hline
\hline 
\end{tabular}
\tablebib{
(1) \citet{Woon_1995}; (2) The dipole moment is assumed to be the same as for the C$_{2}$H; (3) \citet{Pety_2012}; (4) \citet{Yamamoto_1987a}; (5) \citet{Kanata_1987}; (6) \citet{DeFrees_1986}.
}
\end{table}


\subsection{Maps}

The line survey was complemented with maps of the line emission distribution of different species. In particular we 
present maps of the C$_{2}$H (N=1$\rightarrow$0 and 3$\rightarrow$2) and $c$-C$_{3}$H$_{2}$ (J$_{\rm K_a,K_c}$=2$_{1,2}$$\rightarrow$1$_{0,1}$ and 6$_{1,6}$$\rightarrow$5$_{0,5}$) line emission. The spectral mapping observations were also
obtained at the IRAM 30m telescope in two separate runs. The 3~mm maps were obtained in July 2012, in approximately
\mbox{3 hours} integration time for each configuration. We used the EMIR receivers and the FFTS spectrometers 
at \mbox{50 kHz} resolution to get accurate velocity information. Because of the limited bandwidth of these high 
resolution spectrometers, we used two different configurations for C$_{2}$H and $c$-C$_{3}$H$_{2}$. The 1~mm maps
were obtained in December 2012, using the FFTS at 200~kHz resolution. The maps were obtained using the on-the-fly 
observing mode, with an OFF position at \mbox{(-600", 0")} relative to the map centre. This position is free 
of any emission in these two tracers. 
Data processing consisted in a linear baseline subtraction in each observed spectra. 
The resulting spectra were finally gridded through convolution by a Gaussian.

    
\section{Results: small hydrocarbon detections}

The millimetre molecular line survey of the Orion Bar covers a bandwidth of $\sim$220 GHz in which more than 200 lines
 from small hydrocarbon molecules have been identified. The detected lines were attributed to nine different molecules, from the simplest carbon-chain molecule,
  C$_{2}$H, to more complex molecules with five atoms such as 
 C$_{3}$H$_{2}$. Lines from two isotopologues ($^{13}$CCH and C$^{13}$CH) and one cation ($l$-C$_{3}$H$^{+}$) were 
 detected. No lines from anions or vibrationally excited states of hydrocarbons were identified. Line assignment 
 was carried out using J. Cernicharo's own spectral catalogue \citep[MADEX\footnote{MADEX was extensively used in the line identification and excitation analysis of
several line surveys conducted by our team: e.g. IRC+10216 \citep{Cernicharo_2000}, Sgr~B2 \citep{Goicoechea_2004},
CRL618 \citep{Pardo_2007a}, TMC-1 \citep{Marcelino_2007}  or Orion BN/KL \citep{Tercero_2010}},][]{Cernicharo_2012} and two molecular databases 
 with public access: JPL \citep{Pickett_1998}\footnote{http://spec.jpl.nasa.gov/} and CDMS \citep{Muller_2001,Muller_2005}\footnote{http://www.astro.uni-koeln.de/cdms/}.

\begin{figure*}[t]
\centering
 \includegraphics[scale=0.62,angle=0]{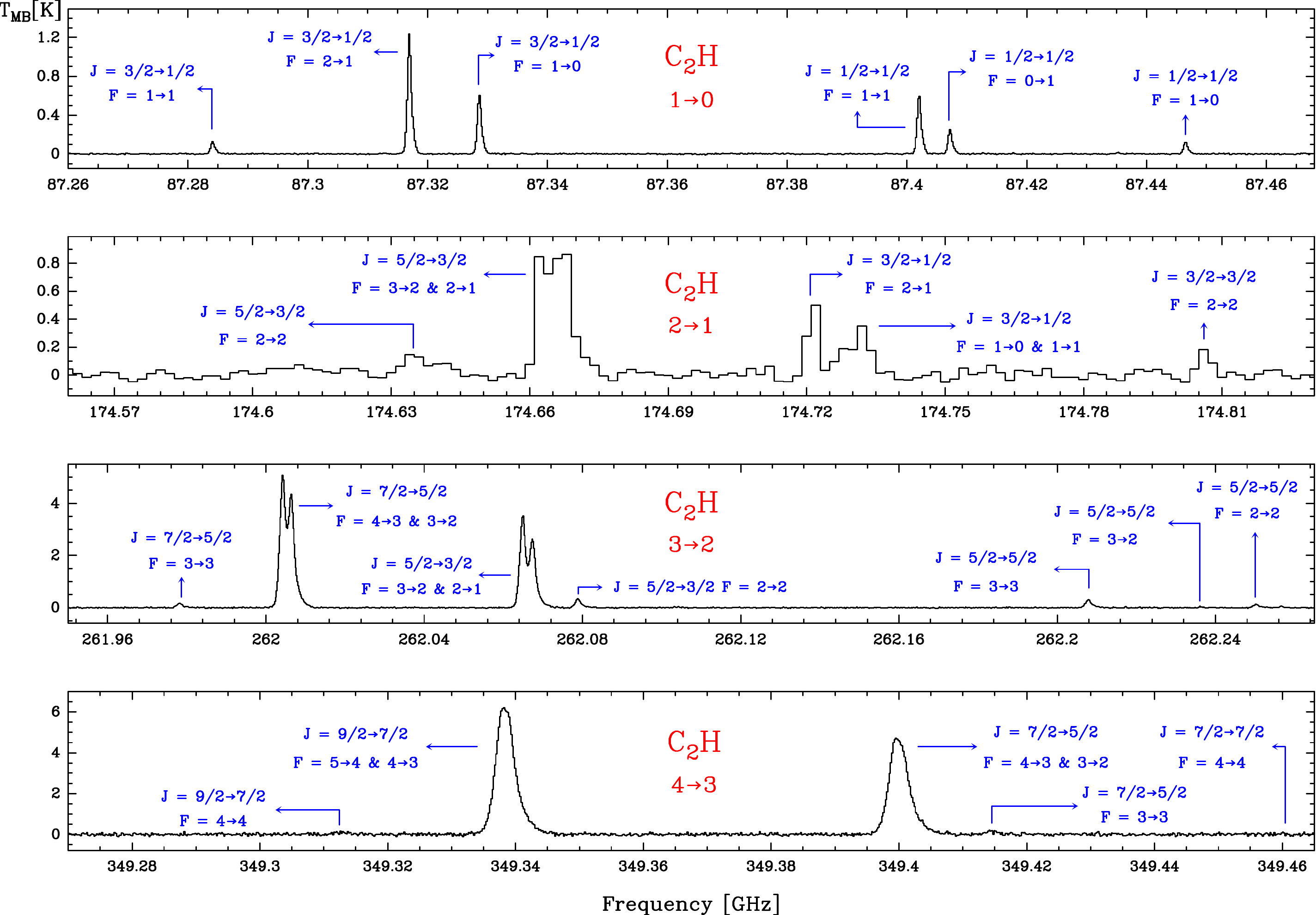} \\
 \caption{Detected C$_{2}$H hyperfine structure lines of the N=1$\rightarrow$0, 2$\rightarrow$1, 3$\rightarrow$2, and
4$\rightarrow$3 rotational transitions. The N=2$\rightarrow$1 transition was observed at 2~MHz resolution while the N=1$\rightarrow$0, 3$\rightarrow$2, and 4$\rightarrow$3 transitions were detected at 200~kHz.
Hyperfine structure are only fully resolved in the N=1$\rightarrow$0 transition. The fine and hyperfine components are 
labelled with the quantum numbers J and F. For clarity, we have not included the labels of the transitions that were 
too weak to be detected.}
\label{fig:C2H_completo}
\end{figure*}

In this section we individually introduce the detected hydrocarbons and summarise their most important spectroscopic features. The studied species are classified according to their empirical formula: C$_{2}$H, C$_{3}$H$^{+}$, C$_{3}$H, C$_{3}$H$_{2}$, and C$_{4}$H. Table~\ref{Table_detected_hydrocarbon} summarises the dipole moments and electronic ground states of the detected species. The spectroscopic and observational line parameters of the detected hydrocarbons are given in Appendix~B. 


{\bf \subsection{C$_{\bf 2}$H}}
	
The C$_{2}$H radical was first detected in the ISM by \citet{Tucker_1974} who detected four components of the N=1$\rightarrow$0 rotational transition in several sources associated with massive star-forming regions. Since then, C$_{2}$H has been detected in a wide variety of sources including the circumstellar envelopes around carbon-rich evolved stars \citep[][and references therein]{DeBeck_2012}, diffuse clouds \citep{Lucas_2000}, cold dark clouds \citep{Wootten_1980}, and even in extragalactic sources \citep{Martin_2006}. The isotopologues of the ethynyl radical, $^{13}$CCH, C$^{13}$CH, and CCD, have also been detected in the ISM, first observed towards the Orion A ridge and Orion KL star forming core \mbox{(3'N, 1'E)} \citep{Saleck_1992,Saleck_1994,Combes_1985,Vrtilek_1985}.\\

\begin{table}
\centering 
\caption{Line intensities of the detected C$_{2}$H N=1$\rightarrow$0 lines.}  
\label{Table_intensities_C2H}     
\begin{tabular}{c c c c c@{\vrule height 10pt depth 5pt width 0pt}}     
\hline\hline      
\bf Transition & \bf S$_{_{\mathrm{\bf ij}}}$\tablefootmark{a} & \bf T$_{_{\mathrm{\bf MB}}}$\tablefootmark{b} &  \multicolumn{2}{c}{{\bf Relative I}\tablefootmark{c} [x100]} \rule[0.15cm]{0cm}{0.2cm}\  \\ \cline{4-5} \cline{1-1}
${\mathrm{(J, F)_{_{N=1}} \rightarrow (J, F)_{_{N=0}}}}$ & & [K] & LTE\tablefootmark{d} & Obs.\tablefootmark{e} \\
\hline     
$(3/2, 1)\rightarrow (1/2, 1)$ 	& 0.17 & 0.14 & \,4.25 & \,4.20  \\	
$(3/2, 2)\rightarrow (1/2, 1)$  & 1.67 & 1.39 & 41.72  & 41.91   \\	
$(3/2, 1)\rightarrow (1/2, 0)$ 	& 0.83 & 0.69 & 20.73  & 20.73   \\	
$(1/2, 1)\rightarrow (1/2, 1)$  & 0.83 & 0.68 & 20.73  & 20.56   \\	
$(1/2, 0)\rightarrow (1/2, 1)$  & 0.33 & 0.28 & \,8.32 & \,8.42  \\	
$(1/2, 1)\rightarrow (1/2, 0)$  & 0.17 & 0.14 & \,4.25 & \,4.19  \\	
\hline 
\hline    
\end{tabular}
\tablefoot{
\tablefoottext{a}{Theoretical line strengths.} 
\tablefoottext{b}{Observed line intensities in mean beam temperature.}
\tablefoottext{c}{Relative intensities.}
\tablefoottext{d}{Expected relative intensities ($S_{\rm ij}$/$\sum S_{\rm ij}$), assuming that lines are optically thin ($\tau<1$).}
\tablefoottext{e}{Observed relative intensity ($T_{\rm MB_{ij}}$/$\sum T_{\rm MB_{ij}}$).}
}
\end{table}

\subsubsection{Ethynyl: $^{12}$C$_{2}$H}

The C$_{2}$H hydrocarbon is a linear molecule with $^{2}\Sigma^{+}$ electronic ground state, therefore its rotational spectrum shows spin rotation interaction and hyperfine structure. The quantum numbers designating the energy levels are N, J, and F. Spin doubling (J=N+S) is produced by the coupling between the rotational angular momentum, N, and the unpaired electron spin, S, while the hyperfine structure (F=J+I) is due to the coupling of the angular momentum, J, and the spin of the hydrogen nucleus, I. Electric dipole selection rules require $\Delta$F=0 (with 0 $\nleftrightarrow$ 0) and  $\Delta$F=$\pm$1, so the N=1$\rightarrow$0, 2$\rightarrow$1, 3$\rightarrow$2, and 4$\rightarrow$3 transitions detected in this work split into 6, 11, 11, and 11 allowed hyperfine components, respectively. The C$_{2}$H  spectroscopic constants were obtained from a simultaneous fit of both laboratory and astronomical data by \citet{Gottlieb_1983a} and \citet{Muller_2000}. 

We have identified a total of 25 lines of C$_{2}$H. They consist of four sets of rotational transitions 
corresponding to the hyperfine splitting of the N=1$\rightarrow$0 to 4$\rightarrow$3 transitions. The six hyperfine 
components of the lowest energy rotational transition (N=1$\rightarrow$0) are well separated in frequency. There are no significant differences in the line widths \mbox{($\Delta$v$\approx$3 km s$^{-1}$)} and line peak velocities. As shown in Table~\ref{Table_intensities_C2H}, 
the relative intensities of the observed N=1$\rightarrow$0 hyperfine components agree with the expected relative
 intrinsic intensities. This shows that the lines are optically thin and do not show hyperfine emission anomalies.

Nine and five spectral lines of each N=3$\rightarrow$2 and 4$\rightarrow$3 hyperfine transition have also been detected at high resolution, but some hyperfine components are partially or fully overlapped. Five lines of
the N=2$\rightarrow$1 transition, corresponding to seven hyperfine rotational transitions, were detected at 2 MHz 
spectral resolution. The hyperfine structure is not fully resolved and some lines are overlapped. The quantum numbers 
of the detected C$_{2}$H transitions, their spectroscopic parameters, and the results from fitting the line profiles with Gaussians are listed in Table~\ref{Table_C2H}. In Fig.~\ref{fig:C2H_completo} we present the spectra of the  
N=1$\rightarrow$0 to 4$\rightarrow$3 rotational lines.

\subsubsection{$^{13}$CCH and C$^{13}$CH}

\begin{figure}
 \centering
 \includegraphics[scale=0.46,angle=0]{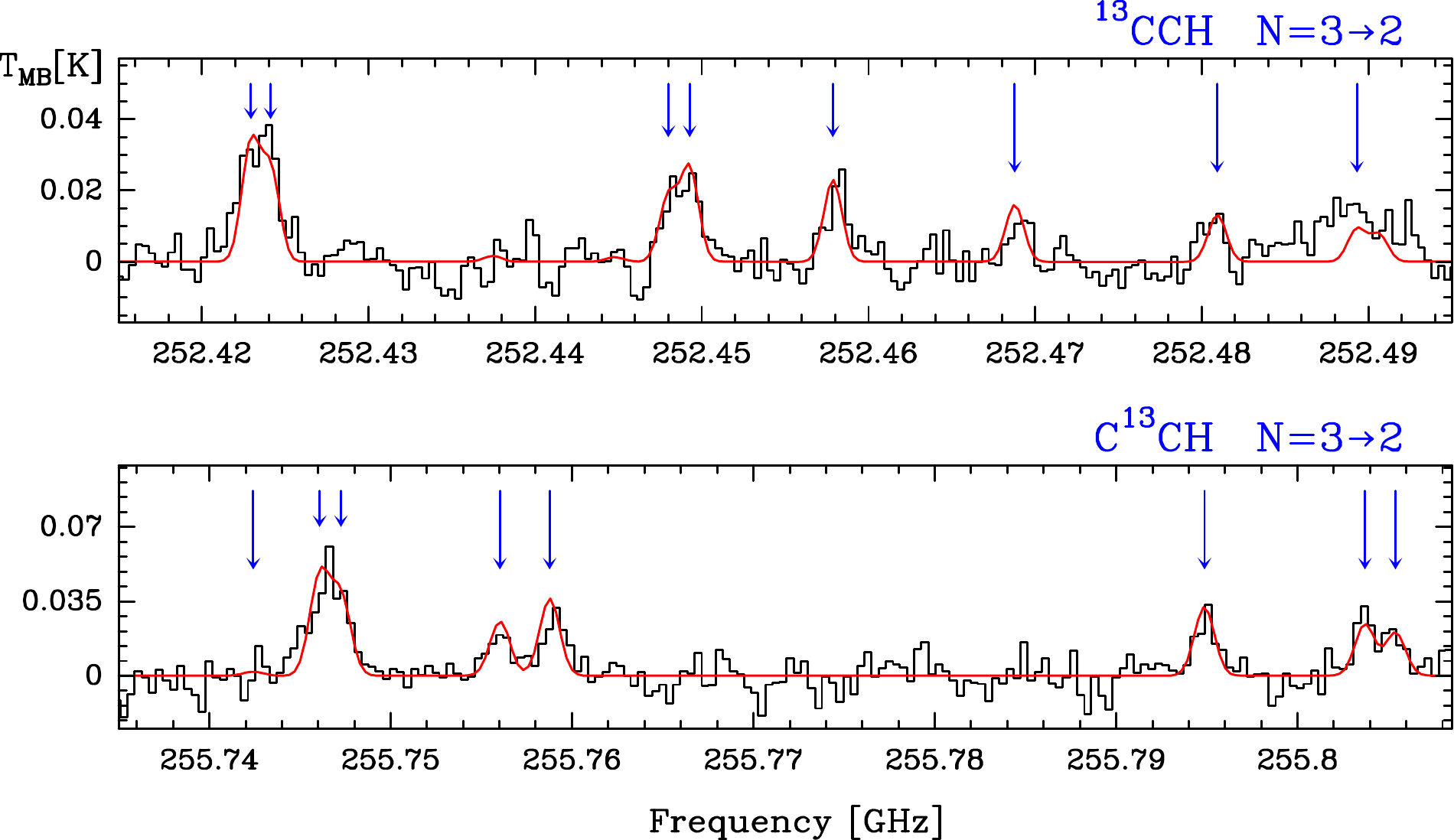} \\ 
 \caption{Detection of the N=3$\rightarrow$2 hyperfine structure of $^{13}$CCH (top) and C$^{13}$CH (bottom) in the Orion Bar PDR (black histogram spectra). A LTE model is overlaid in red (see \mbox{Sect. 5.2)}. The position of hyperfine transitions are indicated by the blue arrows. We note the different abscissa and ordinate axis scales in both spectra.}
\label{fig:13CCH_lines}
\end{figure}

We have detected six lines of $^{13}$CCH and seven lines of its isotopomer C$^{13}$CH in the 1 mm band 
(252.4~GHz and \mbox{255.7 GHz}, respectively), corresponding to the N=3$\rightarrow$2 rotational transition 
(Fig.~\ref{fig:13CCH_lines}). Observations of the $^{13}$C isotopic species of ethynyl have been quite limited. 
To our knowledge, this is the first detection of both isotopologues in a PDR. The rotational spectrum and hyperfine structure
 of both $^{13}$CCH and C$^{13}$CH are described in detail in \citet{McCarthy_1995}. 
 Table~\ref{Table_13CCH} lists the spectroscopic and observational line parameters. The main hyperfine line of
  C$^{13}$CH is more intense than the main hyperfine line of $^{13}$CCH by a factor of \mbox{1.4$\pm$0.1 (3$\sigma$)}, 
  where the quoted uncertainty is three times the standard deviation. 
This difference suggests that fractionation processes differently  affect the two $^{13}$C isotopes of C$_{2}$H (see Sect.~7.2).


{\bf
\subsection{C$_{\bf3}$H$^{\bf+}$}
}
We have also detected a series of lines that \citet{Pety_2012} originally attributed to $l$-C$_{3}$H$^{+}$ in the Horsehead Nebula PDR \citep[see also][]{McGuire_2013}. 
This detection has been recently confirmed in the laboratory \citep{Brunken_2014} and by quantum chemical 
calculations \citep{Botschwina_2014}. The linear C$_{3}$H$^{+}$ is an essential intermediate in the gas-phase 
synthesis of hydrocarbons through ion-molecule reactions. 

\begin{figure}[!b]
 \centering
 \includegraphics[scale=0.67,angle=0]{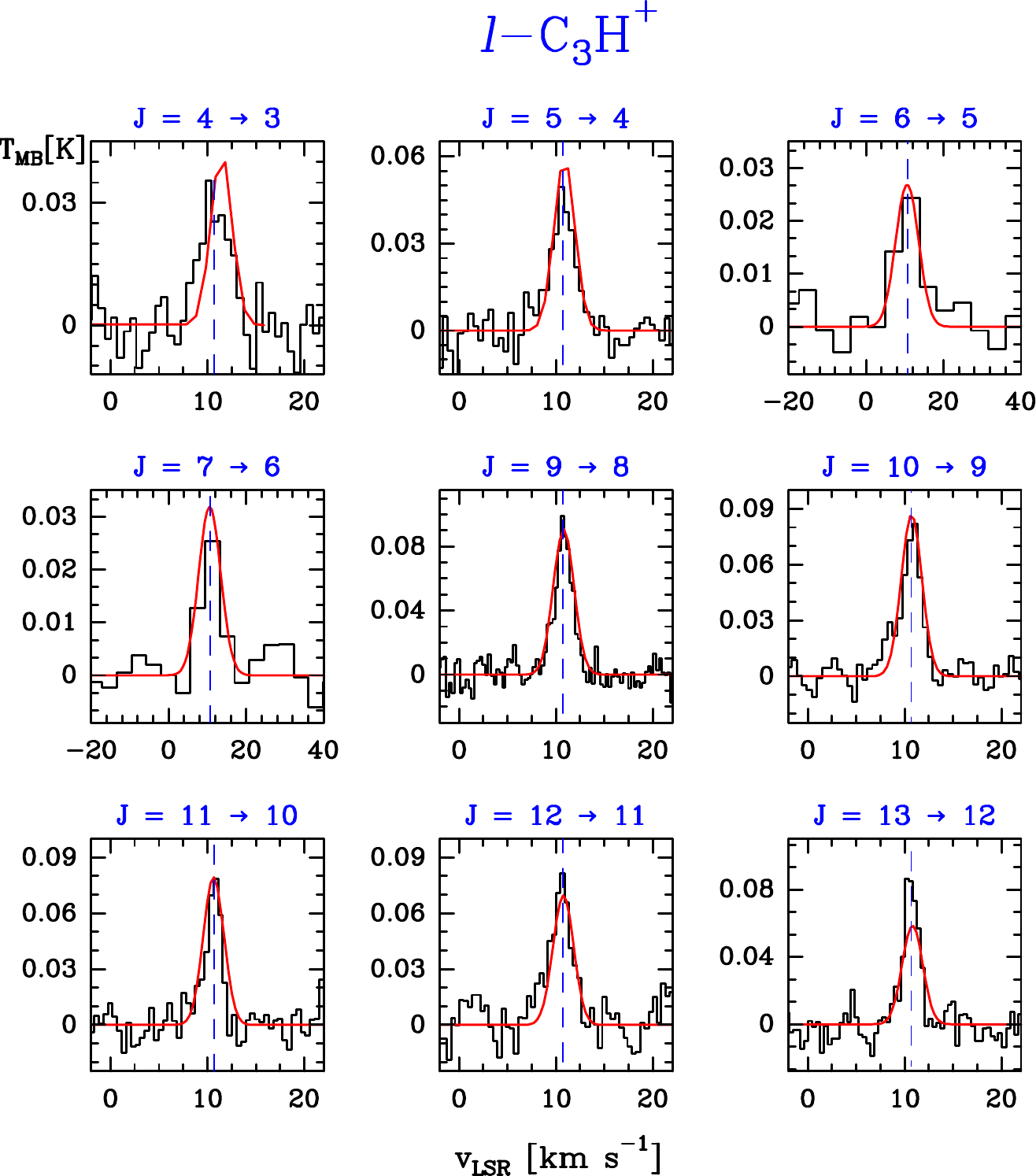} 
 \caption{Series of $l$-C$_{3}$H$^{+}$ lines detected in the Orion Bar PDR. Spectra of the J=4$\rightarrow$3, 5$\rightarrow$4, 9$\rightarrow$8, 10$\rightarrow$9, 11$\rightarrow$10, 12$\rightarrow$11, and 13$\rightarrow$12 rotational transitions of $l$-C$_{3}$H$^{+}$ observed at 200 kHz spectral resolution, and 6$\rightarrow$5  and 7$\rightarrow$6 observed at 2 MHz (black histogram spectra). A LTE model is overlaid in red (see Sect.~5.2). The dashed lines indicate the LSR velocity \mbox{(10.7 km s$^{-1}$)} of the Orion Bar PDR.}\label{fig:C3H+_lines}
\end{figure}

Using the CSO telescope, \citet{McGuire_2014}  searched for $l$-C$_{3}$H$^{+}$ towards a large sample of galactic sources. They only detected the molecule towards the Orion Bar (three rotational lines).
\mbox{$l$-C$_{3}$H$^{+}$} presents a simple rotational spectrum, with J+1$\rightarrow$J 
 transitions. Here we present nine rotational lines,
 from J=4$\rightarrow$3 to 13$\rightarrow$12, the highest frequency
 line detected so far. 
Figure~\ref{fig:C3H+_lines} shows the detected lines and Table~\ref{Table_C3H+} summarises the observed parameters.

\subsubsection{Improved $l$-C$_3$H$^+$ rotational constants}

We have fitted all laboratory lines measured by \citet{Brunken_2014} together with those measured 
towards the Horsehead by \citet{Pety_2012} and those reported here in the Orion Bar. 
Since we detect up to the J=13$\rightarrow$12 line, we can better constrain the distortion constants.
We derive $B$=11244.94793(116)~MHz and $D$=7.66055(853)~kHz with a correlation coefficient
of 0.77 between both constants. The standard deviation of the fit is 47~kHz. 
The rotational constants obtained by merging the laboratory measurements and the astronomical data have
two times better uncertainties for $B$ and almost one order of magnitude better accuracy for $D$.
An attempt to fit the distortion constant $H$ produced a slightly better fit (43 kHz), with 
$H$=(2$\pm$1)$\times$10$^{-7}$~MHz, but with this constant strongly correlated with $D$ and with a significant 
degradation of the uncertainty of the other constants. Higher $J$ lines will be needed to derive the 
sextic order distortion constant of \mbox{$l$-C$_{3}$H$^{+}$}. Nevertheless, it is clear that the molecule is rather 
floppy because the $D$  value is high, even higher than that of C$_{3}$. If the lowest energy bending mode of 
$l$-C$_{3}$H$^{+}$ is as low as that of C$_{3}$, we estimate that this mode can be populated  
for gas temperatures around $\sim$30-40 K. The lines from the bending mode will consist of a series of doublets 
($l$-doubling), blue-shifted with respect to the lines of the ground vibrational mode. Each member of these doublets
 will pertain to a series of harmonically related frequencies.  We have
not found any series of lines with these attributes at the sensitivity level of our survey.
Another related molecule having low lying bending modes is C$_{4}$H
(see discussion in Sect.~7.1). 


{\bf 
\subsection{C$_{\bf 3}$H}
}

\begin{figure}
\centering
\includegraphics[scale=0.45,angle=0]{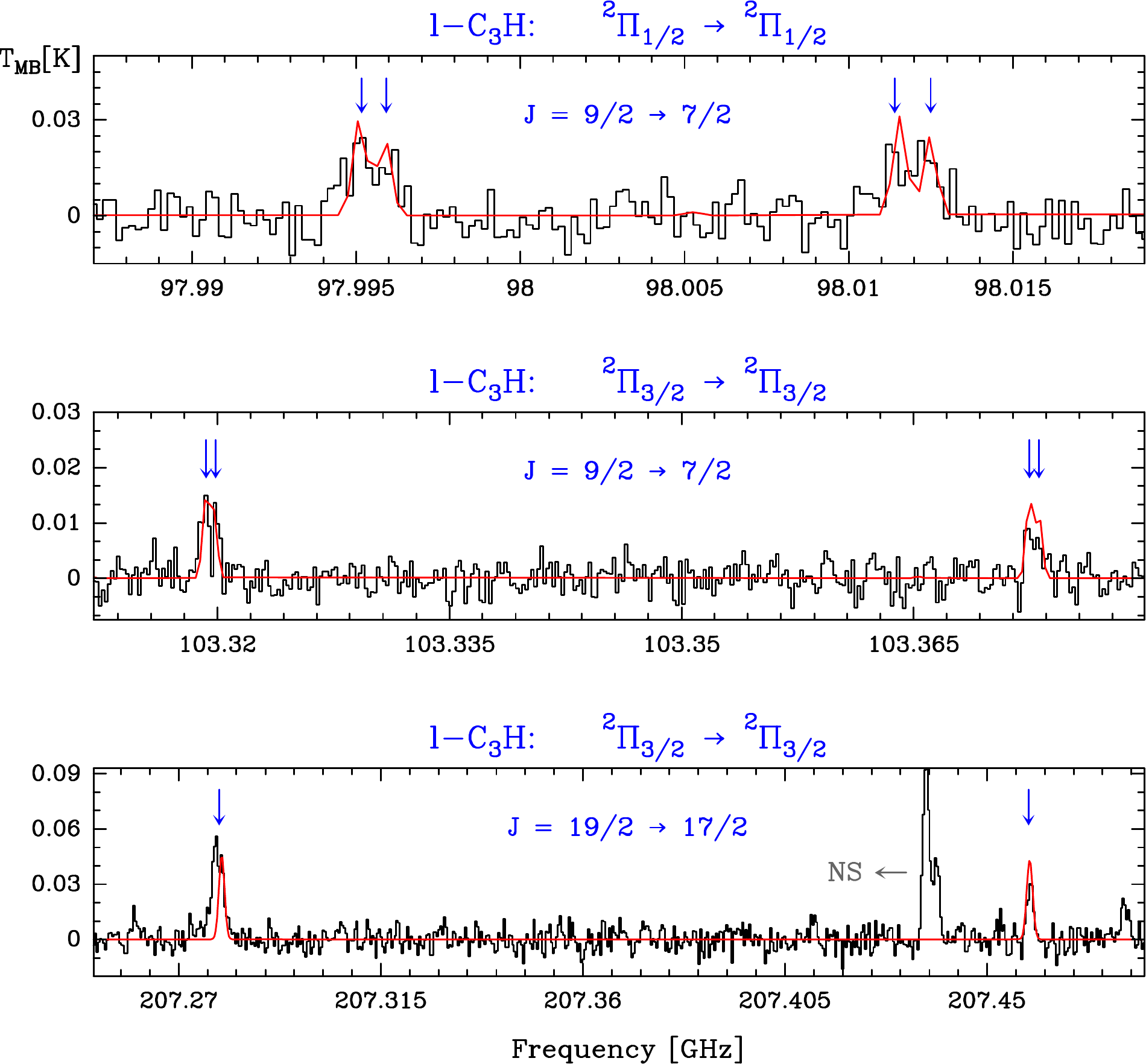} 
\caption{Examples of $l$-C$_{3}$H detected lines (black histogram spectra). A LTE model is overlaid in red
(see Sect.~5.2). $l$-C$_{3}$H lines are indicated by blue arrows. The other spectral features appearing in the selected windows are labelled with their corresponding identification.}\label{fig:$l$-C3H_lines}
\end{figure}	

The C$_{3}$H radical is found in the ISM in two isomeric forms: as a linear chain, $l$-C$_{3}$H, and as a three-carbon ring, $c$-C$_{3}$H. Both isomers were first detected by their rotational transitions at millimetre wavelengths in TMC-1 \citep{Thaddeus_1985a,Yamamoto_1987a}. The $c$-C$_{3}$H radical is more energetically stable than $l$-C$_{3}$H by $\sim$860\,K \citep{Sheehan_2008}.\\

\subsubsection{Propynylidyne: \emph{linear}-C$_{3}$H}

The linear carbon chain, $l$-C$_{3}$H, has a $^{2}\Pi$ electronic ground state. Its laboratory millimetre wave spectrum was reported by \citeauthor{Gottlieb_1985} \citeyearpar{Gottlieb_1985,Gottlieb_1986}. The spin-orbit interaction results in two rotational ladders, $^{2}\Pi_{3/2}$ and $^{2}\Pi_{1/2}$, with intra-ladder and much weaker cross-ladder transitions. Furthermore, each rotational transition is split by $\Lambda$-type doubling due to the nuclei rotation and the unpaired electron motion. In interstellar conditions, only the lower transitions show fully-resolved hyperfine transitions.

We have detected 25 lines of $l$-C$_{3}$H, consisting of \emph{(i)} 12 lines in the $^{2}\Pi_{1/2}$ ladder
\mbox{(E$_{\rm u}$/k $\leq$ 102.2 K)}, and \emph{(ii)} 13 lines in the $^{2}\Pi_{3/2}$ ladder \mbox{(E$_{\rm u}$/k $\leq$ 75 K)}. The $l$-C$_{3}$H lines are rather weak and show very narrow lines profiles \mbox{($\Delta$v $\leq$ 1.6 km s$^{-1}$)}. The hyperfine structure has only been partially resolved in the lowest detected rotational levels. Figure~\ref{fig:$l$-C3H_lines} shows three $l$-C$_{3}$H spectra, which are representative examples of the partially resolved hyperfine structure and the $\Lambda$-doublet components. Spectroscopic and observational parameters of $l$-C$_{3}$H are summarised in Table~\ref{Table_$l$-C3H}.

\begin{figure}
 \centering
 \vspace{0.05cm}
 \includegraphics[scale=0.452,angle=0]{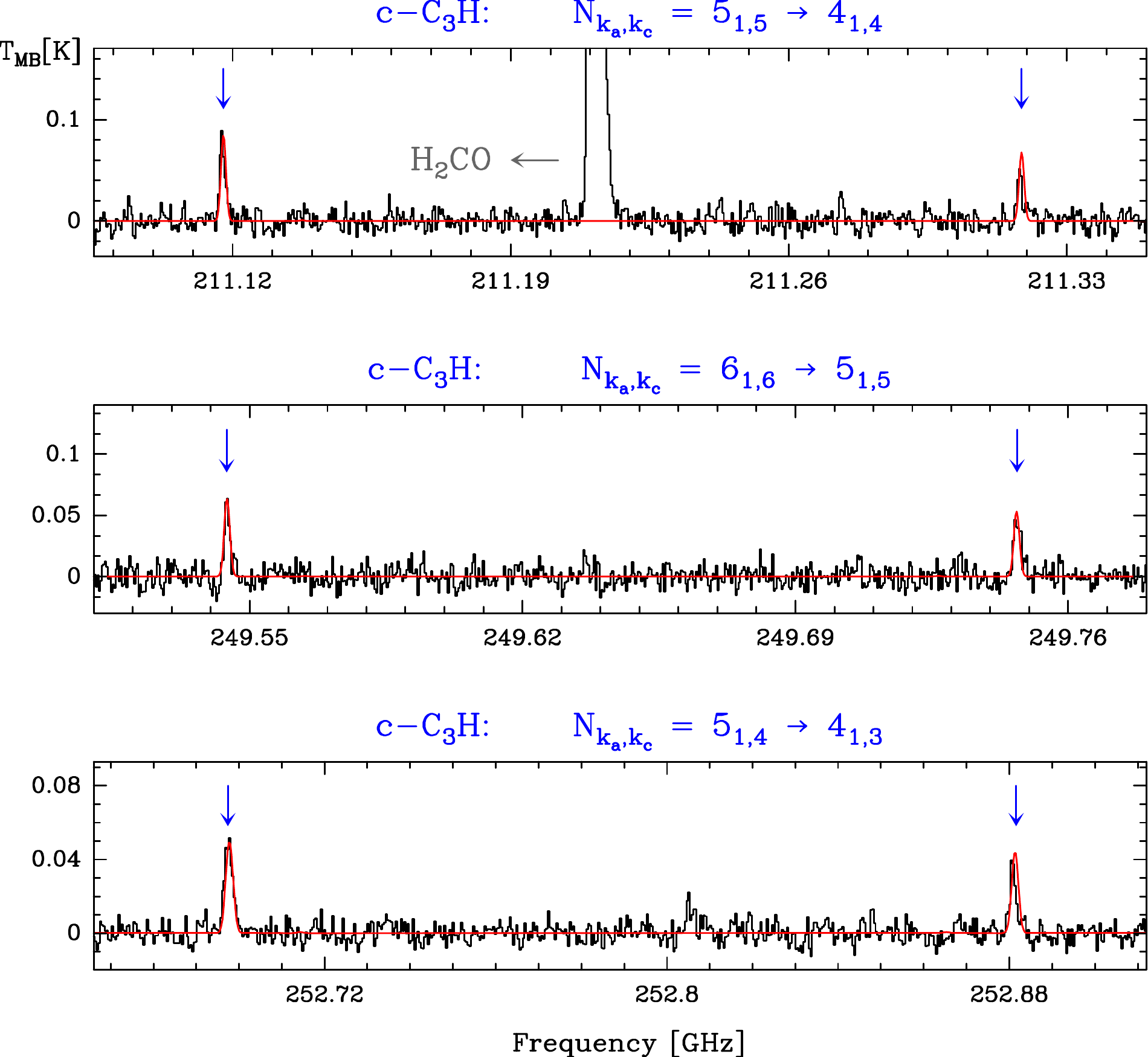} \\
 \caption{$c$-C$_{3}$H spectra observed in the \mbox{1 mm} band (black histogram spectra). A LTE model is overlaid in red 
 (see \mbox{Sect. 5.2}). $c$-C$_{3}$H lines are indicated by blue arrows. The other spectral features appearing in the selected 
 windows are labelled with their corresponding identification.}\label{fig:$c$-C3H_lines}
\end{figure}

\subsubsection{Cyclopropynylidyne: \emph{cyclic}-C$_{3}$H}

The cyclic form, $c$-C$_{3}$H, is an a-type asymmetric top molecule.
 The $c$-C$_{3}$H rotational level transitions, given by N$_{\mathrm{K_{a} K_{c}}}$, are split into fine and hyperfine
  levels (labelled by the quantum numbers J and F, respectively). The rotational transitions are governed by the 
  $\Delta$J=0 and $\pm$1 selection rules. The laboratory spectrum was first measured by \citet{Lovas_1992} 
  and \citet{Yamamoto_1994}. 

We have detected 24 lines of $c$-C$_{3}$H. These consist of ten sets of rotational transitions with fine and hyperfine structure lines \mbox{(E$_{\rm u}$/k $\leq$ 55 K)}. No transition is completely resolved in its hyperfine components. Examples of $c$-C$_{3}$H spectra are shown in Fig.~\ref{fig:$c$-C3H_lines} and the line parameters are listed in Table~\ref{Table_$c$-C3H}.


{\bf
\subsection{C$_{\bf 3}$H$_{\bf 2}$}
}

Interstellar molecules with the elemental formula C$_{3}$H$_{2}$ belong to a type of compounds known as carbenes. They are highly reactive because of the two non-bonded electrons on one of the three carbon atoms.
Two structural isomers are observed: cyclic and linear. The cyclic one,
the first organic ring detected in
the ISM \citep{Thaddeus_1985b,Vrtilek_1987}, is $\sim$6960~K more stable 
than the linear form \citep{Dykstra_1979,DeFrees_1986}. \citet{Cernicharo_1991a} observed one of the C$_{3}$H$_{2}$ linear isomers in TMC-1 
(hereafter $l$-H$_{2}$C$_{3}$) characterised by two or more consecutive double carbon bonds with two non-bonded electrons on the terminal carbon \citep{Cernicharo_1991b,Langer_1997}. Although the cyclic form is more abundant and ubiquitous,
both C$_{3}$H$_{2}$ isomers are widely observed in the ISM. 
The cyclic and linear isomers of C$_{3}$H$_{2}$ have two indiscernible off-axis hydrogen atoms which impose an additional ortho-para symmetry. 
The rotational spectra of both isomers were originally reported by \citet{Vrtilek_1987,Vrtilek_1990}.\\

\begin{figure}
\centering
\includegraphics[scale=0.62,angle=0]{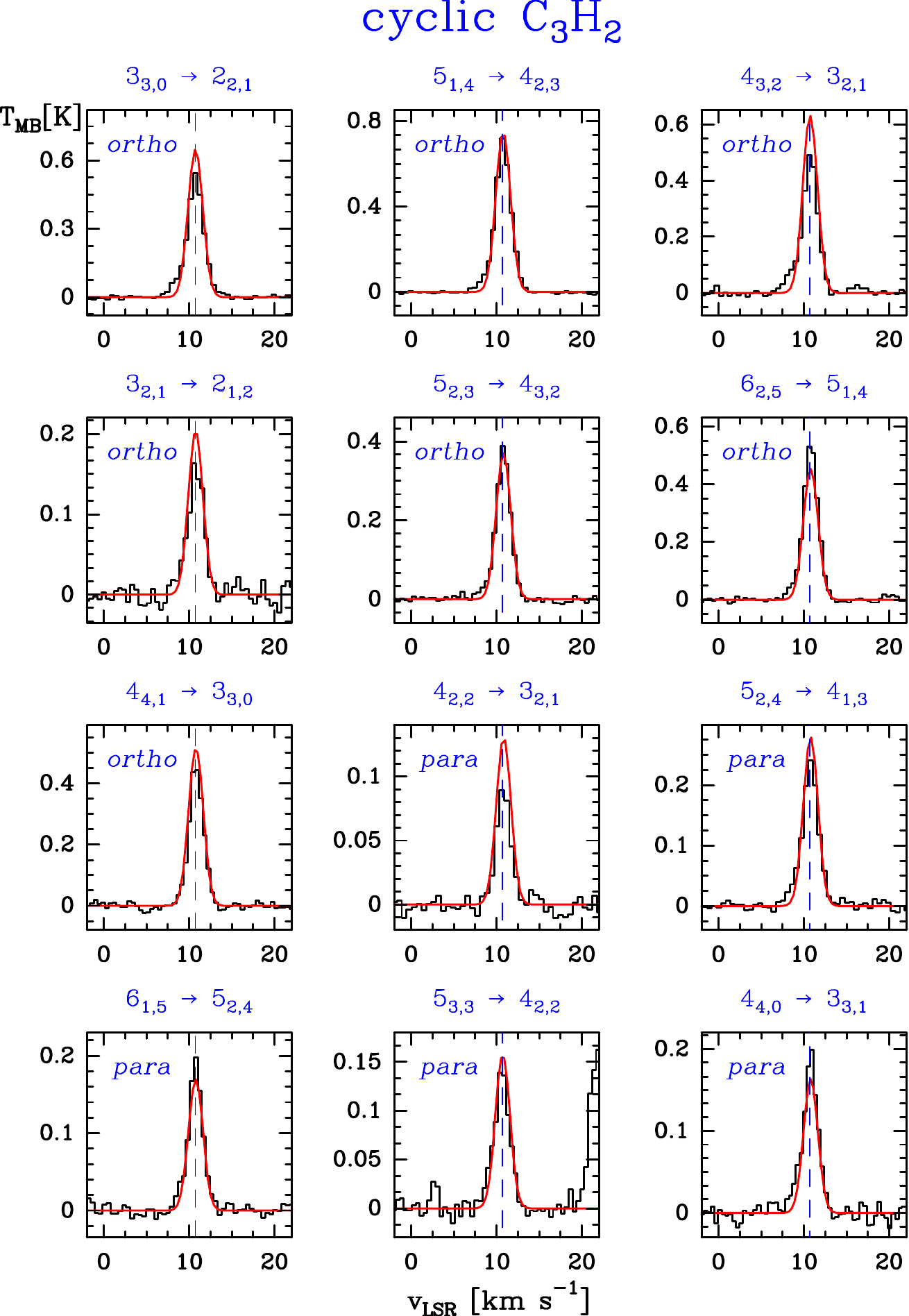}\\
\caption{Observed $c$-C$_{3}$H$_{2}$ (black histogram spectra). Best fit LVG model is shown overlaid in red (see \mbox{Sect. 5.4}). The dashed lines indicate the LSR velocity \mbox{(10.7 km s$^{-1}$)} of the Orion Bar PDR.}
\label{fig:C3H2_lines}
\end{figure}

\subsubsection{Cyclopropenylidene: \emph{Cyclic-C$_{3}$H$_{2}$}}

The cyclic isomer of C$_{3}$H$_{2}$ is a three-carbon ring with \emph{(i)} two carbon atoms linked by a double bond (semirigid structure), \emph{(ii)} one extremely reactive bivalent carbon atom which makes it highly polar (with a large dipole moment of \mbox{3.4 Debye}), and \emph{(iii)} two equivalent off-axis hydrogen atoms (responsible for the ortho-para symmetries).

The $c$-C$_{3}$H$_{2}$ molecule is an oblate asymmetric top with an asymmetric parameter \mbox{$\kappa$=-0.69}, b-type rotational transitions ($\Delta$K$_{\rm a}$ and \mbox{$\Delta$K$_{\rm c}$=$\pm$1} main selection rules), and without fine or hyperfine structure. Ortho and para levels are described by \mbox{K$_{\rm a}$ + K$_{\rm c}$ = odd} and even values, respectively. We have detected 23 rotational transitions of ortho $c$-C$_{3}$H$_{2}$ with \mbox{E$_{\rm u}$/k $\leq$ 84.6 K}, 17 rotational transitions of para $c$-C$_{3}$H$_{2}$ with \mbox{E$_{\rm u}$/k $\leq$ 54.7 K}, and 10 lines corresponding to several fully overlapped ortho-para transitions. The line profiles of spectrally resolved ortho and para lines at \mbox{1 mm} are shown in Fig.~\ref{fig:C3H2_lines}. The line parameters are listed in Table~\ref{Table_$c$-C3H2}.

\subsubsection{Propadienylidene: \emph{Linear-H$_{2}$C$_{3}$}}

In addition to being a carbene, the linear isomer is a cumulene, with a linear and rigid backbone of three carbon atoms linked by adjacent double bonds \mbox{[H$_{2}$--C=(C=)$_{n}$C:]}. Because of the linear symmetry and the low mass of the two off-axis equivalent hydrogen atoms, the rotational spectrum of propadienylidene is a nearly prolate top with an asymmetric parameter \mbox{$\kappa$=-0.997} and a-type R-branch selection rules which involve $\Delta$J=1, $\Delta$K$_{\rm a}$=0, and $\Delta$K$_{\rm c}$=1 transitions. Para levels have even values of K$_{\rm a}$, while ortho levels have odd values.

\begin{figure}
\centering
\includegraphics[scale=0.62,angle=0]{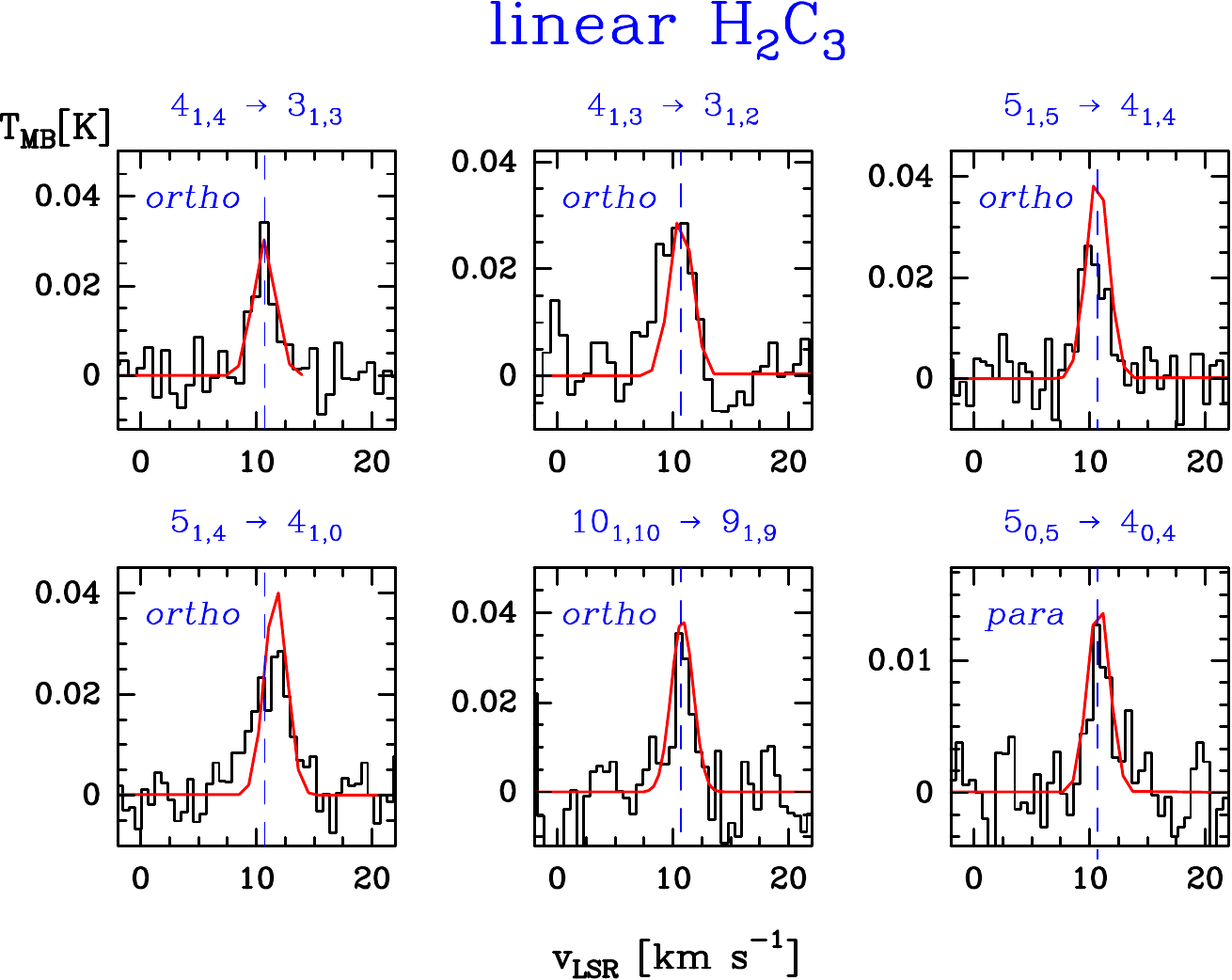}\\
\caption{Observed $l$-H$_{2}$C$_{3}$ (black histogram spectra). Best fit LTE model is shown overlaid in red (see Sect.~5.2). The dashed lines indicate the LSR velocity \mbox{(10.7 km s$^{-1}$)} of the Orion Bar PDR.}
\label{fig:H2C3_lines}
\end{figure}

We observed the lowest para ladder (J$_{0,\rm J}$ $\rightarrow$ J-1$_{0,\rm J-1}$) with rotational transitions separated by \mbox{$\sim$20.8 GHz}, and the lowest ortho ladder (K$_{\rm a}$=1), that shows K-type doublets with the two lines displaced by several GHz. Transitions with K$_{\rm a} \negthickspace > \negthickspace 1$ have not been detected. We have identified 13 lines of ortho species, but only 6 lines for the para species (both with \mbox{E$_{\rm u}$/k $\leq$ 90.8 K)}. The line profiles of spectrally resolved ortho and para lines at \mbox{1 mm} are shown in Fig.~\ref{fig:H2C3_lines}. Table~\ref{Table_$l$-C3H2} gives the observed line parameters.

{\bf
\subsection{C$_{\bf 4}$H}
}
The linear butadiynyl, C$_{4}$H, is a univalent radical and the simplest example of polyynes, a type of organic species in which molecular structure have single and triple bonds in alternate positions \mbox{(--C$\equiv$C--)$_{n}$}, with n$>$1. This molecule was first identified in the circumstellar envelope of IRC+10216 \citep{Guelin_1978}. Its laboratory microwave spectrum in the ground vibrational state was first obtained by \citet{Gottlieb_1983b}.

\begin{figure}
\centering
\vspace{0.62cm}
\includegraphics[scale=0.425,angle=0]{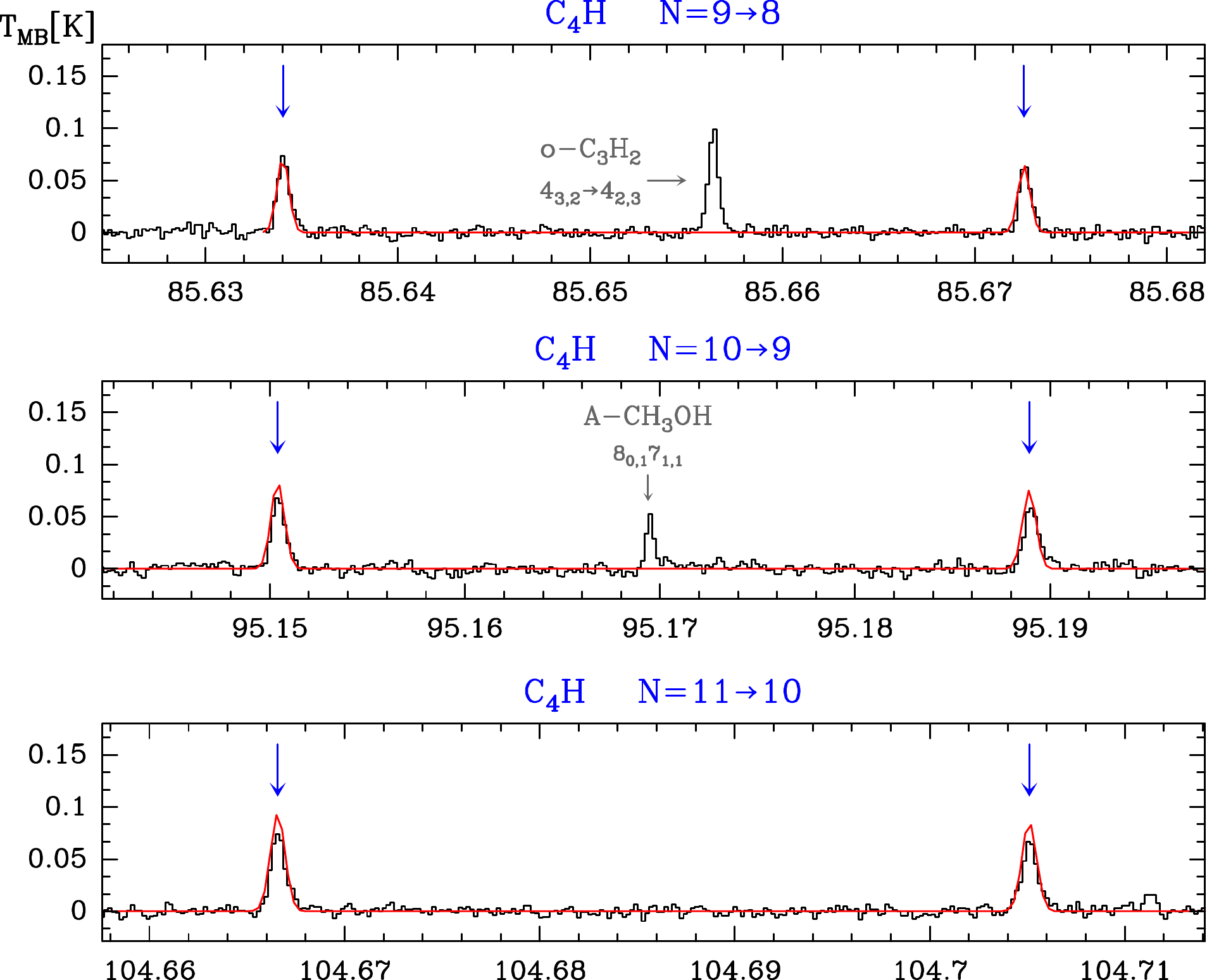} \\
\vspace{0.32cm}
\includegraphics[scale=0.425,angle=0]{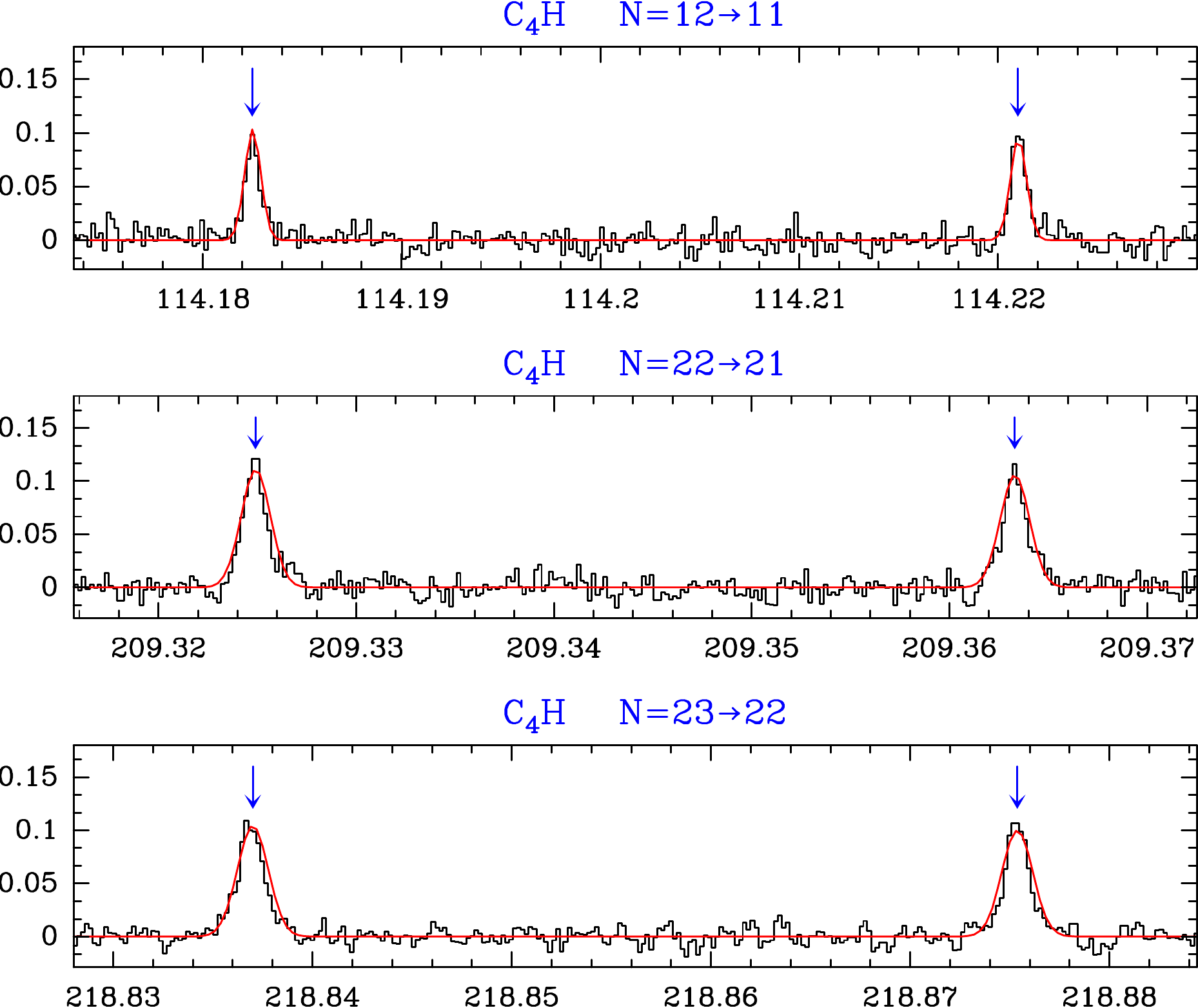} \\
\vspace{0.32cm}
\includegraphics[scale=0.425,angle=0]{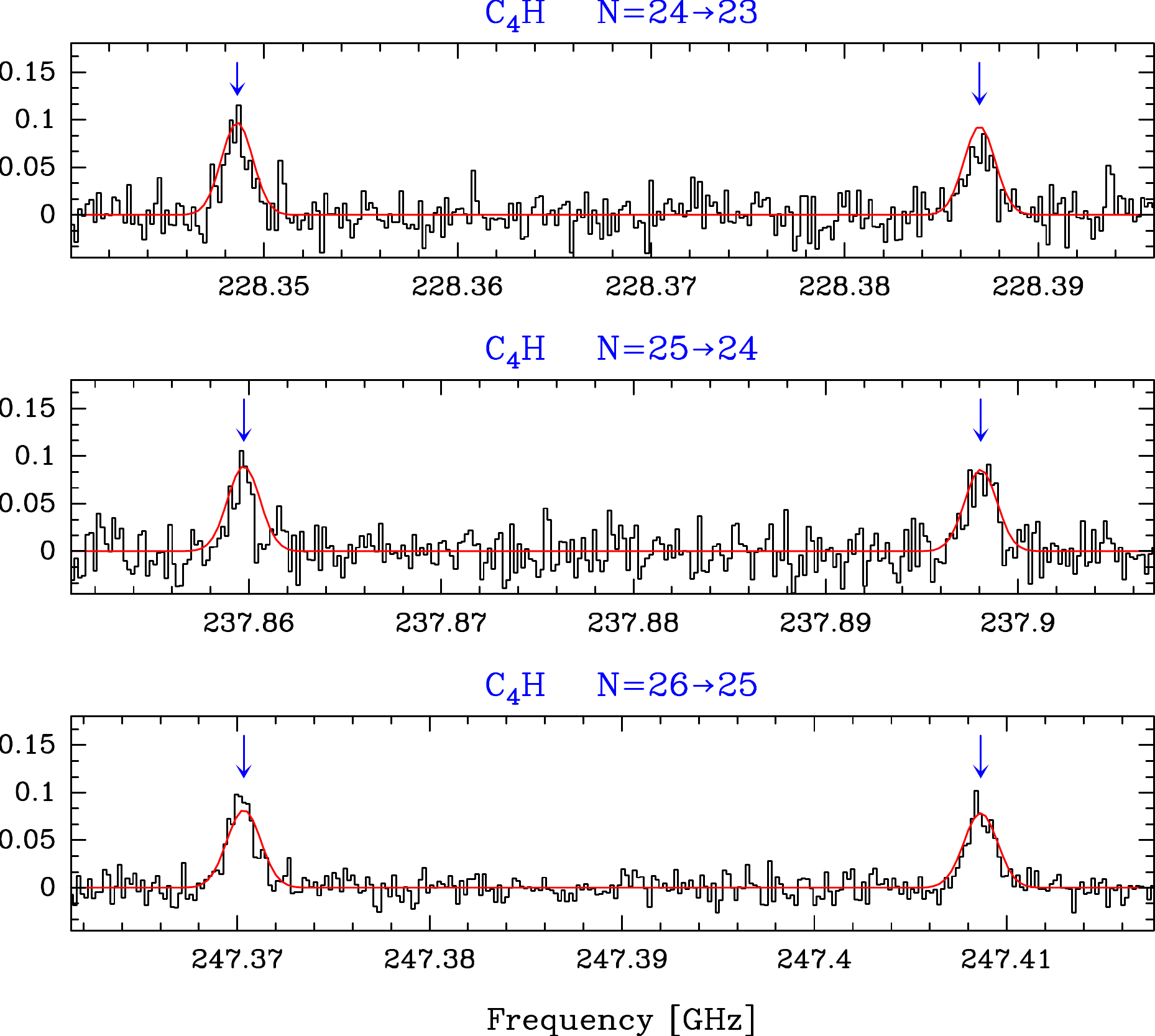} \\
\caption{C$_{4}$H spectra observed in the Orion Bar (black histogram spectra). A LTE model is overlaid in red (see Sect~.5.2). Line doublets are indicated by the blue arrows. The other spectral features appearing in the selected windows are labelled with their corresponding identification.}
\label{fig:C4H_lines}
\end{figure}

The C$_{4}$H radical has the same symmetry and electronic ground state ($^{2}\Sigma^{+}$) as C$_{2}$H, therefore its rotational levels are also split into fine and hyperfine structure. This hyperfine splitting is only fully resolved in the lowest rotational transitions \citep[e.g.][]{Bell_1982,Bell_1983,Guelin_1982}. As a consequence, in the studied spectral range, the strongest rotational transitions are doublets. Forty lines of C$_{4}$H consisting of 20 successive doublets starting from N=9$\rightarrow$8 have been identified in this work, most of them with high spectral resolution. The C$_{4}$H doublets show spectral features that allow us to distinguish them easily:
 \emph{(i)} each pair of lines is separated by $\sim$38.5 MHz, and
 \emph{(ii)} the lowest frequency transition of each doublet is slightly stronger than the second. This is in perfect agreement with the theoretical line strength. Figure~\ref{fig:C4H_lines} shows some of the strongest lines detected at high spectral resolution in the \mbox{3 mm} and \mbox{1 mm} bands. Assignments of the line components, as well as the spectroscopic and observed line parameters of the detected lines are given in Table~\ref{Table_C4H}.\\


\section{Spatial distribution of C$_{\bf 2}$H, $c$-C$_{\bf 3}$H$_{\bf 2}$, and PAHs}

\begin{figure*}
 \centering

\includegraphics[scale=0.425,angle=0]{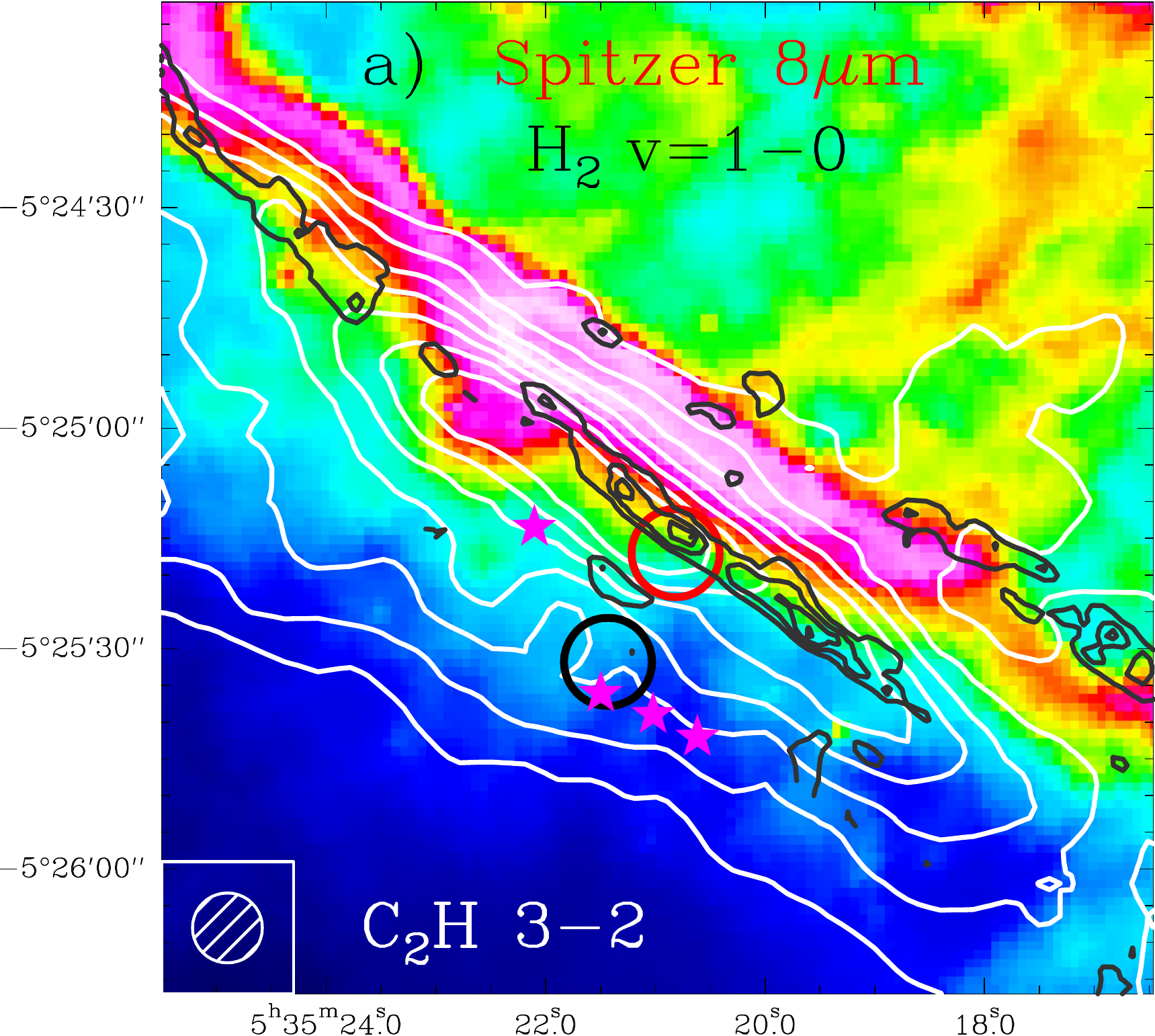} 
\hspace{0.35cm} \vspace{0.2cm}
\includegraphics[scale=0.425,angle=0]{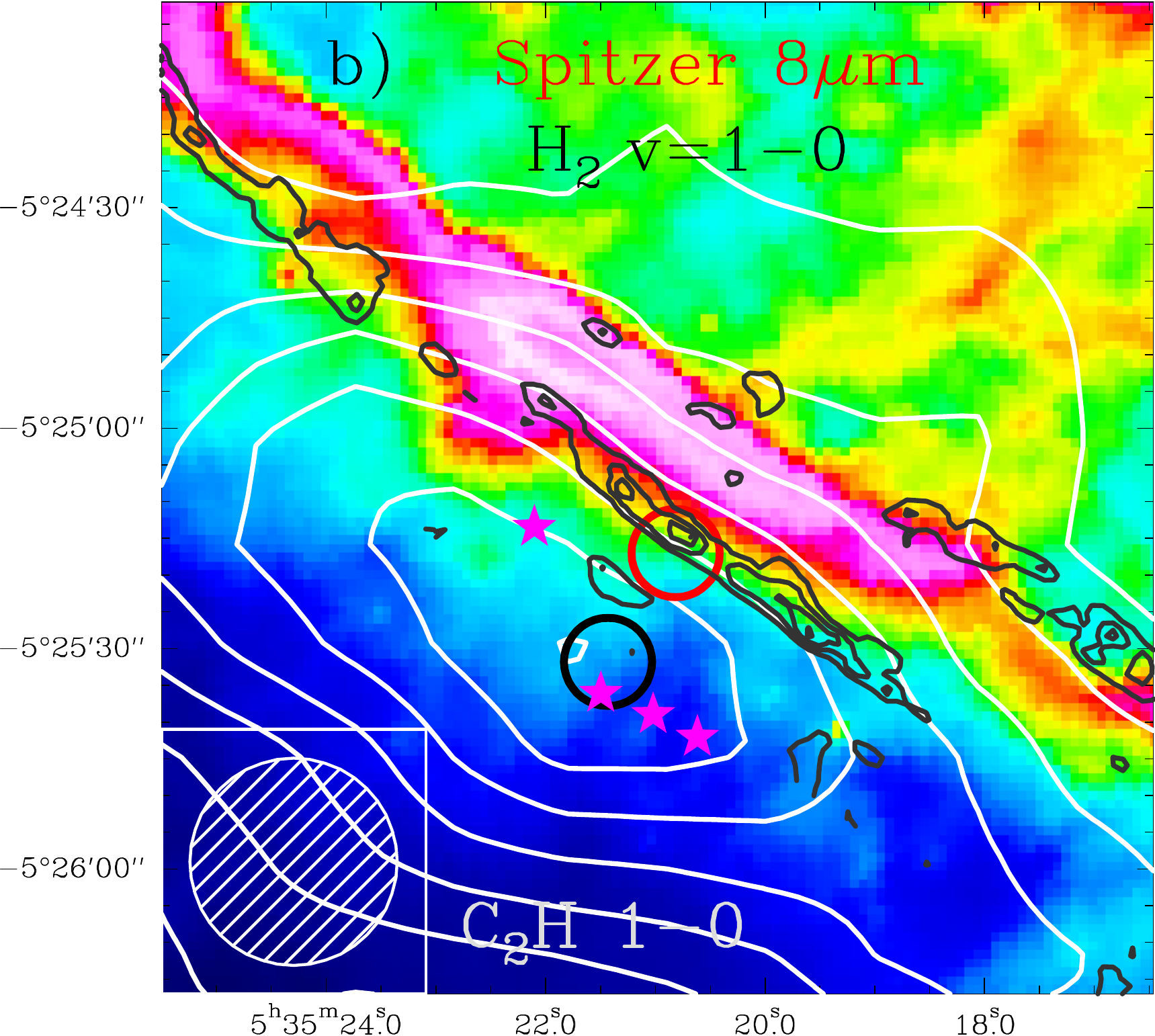} \\
\vspace{0.2cm}
\includegraphics[scale=0.425,angle=0]{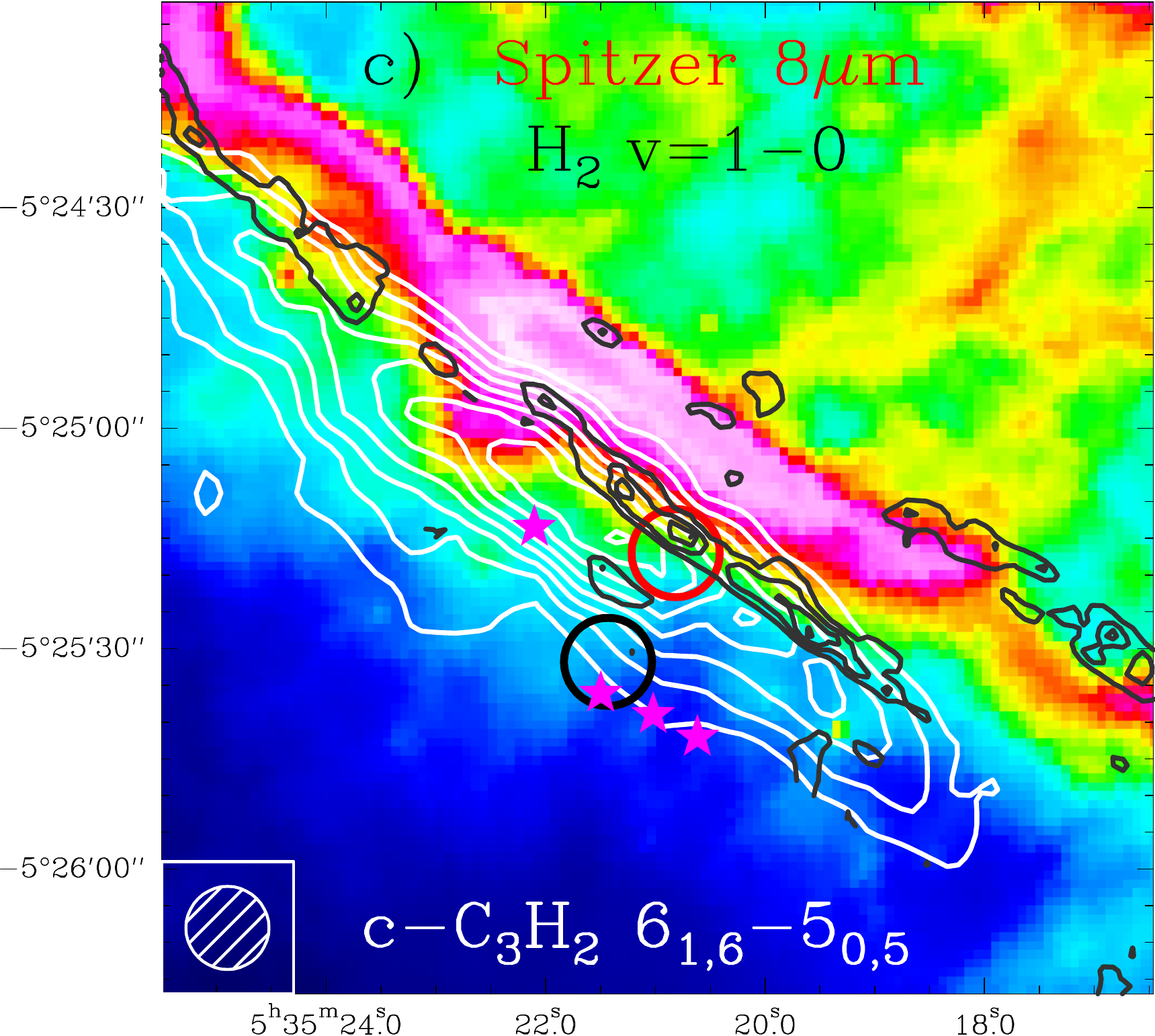} 
\hspace{0.35cm}
\includegraphics[scale=0.425,angle=0]{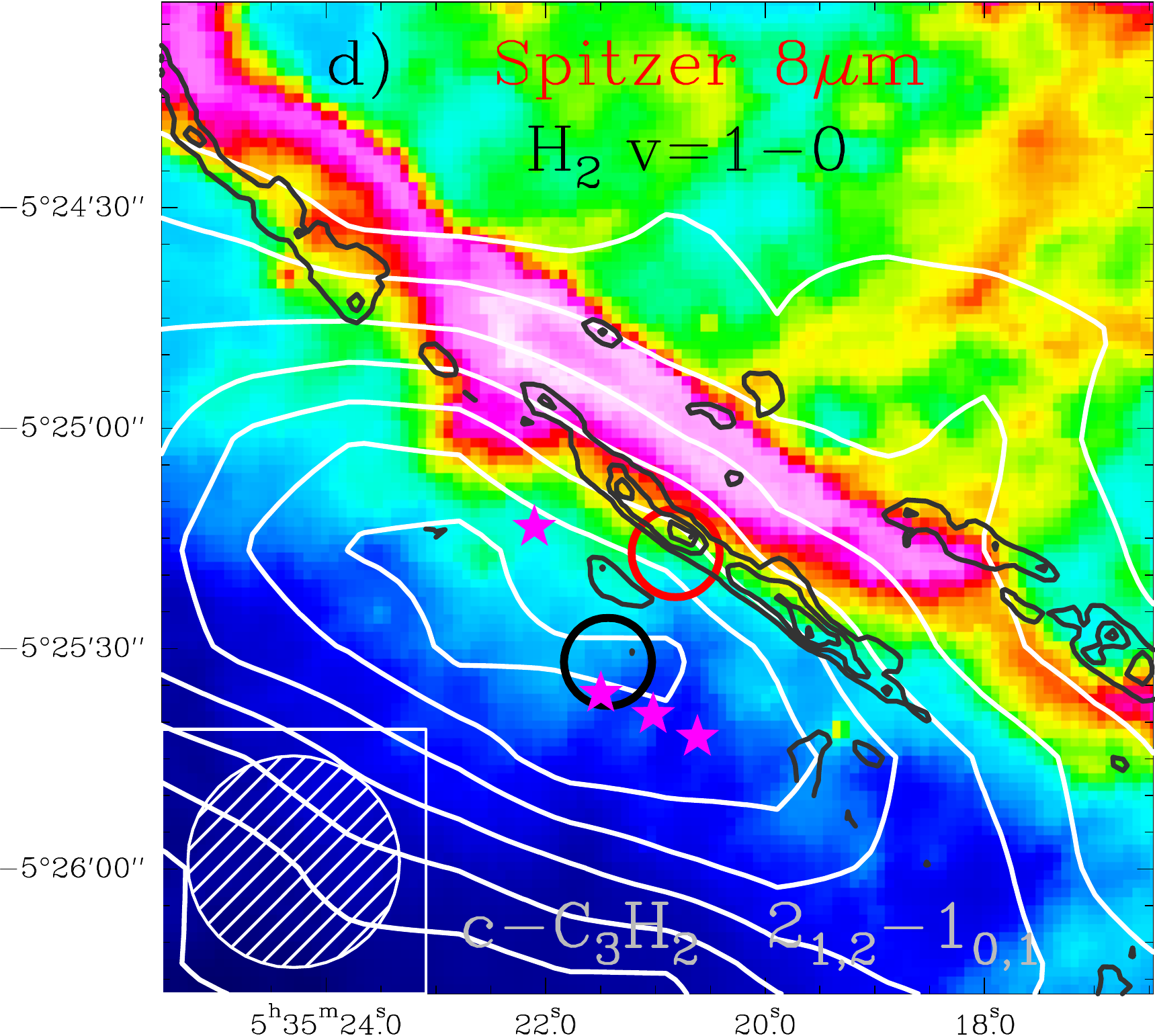} \\
\caption{C$_{2}$H and $c$-C$_{3}$H$_{2}$ integrated line-intensity maps (white contours) of the Orion Bar. \emph{Upper panels:} C$_{2}$H \emph{(a)} N=3$\rightarrow$2 at \mbox{262.0 GHz}, and \emph{(b)} N=1$\rightarrow$0 at \mbox{87.3 GHz}. \emph{Lower panels:} $c$-C$_{3}$H$_{2}$ \emph{(c)} 6$_{1,6}\negthickspace\rightarrow$5$_{0,5}$ at \mbox{217.8 GHz}, and \emph{(d)} 2$_{1,2}\negthickspace\rightarrow$1$_{0,1}$ at \mbox{85.3 GHz}. The Spitzer \mbox{8 $\mu$m} extended emission due to PAHs and very small grains is in colour scale. Black contours are the H$_{2}^{*}$ $\nu$=1$\rightarrow$0 emission \citep{Walmsley_2000}. Stars represent the positions of denser clumps/condensations detected in 
H$^{13}$CN J=1$\rightarrow$0 \citep{Lis_2003}. The IRAM 30m beams at \mbox{1 mm} and \mbox{3 mm} are plotted in the bottom left corner (white striped circle). The target position of the Orion Bar survey, close to the dissociation front, and the molecular peak position beyond the PDR (clump no.~10 of  \citet{Lis_2003}; see text for discussion) are  indicated with a red and a black circle, respectively. The emission of all lines from C$_{2}$H and $c$-C$_{3}$H$_{2}$ is integrated in the \mbox{10-12 km s$^{-1}$} velocity interval in which the Orion Bar shows prominent emission.}\label{fig:maps}
 \end{figure*}

Early mapping observations of the Orion Bar \citep[e.g.][]{Tielens_1993,vanderWerf_1996} confirmed the 
PAH/H$_{2}$/CO spatial stratification predicted by PDR models \citep[e.g.][]{Tielens_1985a}.
Figure~\ref{fig:maps} shows the IRAC \mbox{8 $\mu$m} band emission along the Orion Bar taken from the Spitzer archive (colour scale). 
Analysing the ISOCAM mid-IR spectrum of the Bar with the fitting tool PAHTAT \citep{Pilleri_2012} we estimate that at 
least $\sim$50-80$\%$ of the IRAC \mbox{8 $\mu$m} filter band emission is produced by PAHs. As in other high UV-flux PDRs,
the PAH emission delineates the atomic zone edge of the Orion Bar \citep[see e.g.][for Mon~R2 and NGC7023, respectively]{Berne_2009,Joblin_2010}.
The molecular dissociation front traced by the vibrationally excited
H$_{2}$ $\nu$=1$\rightarrow$0 S(1) emission (H$_{2}^{*}$; black contours in Fig.~\ref{fig:maps};  \citealt{Walmsley_2000}) and also the [\CII]\,158$\mu$m intensity maxima \citep{Bernard-Salas_2012,Ossenkopf_2013} peak slightly deeper in the cloud than the PAHs.

\begin{figure}[!b]
\centering
\includegraphics[scale=0.51,angle=0]{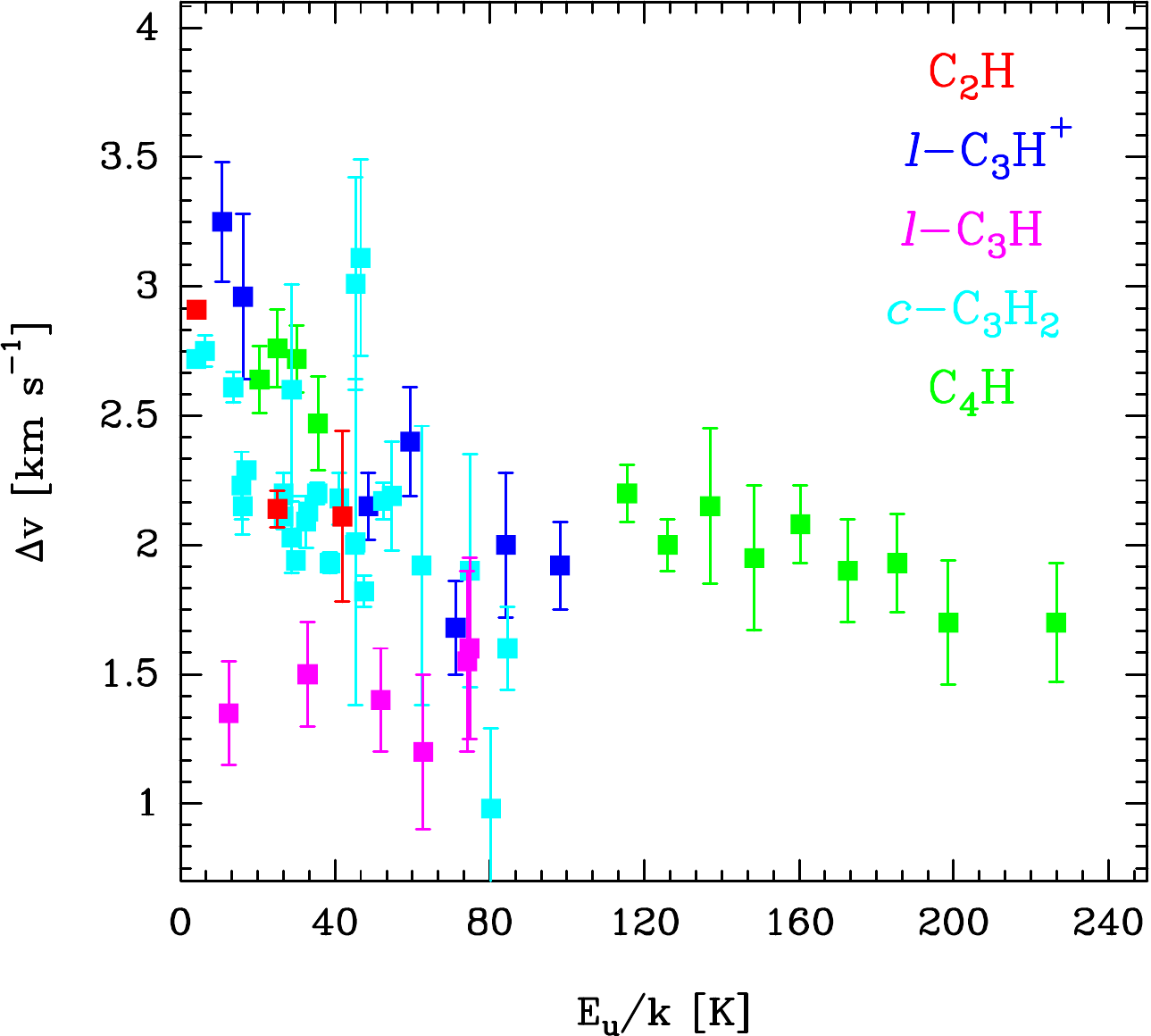} 
\caption{Plot of hydrocarbon line widths versus transition upper level energy for C$_{2}$H, $l$-C$_{3}$H$^{+}$, $l$-C$_3$H, $c$-C$_{3}$H$_{2}$, and C$_{4}$H lines.}\label{fig:anchuras}
\end{figure}

The increased sensitivity and broader frequency coverage of (sub)mm receivers led to the opportunity to 
map the faint emission of trace chemical species. Figure~\ref{fig:maps} also shows the integrated line intensity maps of C$_{2}$H (N=1$\rightarrow$0 and 3$\rightarrow$2 lines at \mbox{87.3 GHz} and \mbox{262.0 GHz}, respectively) and \mbox{$c$-C$_{3}$H$_{2}$} (J$_{\rm K_a,K_c}$=2$_{1,2}\negthickspace\rightarrow$1$_{0,1}$ and 6$_{1,6}\negthickspace\rightarrow$5$_{0,5}$ lines at \mbox{85.3 GHz} and \mbox{217.8 GHz}, respectively) in white contours. The emission from the Orion Bar can be distinguished more easily from the extended OMC1 cloud component by the emission LSR velocity. While OMC1 is brighter in the \mbox{8-10 km s$^{-1}$} velocity range, the Orion Bar emits predominantly in the \mbox{10-12 km s$^{-1}$} range. The C$_{2}$H and $c$-C$_{3}$H$_{2}$ emission contours shown in Fig.~\ref{fig:maps} are integrated in this interval. Both species show a similar distribution delineating the bar structure of the PDR. However, the morphology of the emission depends on the involved transition energy level, with the more excited lines (those at \mbox{1 mm)} peaking closer to the cloud edge. The elongated spatial distribution of the hydrocarbon emission is parallel to the H$_{2}^{*}$ emission, with the \mbox{1 mm} C$_{2}$H and $c$-C$_{3}$H$_{2}$ lines peaking close to the H$_{2}^{*}$ intensity peaks. Despite the similar distance,  
the spatial stratification \mbox{[PAH]/[C$^{+}$-H$_{2}^{*}$-Hydrocarbons]} is more clearly seen towards the Orion Bar
than towards low UV-flux PDRs  seen almost edge on \citep[e.g. the Horsehead;][]{Pety_2005}.
We note that \citet{Wiel_2009} had previously mapped the  C$_{2}$H 
N=4$\rightarrow$3 lines with the JCMT telescope and shown that they
peak closer to the dissociation front than other (higher energy) lines from SO or H$_2$CO. Hence, this is a true chemical stratification effect
 that confirms that C$_2$H is efficiently produced at the edge of the PDR.
The lower excitation lines of both C$_{2}$H and \mbox{$c$-C$_{3}$H$_{2}$} 
(those at \mbox{3 mm)}, however, peak deeper inside the cloud where the gas is less exposed to the FUV-radiation field 
and clumps/condensations of dense material and large $N$(H$_2$) column densities
are known to exist \citep{Lis_2003}.


\section{Analysis}

\begin{figure*}
\centering
\includegraphics[scale=0.4, angle=0]{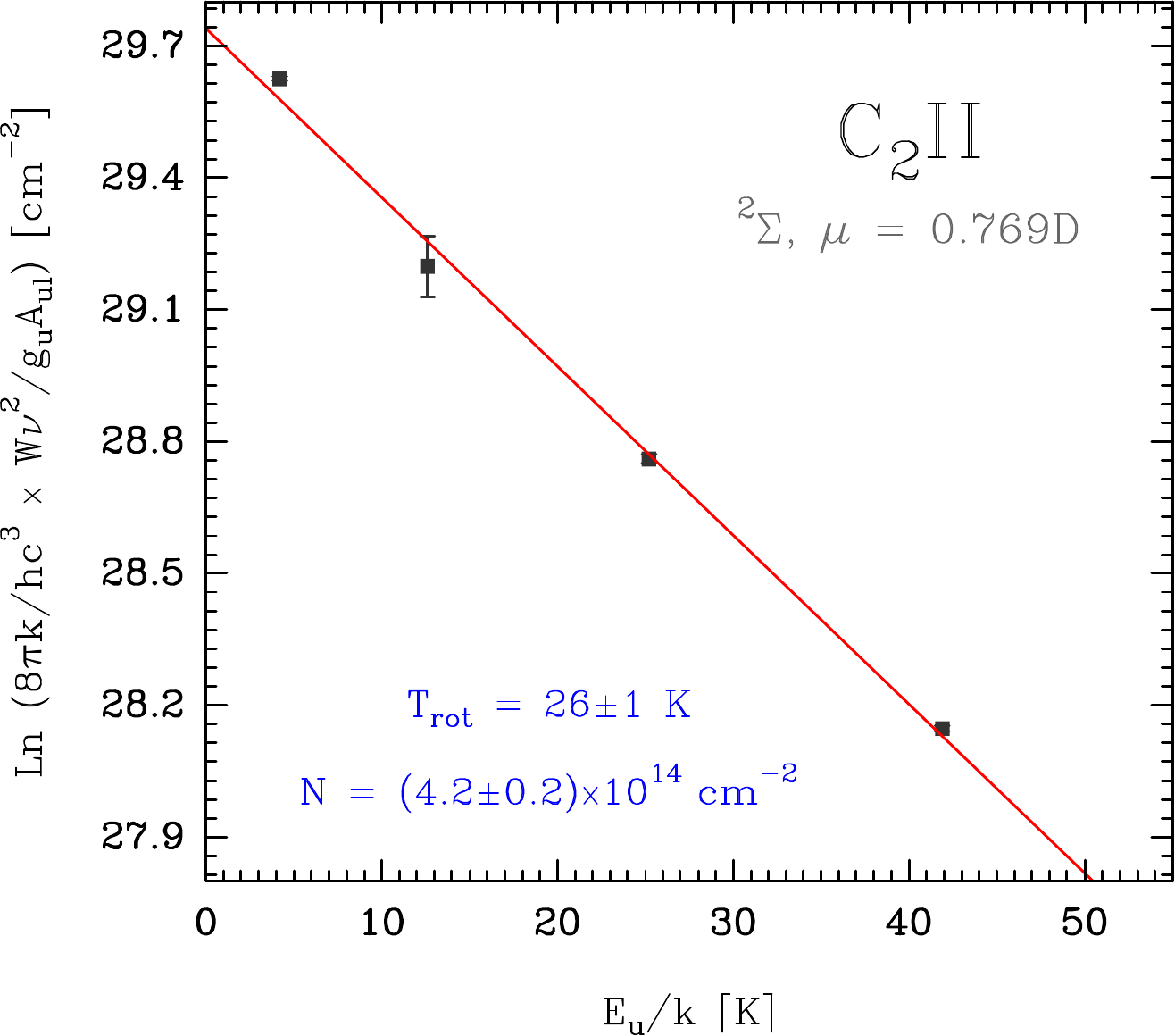} \hspace{0.5cm}
\includegraphics[scale=0.4, angle=0]{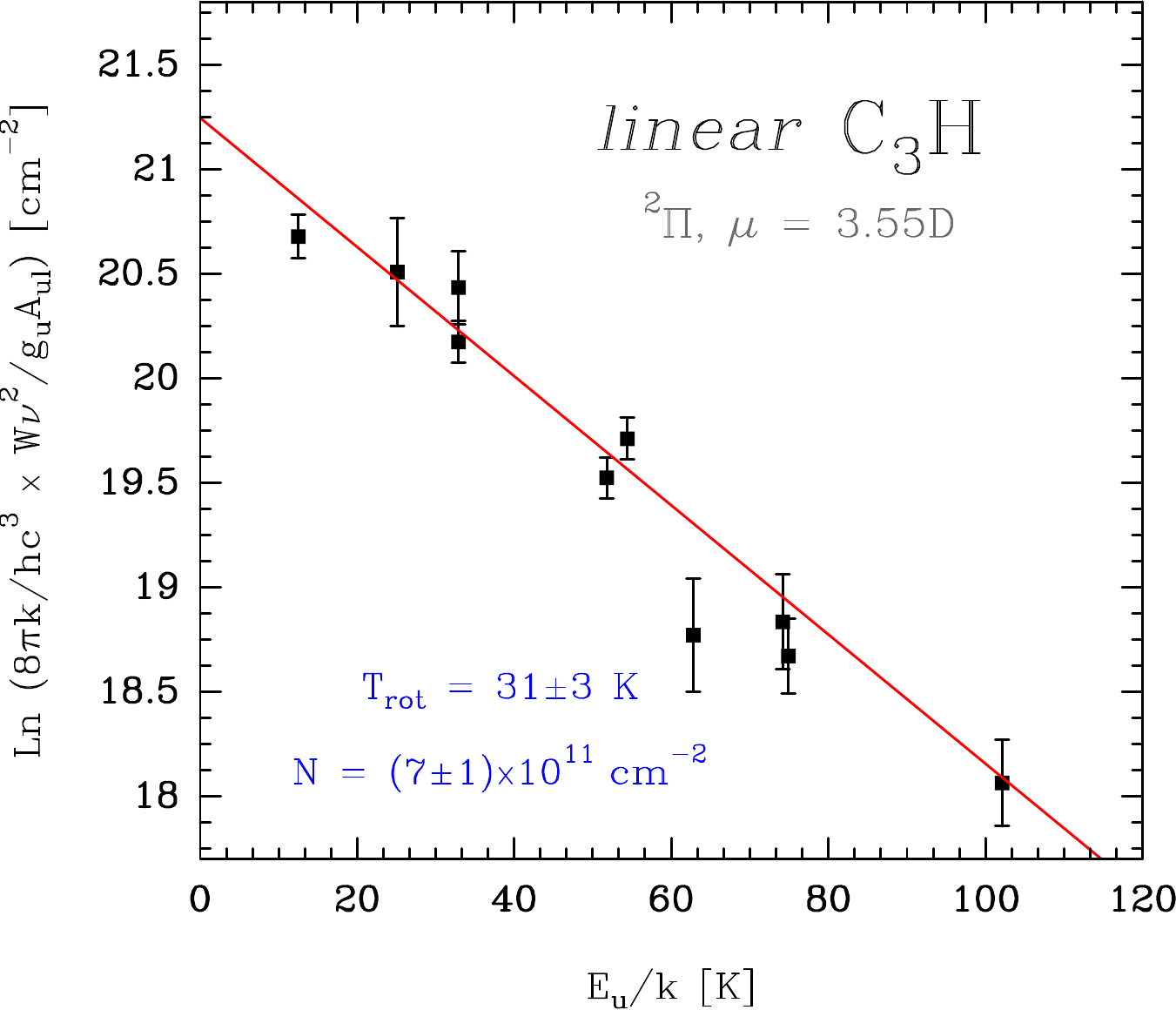}  \hspace{0.5cm}
\includegraphics[scale=0.4, angle=0]{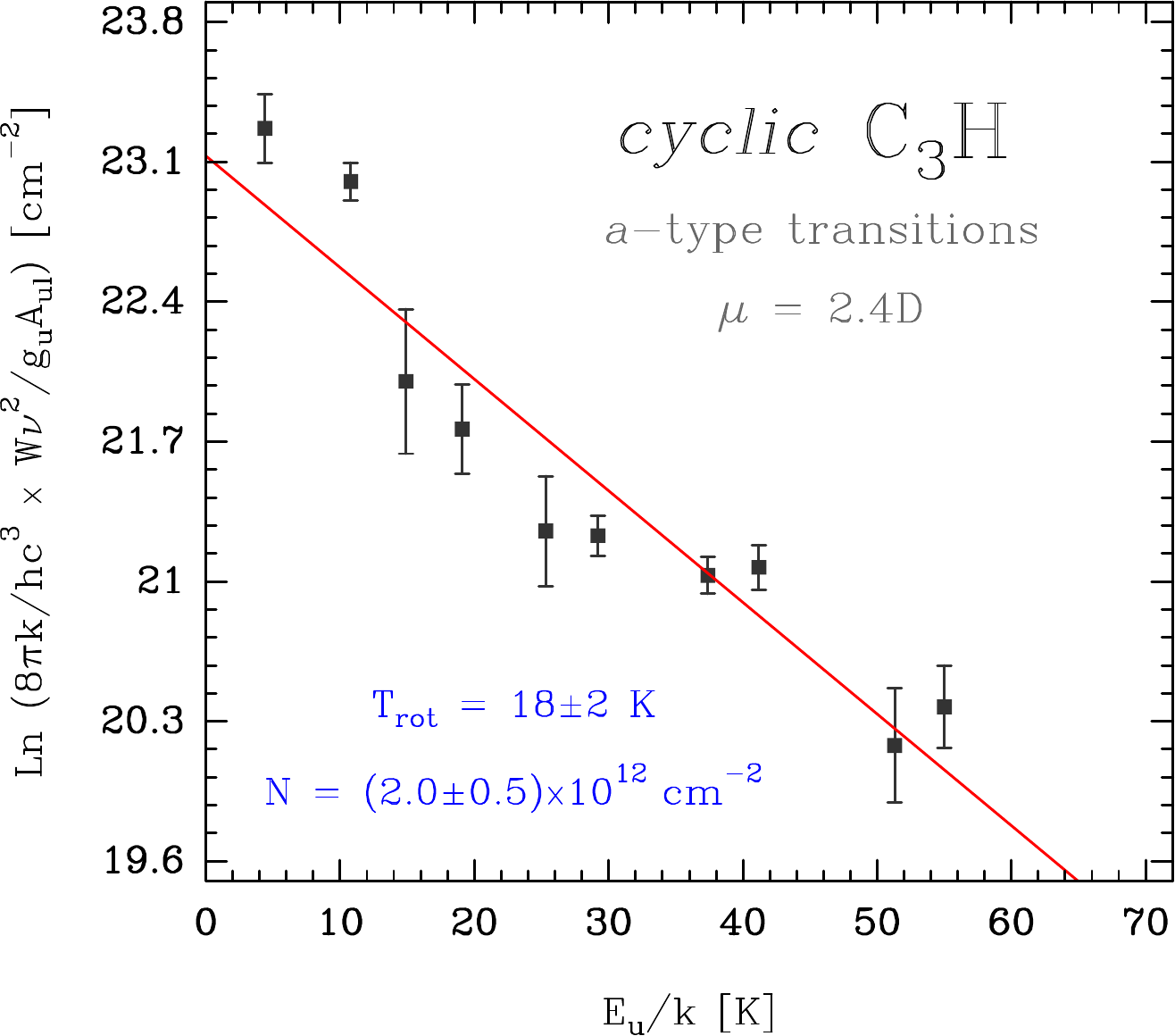}  \\ 
\vspace{0.3cm}
\includegraphics[scale=0.4, angle=0]{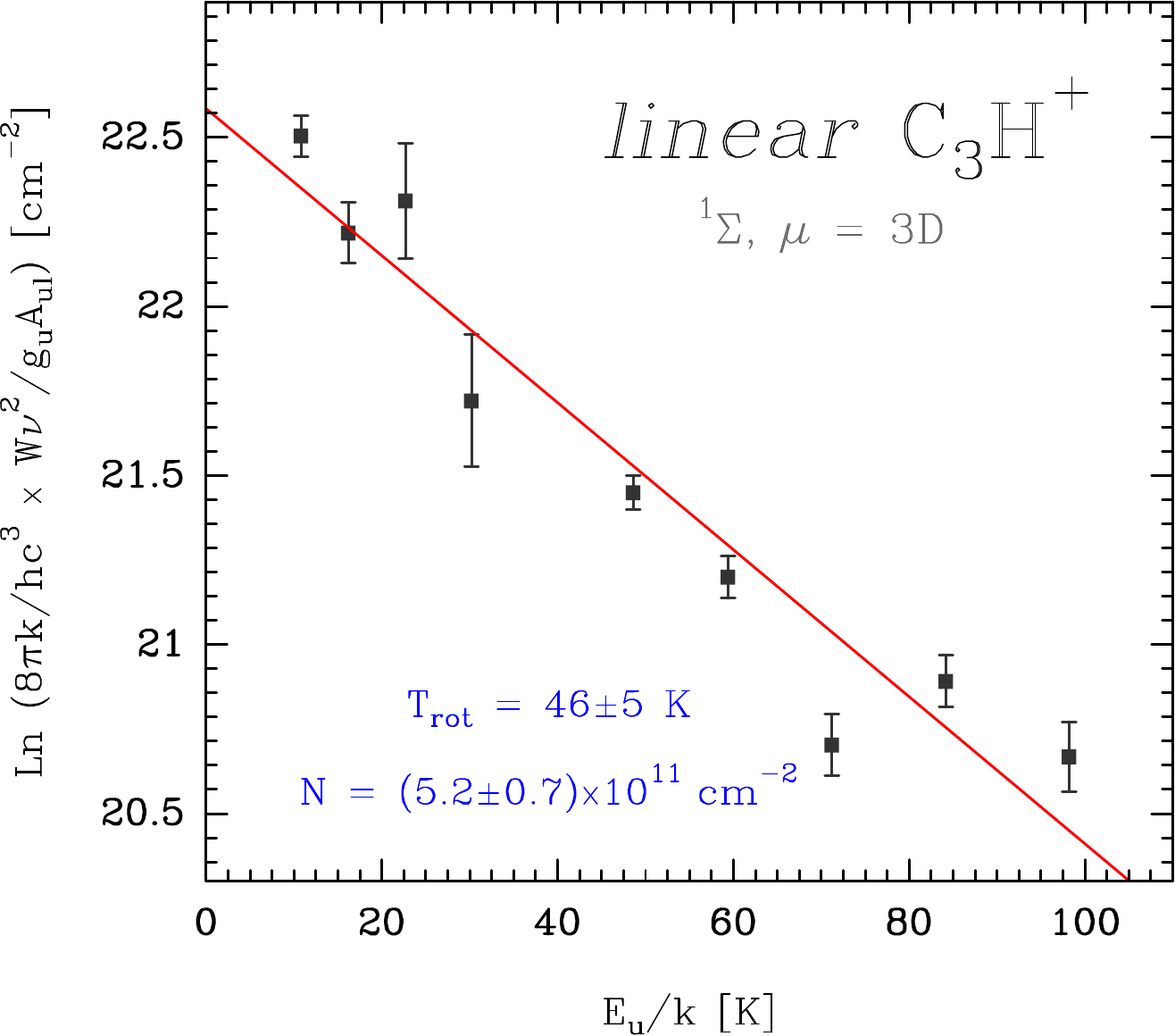} \hspace{0.5cm}
\includegraphics[scale=0.4, angle=0]{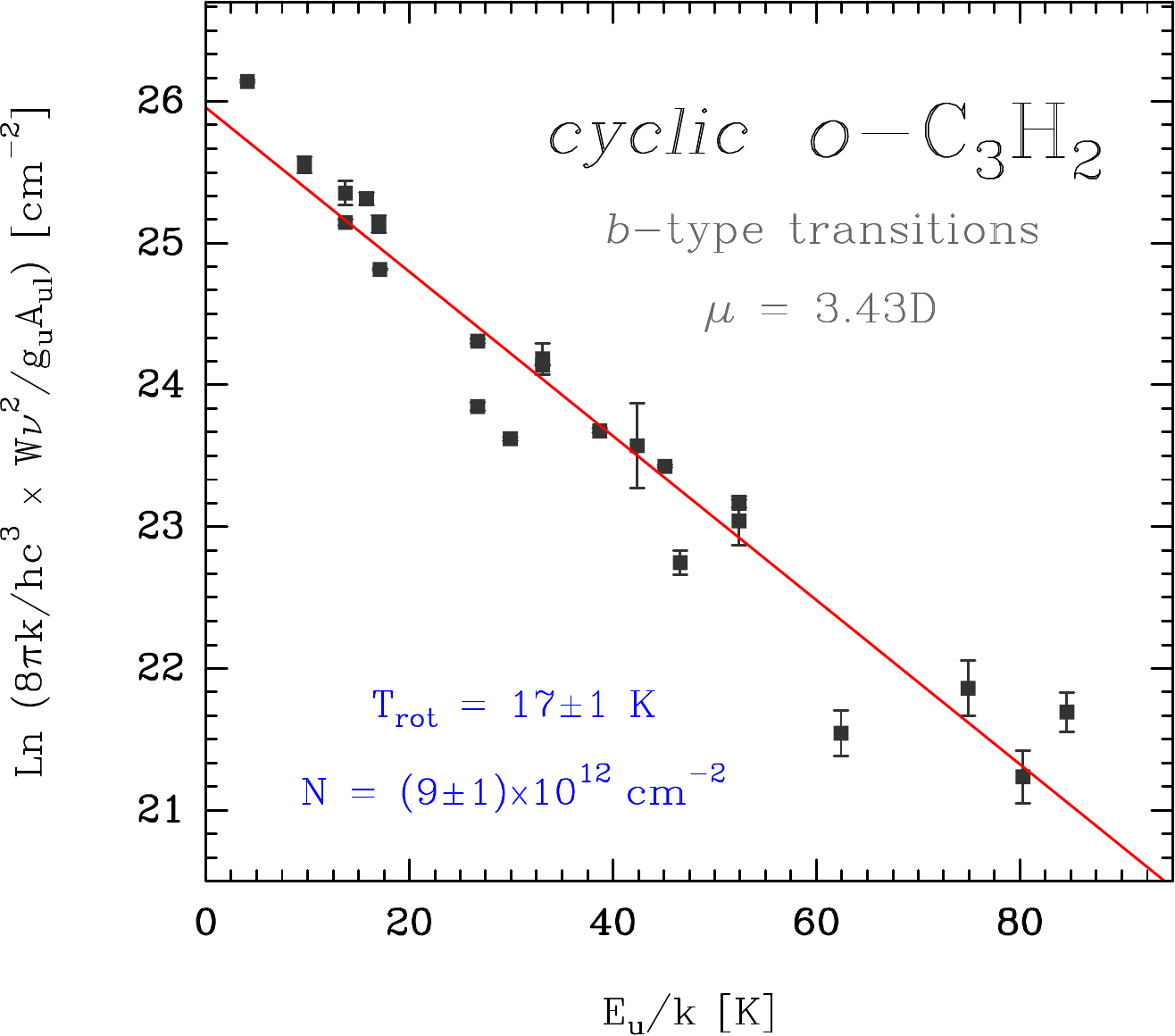} \hspace{0.5cm}
\includegraphics[scale=0.4, angle=0]{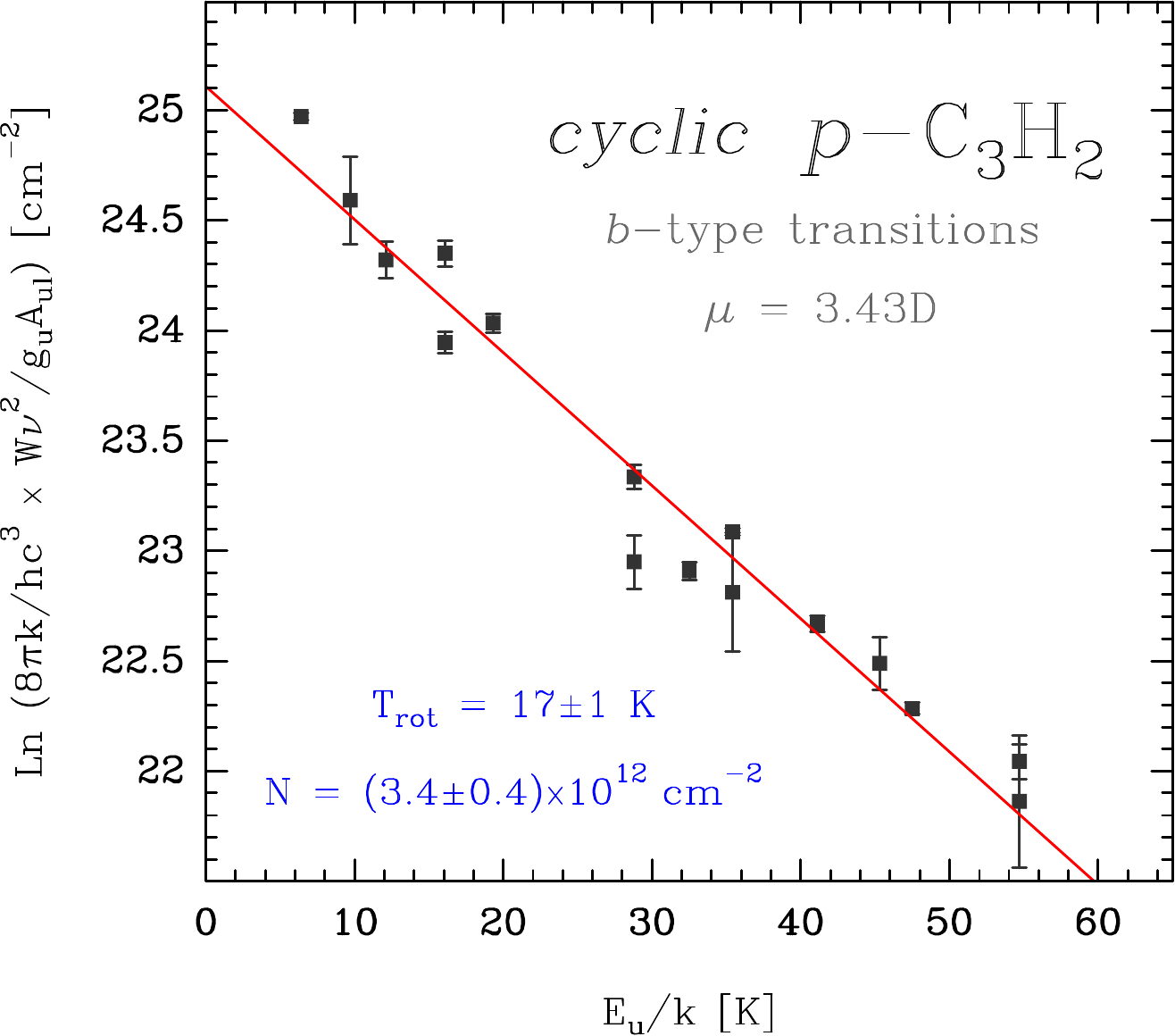}  \\
 \vspace{0.3cm}
\includegraphics[scale=0.4, angle=0]{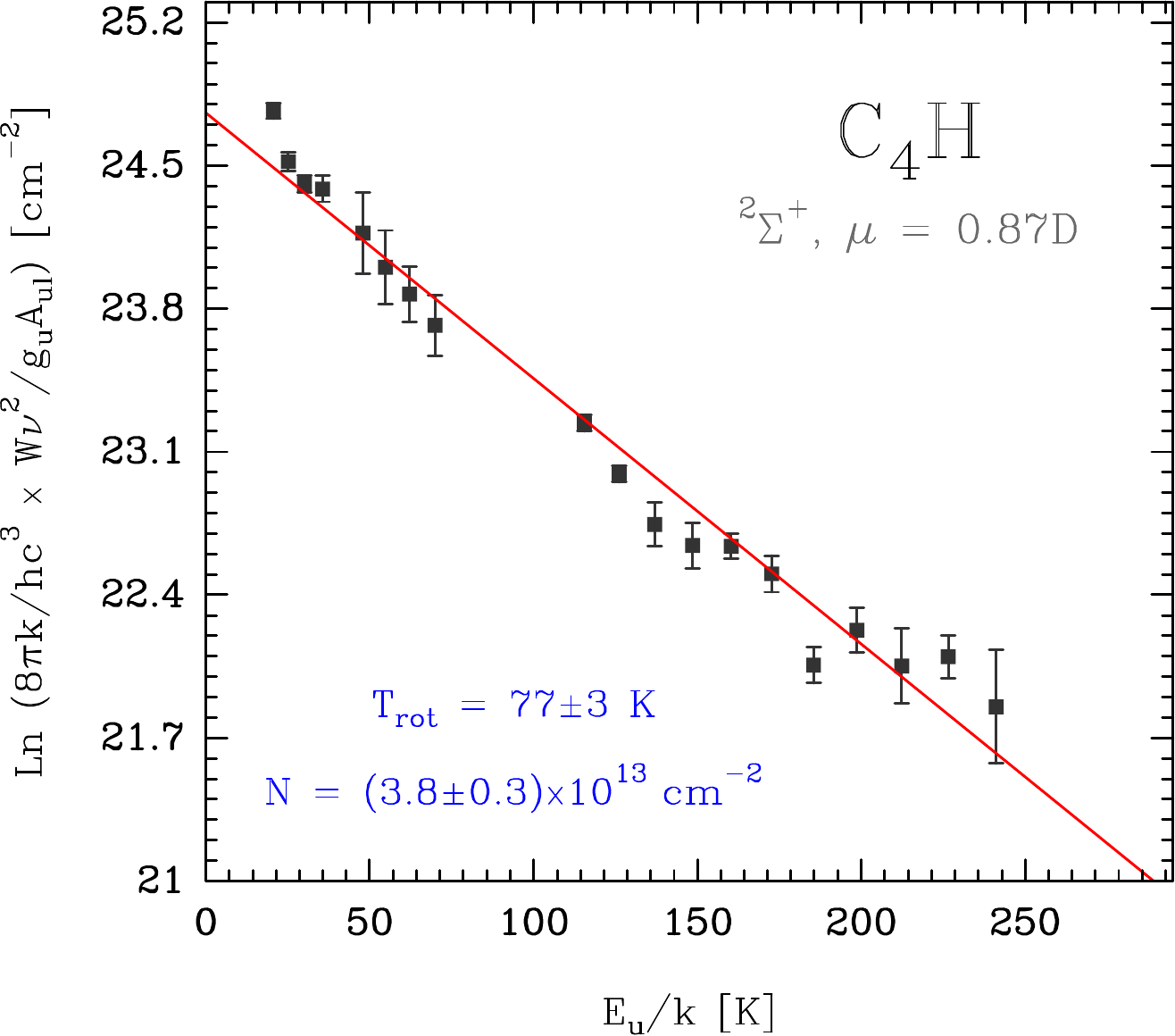} \hspace{0.5cm}
\includegraphics[scale=0.4, angle=0]{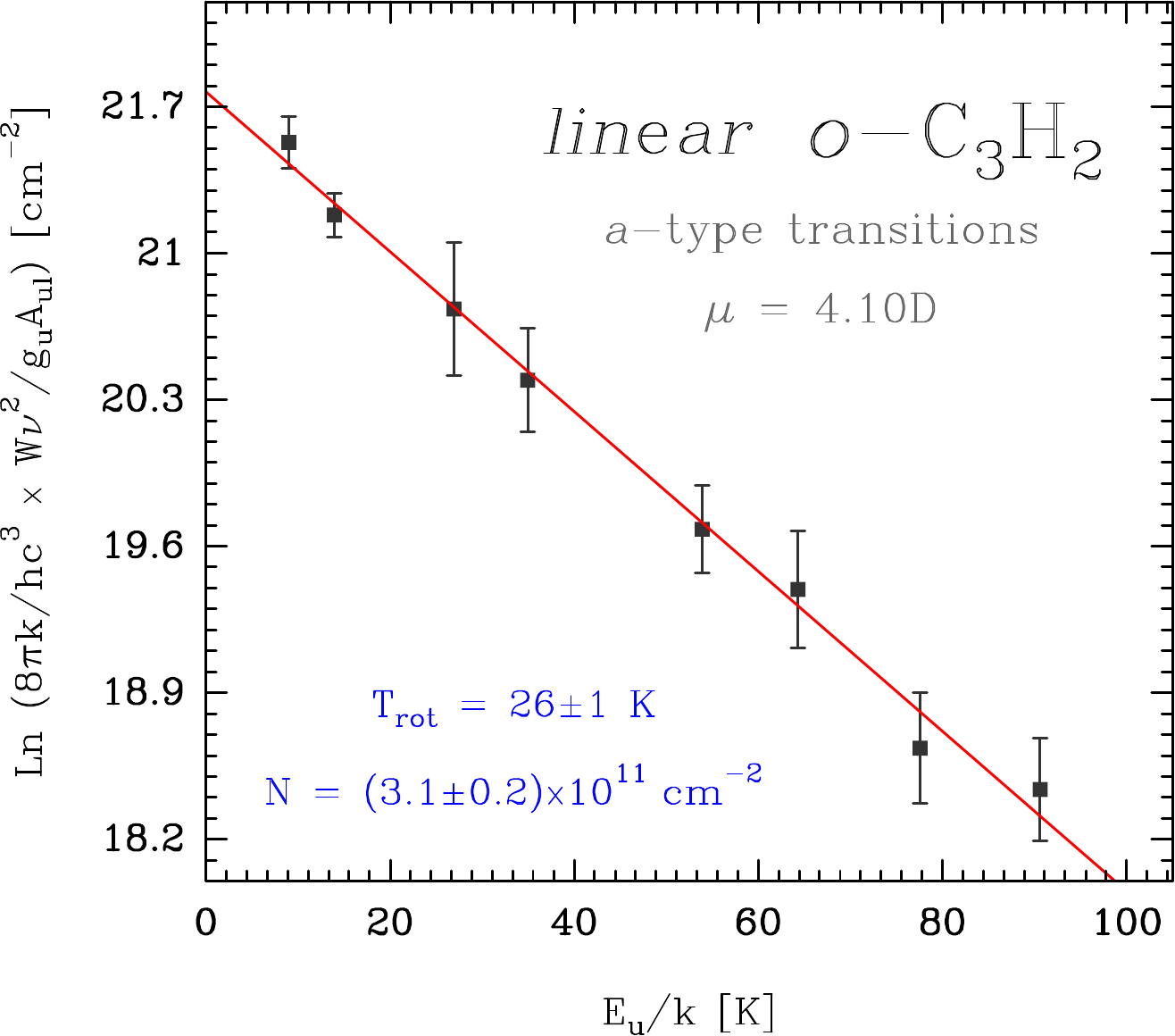}  \hspace{0.5cm}
\includegraphics[scale=0.4, angle=0]{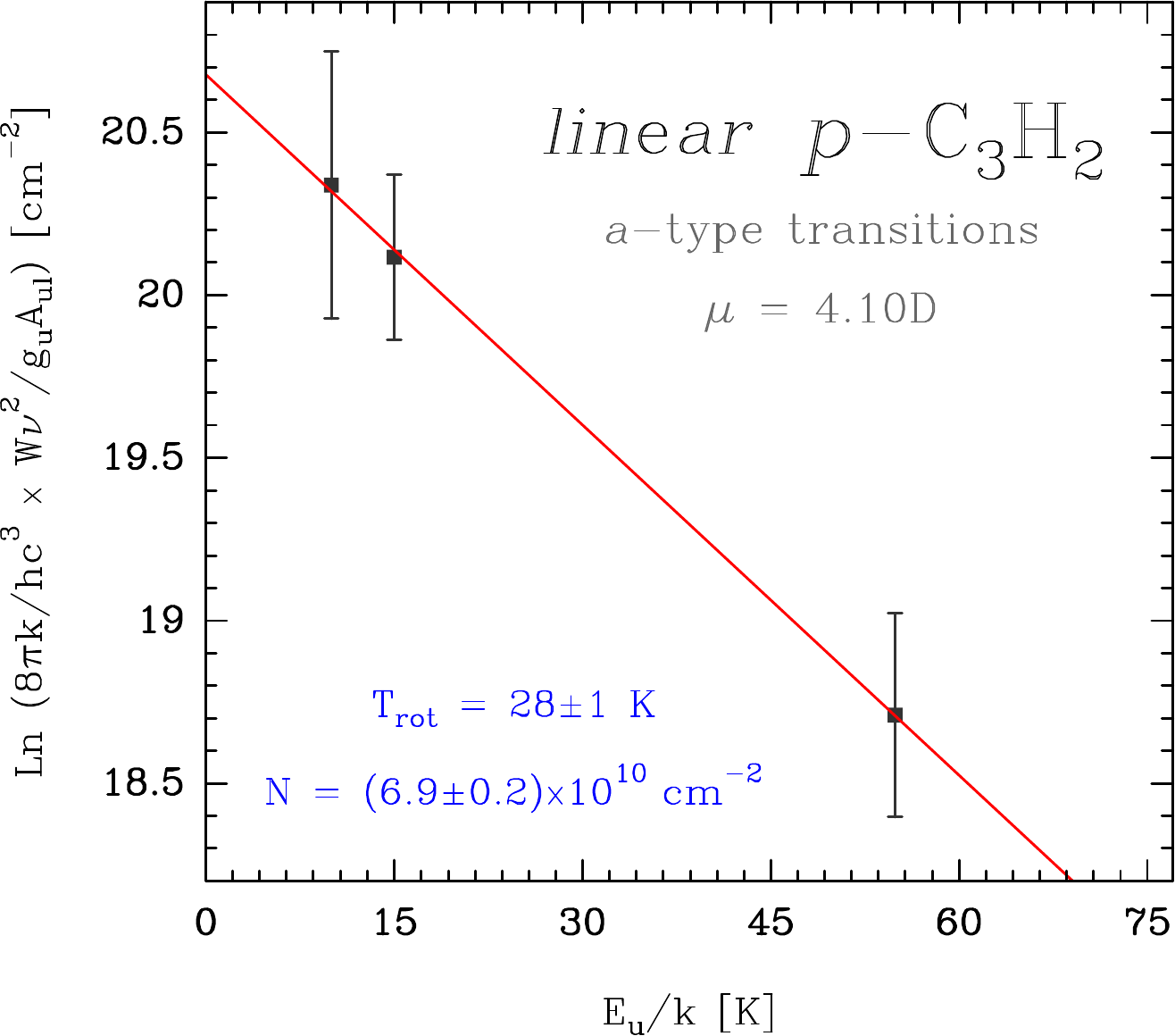}   \\
\caption{Rotational diagrams of the detected hydrocarbon molecules in the Orion Bar PDR. Fitted values of the rotational temperature, $T_{\rm rot}$, column density, $N$, and their respective uncertainties are also indicated for each molecule.}\label{fig:DR}
\end{figure*}

The large number of lines of the different species detected in the 3, 2, 1, and \mbox{0.8 mm} surveys allows us to carry out a detailed analysis of their excitation to \emph{(i)} constrain rotational temperatures and column densities accurately, and \emph{(ii)} investigate the physical conditions for those species with known collisional rates. We first extracted the line profile parameters by Gaussian fits. Second, we calculated rotational temperatures and column densities by constructing rotational population diagrams for each detected molecule. Third, we used the MADEX code 
\citep[][and references given in Sect.~3]{Cernicharo_2012} to take into account radiative transfer effects in the observed lines. 
MADEX computes the intensities of molecular lines in the non-LTE LVG approximation for those molecules whose collisional rates are available, 
or assuming LTE conditions (to be more precise, a single rotational temperature and a Boltzmann distribution for the rotational levels population)
if these rates are not available. See Appendix~A for a short comparison with other publicly available code.


\subsection{Line parameter fitting procedure}

Gaussian profiles were fitted to all detected lines using CLASS. The derived parameters are shown in Appendix~B.
When two or more transitions overlap, the total profile was fitted. 
The overlapped transitions are marked with connecting symbols in the tables. 
Figure~\ref{fig:anchuras} shows the observed line widths of C$_{2}$H, $l$-C$_{3}$H$^{+}$, $l$-C$_3$H $c$-C$_{3}$H$_{2}$, 
and C$_{4}$H versus the energy of the upper level of the transition. Comparing the detected lines,
we conclude that 
\emph{(i)} the hydrocarbon lines have a pure Gaussian emission profile; 
\emph{(ii)} the average LSR velocity towards the Bar for these lines is \mbox{10.7$\pm$0.3 km s$^{-1}$}; 
\emph{(iii)} the intrinsic line widths are typically \mbox{$\sim$2 km s$^{-1}$}; 
\emph{(iv)} the broadest lines correspond to the lowest energy transitions at 3 mm observed with a larger beam size (\mbox{$\sim$30"-20"} at \mbox{3 mm)}. 
This is likely an indication of molecular emission in the extended OMC1 cloud that contributes to broadening 
the observed lines. The $c$-C$_3$H and $l$-C$_3$H isomers, however, show narrower line profiles suggesting that  they 
arise from specific (less turbulent and more compact) regions in the Bar.


\subsection{Rotational population diagrams}

\begin{table*}
\centering 
\caption{Rotational temperatures ($T_{\rm rot}$), column densities ($N$(X)), and abundances with respect
to hydrogen nuclei inferred in the Orion Bar PDR.}
\label{Table_results}     
 \begin{tabular}{c c c c c c c c@{\vrule height 10pt depth 5pt width 0pt}}     
\hline\hline      
 & \multicolumn{2}{c}{{\bf EXTENDED SOURCE}} \rule[0.15cm]{0cm}{0.2cm}\ &  &  \multicolumn{2}{c}{{\bf SEMI-EXTENDED SOURCE} } \rule[0.2cm]{0cm}{0.2cm}\ & &  \\ \cline{2-3} \cline{5-6}
& $\mathrm{\bf T_{rot}}$ & $\mathrm{\bf N(X)}$ &  &  $\mathrm{\bf T_{rot}}$ & $\mathrm{\bf N(X)}$ & \bf Abundance  & \bf Notes \rule[0.4cm]{0cm}{0.1cm}\ \\
& $\mathrm{[K]}$ & $\mathrm{[cm^{-2}]}$ &   & $\mathrm{[K]}$ & $\mathrm{[cm^{-2}]}$ & & \\
\hline   
C$_{2}$H	&	26$\pm$1	&	(4.2$\pm$0.2)$\times 10^{+14}$		&	&	12$\pm$3	&	(1.7$\pm$0.7)$\times 10^{+15}$		& (0.7-2.7)$\times 10^{-8}$    & a	 	\\
$^{13}$CCH 	&	26		&	3.0$\times$10$^{+12}$	   &	&	12		&	4.6$\times 10^{+12}$					&	(4.8-7.3)$\times 10^{-11}$	 & b		\\
C$^{13}$CH 	&	26		&	4.2$\times$10$^{+12}$		&	&	12		&	6.5$\times 10^{+12}$				&	(0.7-1.0)$\times 10^{-10}$		 & b		\\
$c$-C$_{3}$H 	 &	18$\pm$2	&	(2.0$\pm$0.5)$\times 10^{+12}$		&	&	11$\pm$1	&	(9.9$\pm$3.5)$\times 10^{+12}$		& (0.3-1.6)$\times 10^{-10}$	 &  a		\\
$l$-C$_{3}$H 	&	31$\pm$3	&	(7.0$\pm$1.1)$\times 10^{+11}$		&	&	20$\pm$2	&	(3.4$\pm$0.9)$\times 10^{+12}$		& (1.1-5.4)$\times 10^{-11}$	 &  a		\\
$l$-C$_{3}$H$^{+}$ 	&	46$\pm$5	&	(5.2$\pm$0.7)$\times 10^{+11}$	&	&	25$\pm$3	&	(2.5$\pm$0.6)$\times 10^{+12}$	& (0.8-4.0)$\times 10^{-11}$	 &  a		\\
c-$o$-C$_{3}$H$_{2}$ 	&	17$\pm$1	&	(9.4$\pm$1.3)$\times 10^{+12}$	&	&	15$\pm$1	&	(4.1$\pm$1.3)$\times 10^{+13}$	& (1.5-6.5)$\times 10^{-10}$	 & a	 \\
c-$p$-C$_{3}$H$_{2}$ 	&	17$\pm$1	&	(3.4$\pm$0.4)$\times 10^{+12}$	&	&	11$\pm$1	&	(1.8$\pm$0.5)$\times 10^{+13}$	& (0.5-2.9)$\times 10^{-10}$	 & a	 \\
$[$c-($o$+$p$)-C$_{3}$H$_{2}]$	&	---	&	(1.3$\pm$0.2)$\times 10^{+13}$	&	&  ---	& (5.9$\pm$0.9)$\times 10^{+13}$	&	(2.1-9.4)$\times 10^{-10}$	 &  c	\\
l-$o$-C$_{3}$H$_{2}$ 	&	26$\pm$1	 &  (3.1$\pm$0.2)$\times 10^{+11}$	&	&	17$\pm$1	&	(1.9$\pm$0.4)$\times 10^{+12}$	& (0.5-3.0)$\times 10^{-11}$	 &  a		\\
l-$p$-C$_{3}$H$_{2}$ 	&       28$\pm$1   &	(6.9$\pm$0.2)$\times 10^{+10}$	&	&	15$\pm$1	&	(4.0$\pm$0.8)$\times 10^{+11}$	& (1.1-6.4)$\times 10^{-12}$	 &  a	\\
$[$l-($o$+$p$)-C$_{3}$H$_{2}]$	&	--- 	& (3.8$\pm$0.2)$\times 10^{+11}$	&	& ---		&	(2.3$\pm$0.5)$\times 10^{+12}$	& (0.6-3.7)$\times 10^{-11}$	 &  c		\\
C$_{4}$H 		&	77$\pm$3	&	(3.8$\pm$0.3)$\times 10^{+13}$	&	&	49$\pm$3	&	(2.0$\pm$0.3)$\times 10^{+14}$	& (0.6-3.2)$\times 10^{-9}$	 & a	\\
\hline
\hline 
\end{tabular}
\tablefoot{
(a) Rotational temperatures and column densities from rotational diagram analysis. 
(b) Column densities derived from a LTE model assuming  \mbox{$T_{\rm rot}$=26 K} for extended source and \mbox{$T_{\rm rot}$=12 K} for semi-extended source. 
(c) Total column densities calculated as the sum of the ortho and para species. The abundance of each species with respect to H nuclei is given by ${\frac{N(X)}{N_H}  =  \frac{N(X)}{N(H)+2N(H_{2})} }$, with \mbox{$N$(H$_{2}$)$\simeq$3$\times$10$^{+22}$ cm$^{-2}$} (see Sect.~5.2) and \mbox{$N$(H)$\simeq$3$\times$10$^{+21}$ cm$^{-2}$} \citep{vanderWerf_2013}}.
\end{table*}

For each molecule, we computed a representative rotational temperature ($T_{\rm rot}$) and a beam-averaged column 
density ($N$) by constructing a rotational diagram, assuming optically thin emission filling the beam and a single 
rotational temperature for all energy levels  \citep{Goldsmith_1999}. The standard relation for the rotational diagram 
analysis is
		
\begin{equation}
\mathrm{{ln \, \frac{N_{u}}{g_{u}}}=ln \, N-ln \, Q_{_{T_{rot}}}- \frac{E_{u}}{\,k  T_{rot}} \, ,}\label{eq:DR}
\end{equation}
	
\noindent with $\mathrm{N_{u}/g_{u}}$ given by

\begin{equation}
\mathrm{{\frac{N_{u}}{g_{u}}}={\frac{8\, \pi \, k}{h\,c^{3}}} \cdot {\frac{\nu_{ul} ^{2}}{A_{ul} \, g_{u} }} \cdot \eta_{_{bf}}^{-1} \cdot \displaystyle{\int}T_{_{MB}}dv  \, \, \, \, \, \, \, \, \, [cm^{-2}],}
\end{equation}

\noindent where $N_{\rm u}$ is the column density of the upper level in the optically thin limit [cm$^{-2}$], $N$ is the total column density [cm$^{-2}$], g$_{\rm u}$ is the statistical weight of the upper state of each level, Q$_{T_{\rm rot}}$ is the partition function evaluated at a rotational temperature $T_{\rm rot}$, E$_{\rm u}$/k is the energy of the upper level of the transition [K], $\nu \mathrm{_{ul}}$ is the frequency of the $\mathrm{u\rightarrow l}$ transition [s$^{-1}$], \mbox{$\int \negmedspace T_{_{\rm MB}}$dv} is the velocity-integrated line intensity corrected  from beam efficiency \mbox{[K km s$^{-1}$]}, and $\mathrm{\eta_{_{bf}}}$ is the beam filling factor. Assuming that the emission source has a 2D Gaussian shape, $\mathrm{\eta_{_{bf}}}$ is equal to \mbox{$\mathrm{\eta_{_{bf}}=\theta_{_{S}}^{\, 2}/\,(\theta_{_{S}}^{\, 2}+\theta_{_{B}}^{\, 2})}$}, with $\theta_{_{\rm B}}$ meaning the HPBW of the 30m telescope in arcsec and $\theta_{_{\rm S}}$ the diameter of the Gaussian source in arcsec. The values for $\nu \mathrm{_{ul}}$, A$\mathrm{_{ul}}$, g$\mathrm{_{u}}$, \mbox{$\int \negmedspace T_{_{\rm MB}}$dv}, and E$\mathrm{_{u}}$/k for each molecular transition are given in Appendix~B and were taken from the MADEX spectral catalogue.

The condition for optically thin emission is correct in our case because we do not expect excessively large hydrocarbon column densities towards the Orion Bar dissociation front. The most abundant hydrocarbon in the Orion Bar is C$_{2}$H and, as previously mentioned in \mbox{Sect. 3.1.1}, the relative intensities of the hyperfine components show that lines are optically thin. The rotational diagrams were built considering two limiting cases: \emph{(i)} that the detected emission is extended, with $\eta_{_{\rm bf}} \negthickspace = \negthickspace 1$; and \emph{(ii)} that the emission is semi-extended, assuming that $\theta_{_{\rm S}}$=9" (the typical beam at $\sim$1~mm). We considered only lines not blended with other transitions. Ortho and para forms of linear and cyclic C$_{3}$H$_{2}$ are treated as different species because radiative transitions between both states are forbidden. In order to build the rotational diagram for molecules with hyperfine structure, each hyperfine line component was described without splits, only with a single rotational number N. For this purpose, the integrated intensity, W, level degeneracy, g, and line strength, S, of each transition was calculated as the sum of all allowed hyperfine components of each N+1$\rightarrow$N transition. The characteristic frequency, $\nu$, was determined using the weighted average with the relative strength of each line as weight, and the Einstein coefficient, A, was calculated using the usual relation:

\begin{equation}
\mathrm{A={\frac{64\, \pi^{4}}{3\,h\,c^{3}}} \cdot {\frac{\nu^{3}\,S}{g} \cdot \mu^{2}  \, \, \, \, \, \, \, \, \, [s^{-1}].}}
\end{equation}

The resulting rotational diagrams are shown in Fig.~\ref{fig:DR}. Rotational temperatures and column densities obtained by linear least squares fits for extended and semi-extended emission are listed in Table \ref{Table_results}. The uncertainties shown in Table~\ref{Table_results} indicate the uncertainty obtained in the least squares fit of the rotational diagram. The uncertainty obtained in the determination of the line parameters with the Gaussian fitting programme are included in the uncertainty bars at each point of the rotational diagram. Table~\ref{Table_results} 
also shows the estimated abundances with respect to hydrogen nuclei using a line-of-sight \mbox{$N$(H$_{2}$)$\simeq$3$\times$10$^{22}$ cm$^{-2}$} column density towards the line survey position.
This beam-averaged H$_{2}$ column density was 
derived from our observations of the C$^{18}$O lines (J=1$\rightarrow$0, 2$\rightarrow$1 and 3$\rightarrow$2 transitions)
by constructing a rotational diagram, assuming $^{16}$O/$^{18}$O$\approx$500 \citep{Wilson_1994} and a
CO/H$_{2}$ abundance of $\sim$10$^{-4}$ (lower than the canonical value due to dissociation).
The resulting H$_2$ column density is in good agreement with previous estimations of
$N$(H$_2$) close to the dissociation front \citep[see e.g.][]{Hogerheijde_1995}.
In addition, a  $N$(H)$\simeq$3$\times$10$^{21}$ cm$^{-2}$ column density of hydrogen atoms 
has been inferred from \HI\, observations towards the Orion Bar 
\citep[][]{vanderWerf_2013}. Hence, a small fraction of free H atoms exist.

Rotational temperatures and column densities derived from the rotational diagram of each molecule (assuming extended source) were used as input parameters to model the line profiles using the MADEX radiative transfer model under LTE conditions. Figures \ref{fig:C3H+_lines}, \ref{fig:$l$-C3H_lines}, \ref{fig:$c$-C3H_lines}, \ref{fig:H2C3_lines}, and \ref{fig:C4H_lines} show the observational spectra (black histograms) and the modelled spectra (red lines) of $l$-C$_{3}$H$^{+}$, $l$-C$_{3}$H, $c$-C$_{3}$H, $l$-H$_{2}$C$_{3}$, and C$_{4}$H, respectively. The obtained fits using LTE models agree with the observations. 
The computed optical depths at line centre are $\tau$<0.1 for C$_{2}$H, \mbox{$o/p$-C$_{3}$H$_{2}$}, and C$_{3}$H$^{+}$ lines, and $\tau$<0.01
for the rest of the hydrocarbons. Hence, there are no noticeable line opacity effects in the calculated rotational temperatures and column densities.

For those species for which only one rotational line was detected, we estimated their column densities using as an input parameter the rotational temperatures derived from another molecule with similar structure and rotational constants. In particular, the $^{13}$CCH and C$^{13}$CH column densities were estimated assuming that the rotational temperature is similar to that obtained for C$_{2}$H. Figure~\ref{fig:13CCH_lines} shows the observational (black histogram) and modelled (red lines) spectra of $^{13}$CCH and C$^{13}$CH.

Rotational temperatures range from 17 to \mbox{77 K}, and column densities from 10$^{11}$ to \mbox{10$^{14}$ cm$^{-2}$}. 
The C$_{2}$H radical is by far the most abundant of the detected hydrocarbons, followed by C$_{4}$H and \mbox{c-(\emph{o}+\emph{p})-C$_{3}$H$_{2}$}. The linear three-carbon species (\mbox{$l$-C$_{3}$H} and \mbox{$l$-H$_2$C$_{3}$)} are less abundant than their cyclic isomers. The ortho-to-para ratio obtained from the $c$-C$_{3}$H$_{2}$ column density is 2.8$\pm$0.6, similar to the expected value at high temperature. The [C$^{13}$CH]/[$^{13}$CCH] ratio is \mbox{1.4$\pm$0.1 (3$\sigma$)}. 
Comparing the rotational temperatures, we conclude that \emph{(i)} almost all species have \mbox{$T_{\rm rot}$<30 K}, but \mbox{$l$-C$_{3}$H$^{+}$} and C$_{4}$H are rotationally hotter than the other species, 
reaching \mbox{$T_{\rm rot}$=46 K} and \mbox{77 K}, respectively; 
\emph{(ii)} although cyclic forms have lower dipole moments, the rotational temperatures of the cyclic species 
($c$-C$_{3}$H$_{2}$ and \mbox{$c$-C$_{3}$H}) are smaller than their respective linear isomers ($l$-H$_2$C$_{3}$ and 
\mbox{$l$-C$_{3}$H});
\emph{(iii)} we obtain similar rotational temperatures for the ortho and para forms of the same molecule; 
\emph{(iv)} C$_{2}$H and C$_{4}$H have similar dipole moments and rotational spectroscopy, but show quite different 
rotational temperatures (\mbox{26 K} and \mbox{77 K}, respectively). This low temperature for C$_{2}$H means that in the millimetre domain we detect the warm C$_2$H (containing most of the column density), but
not the hotter C$_2$H recently detected by \textit{Herschel}/HIFI (higher energy transitions from N=6$\rightarrow$5 to
10$\rightarrow$9; see Nagy et al. in prep.). We estimate that this hotter C$_2$H only contributes to $\sim$5\%~of
 the total C$_2$H column. The rotational temperature deduced for C$_{4}$H will be discussed in Sect.~7.1.


\subsection{Spatial variation of the C$_{2}$H and $c$-C$_3$H$_2$ abundance}

\begin{figure}
\centering
\includegraphics[scale=0.62,angle=0]{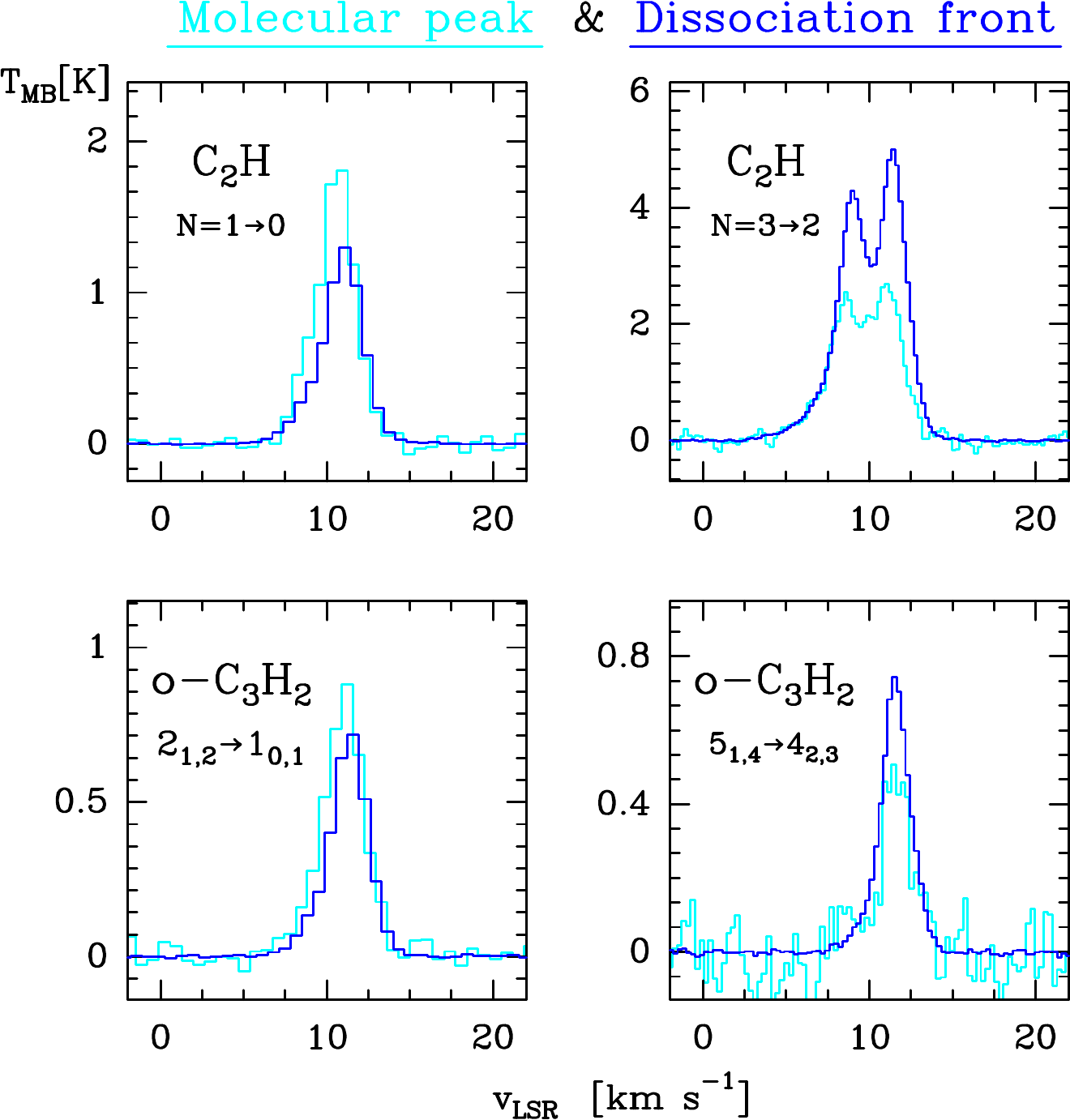} 
\caption{C$_{2}$H (\emph{upper panels}) and $c$-C$_{3}$H$_{2}$ (\emph{lower panels}) spectra towards two different 
positions in the Orion Bar: one in the dissociation front where we have carried out a line survey (dark blue spectrum), and the other towards clump no.~10 of  \citet{Lis_2003} that we call the molecular peak (light blue spectrum). The spectra towards the molecular peak were extracted from the 1~mm and 3~mm reprojected maps.}\label{fig:C2H_2zones}
\end{figure}

\begin{table} 
\centering
\caption{Rotational temperatures ($T_{\rm rot}$), column densities ($N$(X)), and abundances of C$_{2}$H and 
$c$-$o$-C$_{3}$H$_{2}$ with respect to hydrogen nuclei inferred in the dissociation front (DF) and in the molecular peak (MP).}
\label{Table_df_mp}
\begin{tabular}{c c c c c @{\vrule height 10pt depth 5pt width 0pt}}
\hline
\hline
&	 & \bf T$ _{\rm \bf rot}$ & \bf N(X) & \bf Abundance \tablefootmark{a} \\
&	 &	 [K]	 &	 [cm$^{-2}$]	 &	 \\
\hline
\multirow{2}{*}[-0.1cm]{\bf C$_{\bf 2}$H} & DF	& 26$\pm$1 & (4.2$\pm$0.2)$\times$10$^{+14}$ & $\sim$6.7$\times$10$^{-9}$	\ \ \\
& MP	& 13$\pm$2 & (3.7$\pm$0.6)$\times$10$^{+14}$ & $\sim$7.4$\times$10$^{-10}$	 \\
\hline
\multirow{2}{*}[-0.1cm]{\bf c-o-C$_{\bf 3}$H$_{\bf 2}$}	& DF	& 17$\pm$1	 & (9.4$\pm$1.3)$\times$10$^{+12}$	& $\sim$1.5$\times$10$^{-10}$	 \\
& MP	& 11$\pm$2	 & (1.2$\pm$0.2)$\times$10$^{+13}$	& $\sim$2.4$\times$10$^{-11}$	 \\
\hline
\hline
\end{tabular} 
\tablefoot{
\tablefoottext{a}{With respect to H nuclei, using $N_{\rm H}$$\simeq$3.3$\times$10$^{+22}$ cm$^{-2}$ 
(see Sect.~5.2) towards the dissociation front (DF) and $N_{\rm H}$$\simeq$2.5$\times$10$^{+23}$ cm$^{-2}$ \citep{Lis_2003} towards the molecular peak (MP).
}
}
\end{table}

In order to investigate the C$_{2}$H and $c$-C$_{3}$H$_{2}$ column density and abundance variations along the Orion Bar, we reprojected the 
3~mm and \mbox{1 mm} IRAM 30m maps to a common grid.  
We selected two representative positions, one towards the line survey position at the PDR dissociation front 
(near the C$_{2}$H N=3$\rightarrow$2 line emission peak) and the other deeper inside the cloud where the 
C$_{2}$H N=1$\rightarrow$0 line peaks, that we call the molecular peak. This latter  position coincides with 
the dense clump/condensation no.~10 detected by \citet{Lis_2003} in H$^{13}$CN J=1$\rightarrow$0 emission. 
We extracted the C$_{2}$H and \mbox{$c$-$o$-C$_{3}$H$_{2}$} column densities towards these two positions by constructing reduced rotational diagrams with the line 
intensities extracted from the maps. 
Figure~\ref{fig:C2H_2zones}  shows the C$_{2}$H (N=1$\rightarrow$0 and 3$\rightarrow$2) and \mbox{$c$-$o$-C$_{3}$H$_{2}$} (\mbox{J$_{\rm K_a,K_c}$=2$_{1,2}$$\rightarrow$1$_{0,1}$} and 5$_{1,4}$$\rightarrow$4$_{2,3}$) spectra towards the two selected 
positions. The dark blue spectrum shows the emission in the PDR dissociation front and the light blue spectrum shows the
emission deeper inside the cloud. Table~\ref{Table_df_mp} shows the C$_2$H and \mbox{$c$-$o$-C$_3$H$_2$} rotational temperatures 
and column densities towards the two positions. Close to the dissociation front, where the gas is hotter, the inferred
rotational temperatures are higher. Despite the similar 
C$_2$H and $c$-$o$-C$_{3}$H$_{2}$  column densities towards both positions, the $N_{\rm H}$ column density towards 
the cloud edge is necessarily smaller than towards clump/condensation no.~10.
This is readily seen in C$^{18}$O line maps that show the brightest emission towards the molecular peak position
deeper inside the Bar and a faint emission level towards the dissociation front \citep[e.g.][Cuadrado et al. in prep.]{Wiel_2009}.
The variation of the C$_{2}$H and $c$-$o$-C$_{3}$H$_{2}$  abundances can then be estimated using 
\mbox{$N_{\rm H}$$\simeq$3.3$\times$10$^{22}$ cm$^{-2}$} towards the dissociation front (see Sect.~5.2) and 
\mbox{$N_{\rm H}$$\simeq$2.5$\times$10$^{23}$ cm$^{-2}$} towards clump no.~10 (the median column density inferred  by  \citet{Lis_2003} towards the dense H$^{13}$CN clumps).
This factor of $\sim$10 difference is consistent with the expected increase of line-of-sight material with distance from the PDR edge to the density peak ($d$), and estimated as $d^2$ from detailed  dust
radiative transfer models \citep{Arab_2012}.
Therefore, the C$_{2}$H abundance with respect to hydrogen nuclei is higher towards the PDR edge
($\sim$10$^{-8}$) than deeper inside the cloud ($\sim$10$^{-9}$). \mbox{$c$-C$_{3}$H$_{2}$} shows a similar trend and
therefore both hydrocarbons show enhanced abundances towards the UV-illuminated edge of the cloud, but they
are also moderately abundant in the more shielded cloud interior. 
Although higher angular resolution observations will be needed to accurately constrain the abundance gradients in more detail, 
we note that the \mbox{[C$_{2}$H]/[$c$-C$_3$H$_2$]} column density ratio decreases from the illuminated cloud edge 
to the cloud interior.


\subsection{Non-LTE excitation analysis}

\begin{figure}[!b]
\centering
\includegraphics[scale=0.45,angle=0]{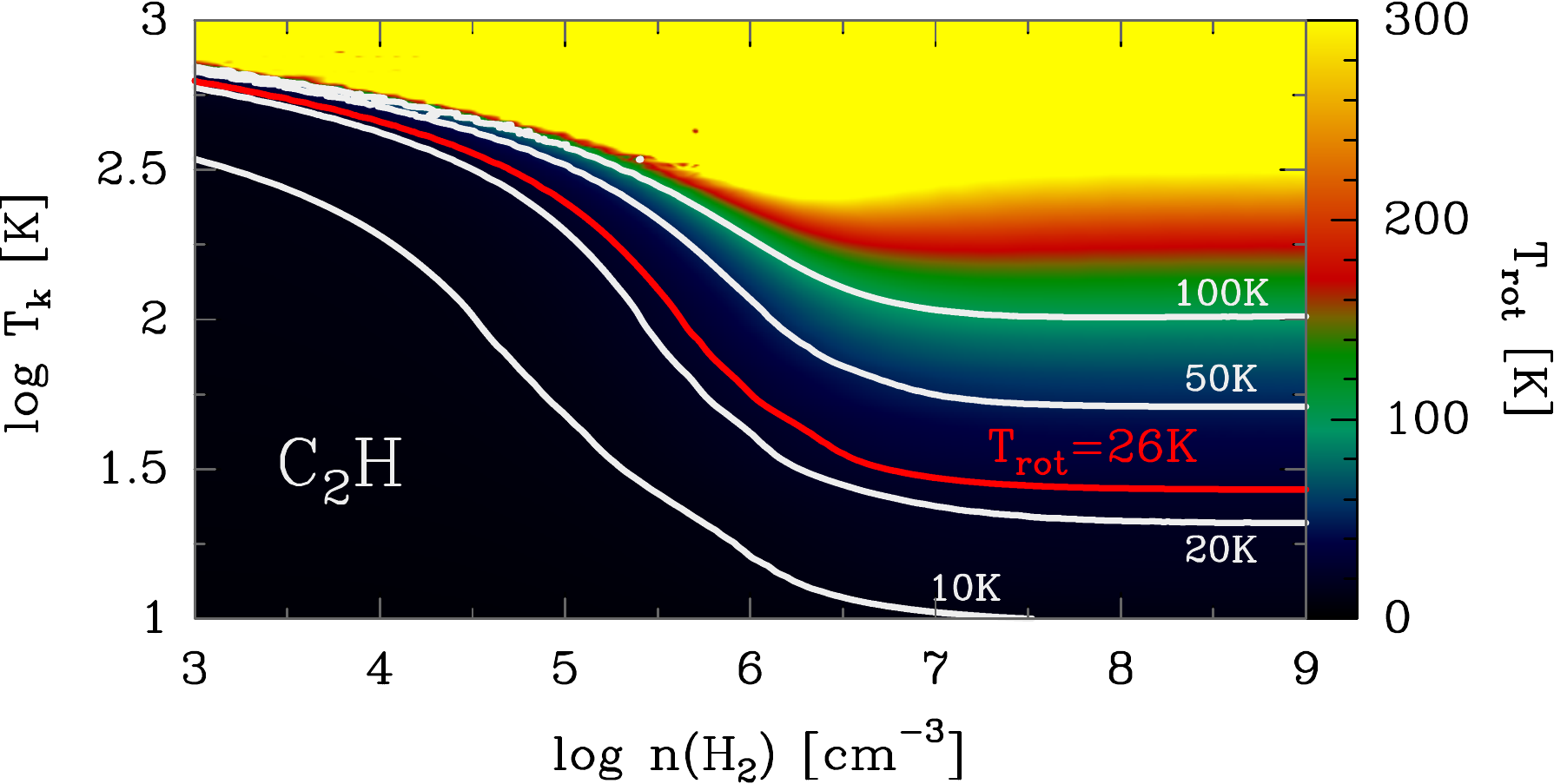}\\ \vspace{0.3cm}
\includegraphics[scale=0.45,angle=0]{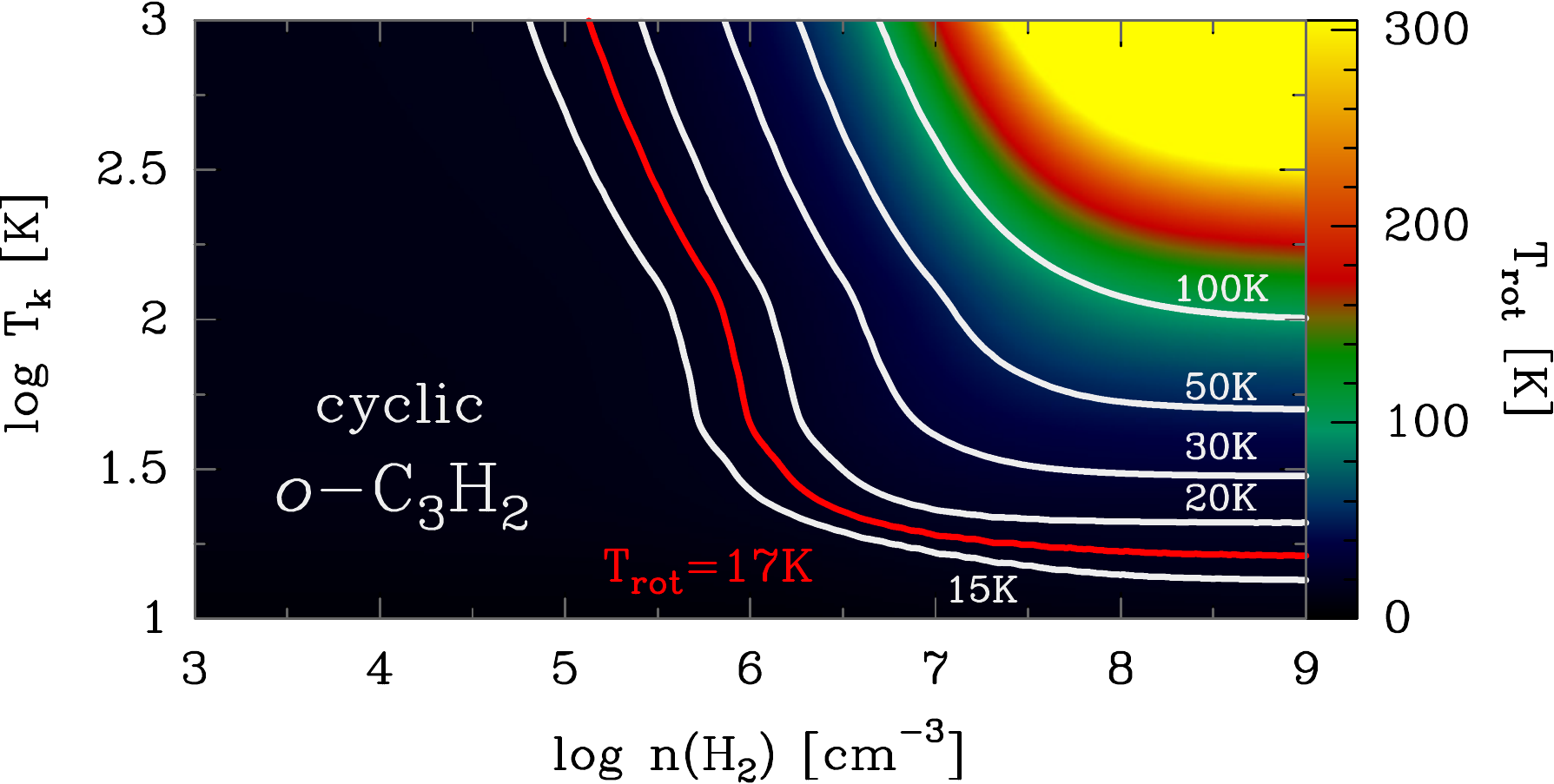}\\ \vspace{0.3cm}
\includegraphics[scale=0.45,angle=0]{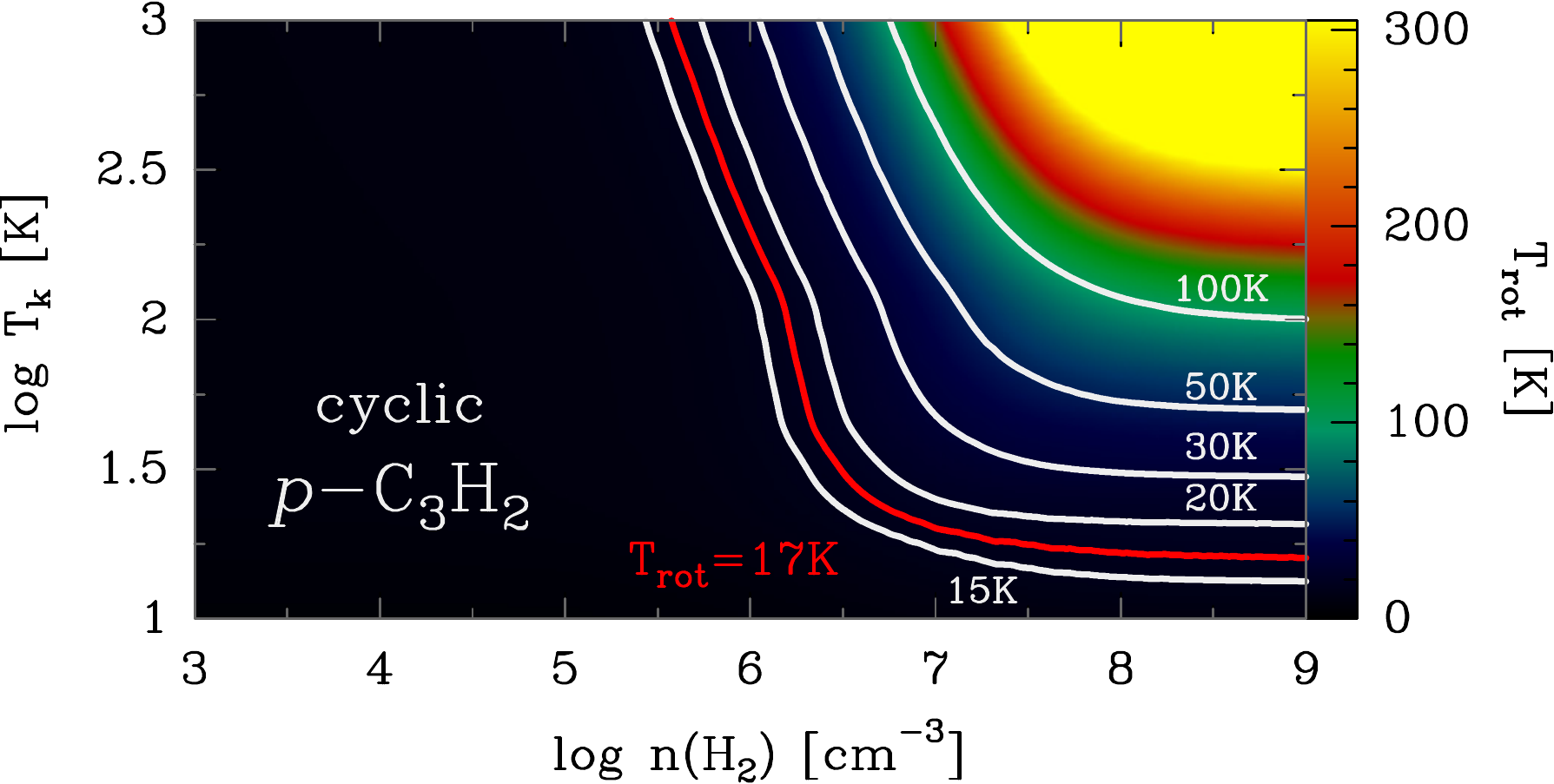}\\
\caption{Grid of C$_{2}$H, o-C$_{3}$H$_{2}$, and p-C$_{3}$H$_{2}$ LVG models for different gas temperatures and densities. White contour levels represent iso-rotational temperatures. The red curves show the rotational temperature derived from our observations towards the survey position.}\label{fig:Trot}
\end{figure}

In order to derive the beam-averaged physical conditions in the observed region we studied the non-LTE excitation of the C$_{2}$H and \mbox{$c$-C$_{3}$H$_{2}$} molecules for which accurate collisional rates exist. We used \mbox{C$_{2}$H-He} collisional rates from \citet{Spielfiedel_2012} and C$_{3}$H$_{2}$-He from \citet{Chandra_2000} for C$_{2}$H and \mbox{$c$-C$_{3}$H$_{2}$}, respectively. 
A large grid of LVG models for a broad range of column densities, H$_{2}$ densities (\mbox{$n$(H$_{2}$)=10$^{3-9}$ cm$^{-3}$}), and kinetic temperatures \mbox{($T_{\rm k}$=10-1000 K)} values were run to obtain synthetic line intensities for C$_{2}$H and cyclic \mbox{$o/p$-C$_{3}$H$_{2}$}. 
The best fit models have column densities within a factor of 2 of the inferred value from the rotational diagram analysis (see Sect.~5.2).
Hence, we used constant beam-averaged column densities for further analysis
(\mbox{$N$(C$_{2}$H)=4.2$\times 10^{+14}$ cm$^{-2}$}, 
\mbox{$N$($c$-$o$-C$_{3}$H$_{2}$)=9.4$\times 10^{+12}$ cm$^{-2}$}, and 
\mbox{$N$($c$-$p$-C$_{3}$H$_{2}$)=3.4$\times 10^{+12}$ cm$^{-2}$)}.
The C$_{2}$H models were fitted in the \mbox{N$_{\rm up}$=1 to 5} transition range, and $c$-C$_{3}$H$_{2}$ models were fitted within the J$_{\rm up}$=2-8 range. We used \mbox{$\Delta$v=2 km s$^{-1}$} line widths.
Figure~\ref{fig:Trot} shows the LVG model results in the form of iso-$T_{\rm rot}$ contours. Each $T_{\rm rot}$ was calculated by building rotational diagrams with the synthetic line intensities obtained from each model. 
Figure~\ref{fig:Trot} shows that the $T_{\rm rot}$ inferred from our survey data (the red contours) can be obtained for different combinations of gas density ($n$(H$_{2}$)) and temperature ($T_{\rm k}$).

In order to constrain accurately the range of physical conditions that reproduce the observed intensities towards the dissociation front, we compared the C$_{2}$H detected lines to the synthetic line intensities obtained in the grid of LVG models. 
We only used  the C$_{2}$H N=3$\rightarrow$2 and 4$\rightarrow$3 hyperfine components in the analysis 
as they were observed with higher angular and spectral resolution.
In addition, the extended emission from OMC1 likely contributes to the  observed \mbox{3 mm} lines, both in amplitude and in line-broadening
(see Sect.~5.1). Following \citet{Neufeld_2014}, the best fit model was obtained by finding the minimum root mean square (rms) value of
 \mbox{log$_{10}(I_{\rm obs}/I_{\rm mod})$}.  This is defined as
\begin{equation}
\mathrm{rms={\sqrt{\frac{1}{n} \,\sum_{i=1}^{n} \, \left(log _{_{10}}\, \frac{I^{\, i}_{obs}}{I^{\, i}_{mod}}\right)^{2}}} \, ,}
\end{equation}

\noindent where $n$ is the number of observed lines, $I^{\, \rm i}_{\rm obs}$ is the observed line intensity calculated 
from Gaussian fits to the lines listed in Table~\ref{Table_C2H}, and $I^{\, \rm i}_{\rm mod}$ is the model line 
intensity using MADEX.

\begin{figure*}
\centering
\includegraphics[scale=0.42,angle=0]{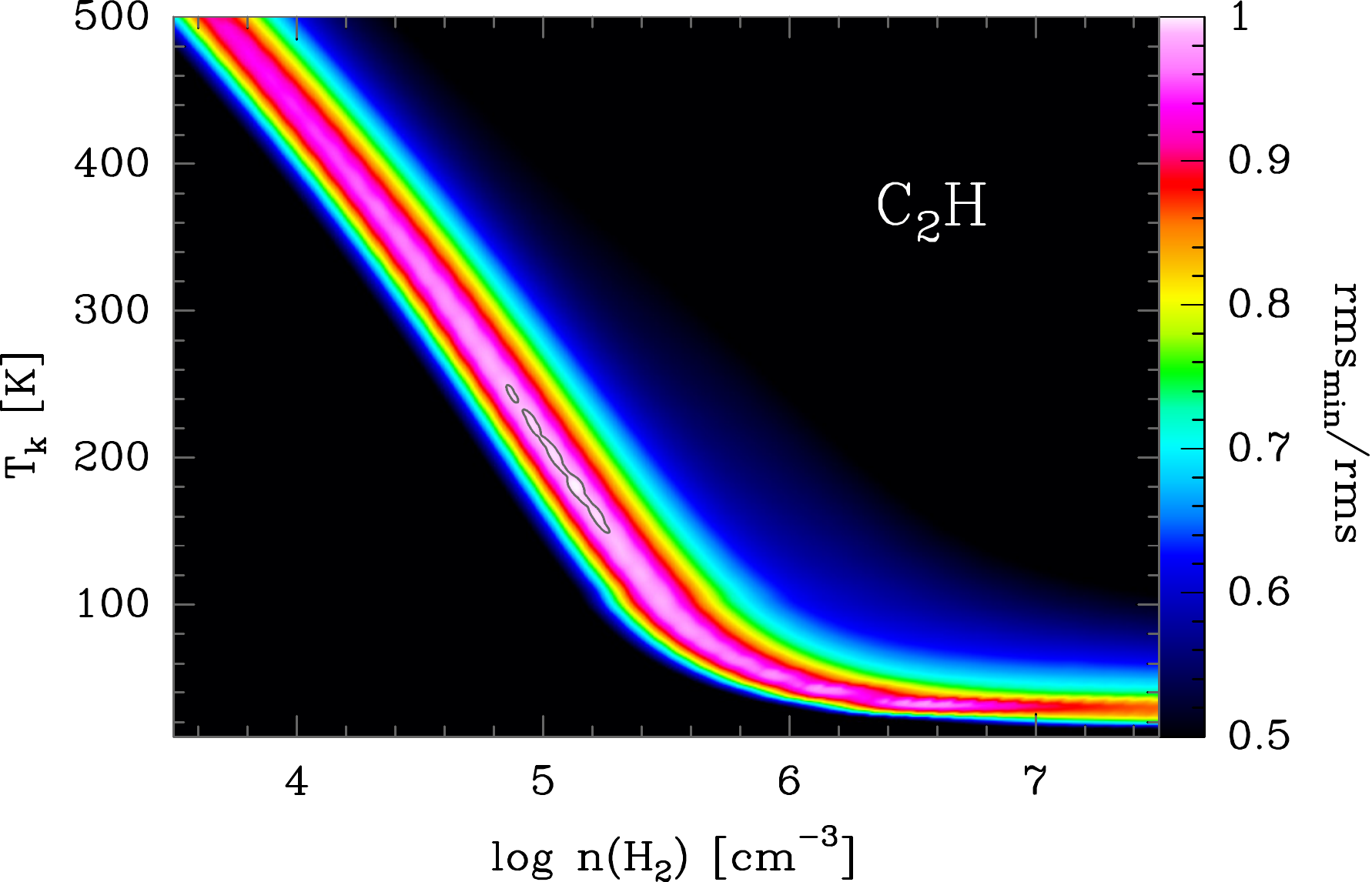} \hspace{0.3cm}
\includegraphics[scale=0.47,angle=0]{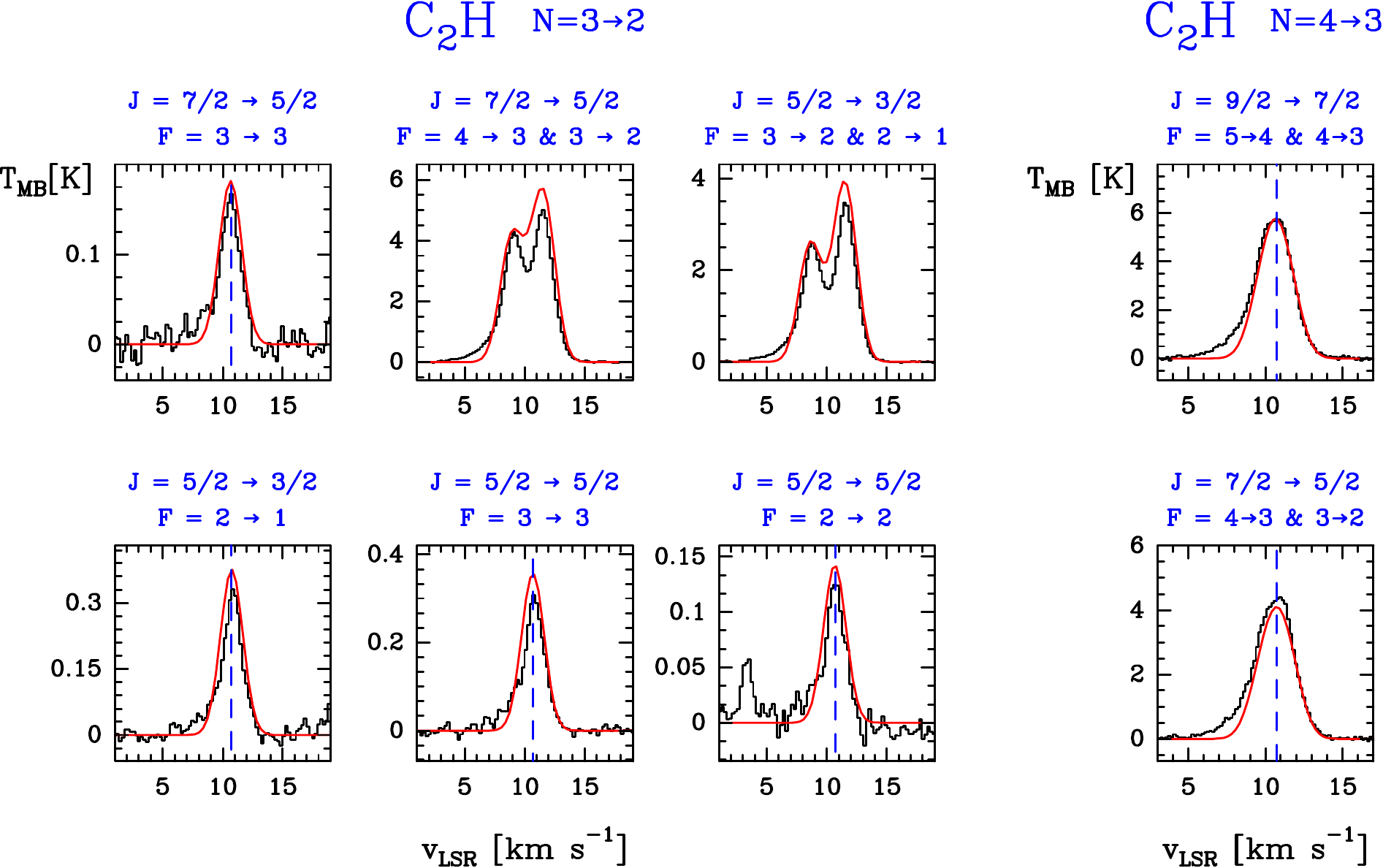} 
\caption{\emph{Left panel:} rms$_{min}$ / rms ratio as a funtion of $T_{\rm k}$ and $n$(H$_{2}$) for a grid of LVG models. 
The best fit models were obtained with \mbox{$T_{\rm k}$=150-250 K} and \mbox{$n$(H$_{2}$)=(0.7-1.7)$\times 10^{5}$ cm$^{-3}$}.
 \emph{Right panel:} Observed C$_{2}$H N=3$\rightarrow$2 and 
4$\rightarrow$3  spectra. 
The best fit LVG model (\mbox{$T_{\rm k}$$\simeq$150 K}, \mbox{$n$(H$_{2}$)$\simeq$1.5$\times10^{5}$ cm$^{-3}$}, and
 \mbox{$N$(C$_{2}$H)=4.2$\times 10^{14}$ cm$^{-2}$})  is shown overlaid in red.}
\label{fig:C2H_rms}
\end{figure*}

Figure~\ref{fig:C2H_rms} (left) represents the rms$_{\rm min}$/rms ratio as a function of 
$T_{\rm k}$ and $n$(H$_{2}$) for a grid of excitation models trying to fit the C$_{2}$H N=3$\rightarrow$2 and 4$\rightarrow$3 lines towards the PDR position. The set of physical conditions that best fit the lines lie within \mbox{$T_{\rm k}$=150-250 K}, \mbox{$n$(H$_{2}$)=(0.7-1.7)$\times 10^{5}$ cm$^{-3}$}. Figure~\ref{fig:C2H_rms} (right) also shows the best fit C$_{2}$H model (\mbox{$T_{\rm k}$$\simeq$150 K}, \mbox{$n$(H$_{2}$)$\simeq$1.5$\times10^{5}$ cm$^{-3}$)} overlaid over the observed lines. 
The computed opacities at line centre  are $\tau$<0.1. We note that the $^{12}$CO line intensity peaks ($T^{\rm p}_{_{\rm MB}}$; in main beam temperature) towards the line survey position goes from $\sim$115 K to $\sim$160\,K (for the $J$=1$\rightarrow$0 and 3$\rightarrow$2 lines, respectively). Since low-$J$ $^{12}$CO lines are 
optically thick (\mbox{W[$^{12}$CO 3$\rightarrow$2]/W[$^{13}$CO 3$\rightarrow$2]$\simeq$5}, much lower than 
the $^{12}$C/$^{13}$C isotopic ratio of 60), their intensity peak provides a good lower limit to the gas temperature \mbox{($T^{\rm p}_{_{\rm MB}}\lesssim T_{\rm ex}\lesssim T_{\rm k}$)}. Therefore, the temperatures inferred from 
C$_2$H and $^{12}$CO  are in good agreement and consistent with 
$T_{\rm k}$$\simeq$150 K.
Nevertheless, the gas temperature and density in a PDR  are 
expected to vary at spatial scales that cannot be resolved with our single-dish observations. Therefore, these 
values should be considered as the averaged conditions towards the Orion Bar dissociation front within a $\sim$30$"$-10$"$ beam.

In the case of $c$-($o/p$)-C$_{3}$H$_{2}$, as shown in the plot of \mbox{iso-$T_{\rm rot}$} (Fig.~\ref{fig:Trot}), the excitation of the millimetre lines provides a lower limit to the H$_2$ density of a few 10$^5$\,cm$^{-3}$.
The statistical analysis to obtain the best fit of the $c$-($o/p$)-C$_{3}$H$_{2}$ lines was not conclusive because several sets of these two parameters fit the lines. However, for the physical conditions expected in the Orion Bar, the $c$-C$_{3}$H$_{2}$ lines do not exactly fit with the same physical conditions as C$_{2}$H. Slightly  denser gas and a lower temperature are needed to obtain a satisfactory fit of $c$-($o/p$)-C$_{3}$H$_{2}$ lines (\mbox{$T_{\rm k}$$\simeq$100 K} and \mbox{$n$(H$_{2}$)$\simeq$4.0$\times$10$^{5}$ cm$^{-3}$}; see Fig.~\ref{fig:C3H2_lines}). 
As we see later, this is roughly consistent with our chemical models (Sect.~6) which predict that the $c$-C$_{3}$H$_{2}$  
abundance peaks slightly deeper inside the cloud than C$_{2}$H.


\subsection{Undetected hydrocarbons}

The broadband coverage of the Orion Bar survey allowed us to obtain upper limits for other chemically interesting
carbon-bearing molecules that have not been detected in the PDR: longer carbon chains, anions, and deuterated 
isotopologues. In particular, we searched for C$_{2}$D, C$_{5}$H, C$_{6}$H, C$_{2}$H$^{-}$, C$_{4}$H$^{-}$, C$_{6}$H$^{-}$, \mbox{H$_{2}$C$_{4}$}, and 
C$_{2}$H$_{3}^{+}$. First, we estimated 3$\sigma$ line intensities using the relation \citep[e.g.][]{Coutens_2012}
\begin{equation}
\mathrm{\displaystyle{\int} T_{_ {MB}} dv=3\sigma \sqrt{2 \, \delta v \, \Delta v}  \, \, \, \, \, \, \, \, \, [K \, km \, s^{-1}],}
\end{equation}
\noindent where $\sigma$ is the rms of the observations [K], $\delta v$ is the velocity-spectral resolution 
\mbox{[km s$^{-1}$]}, and $\Delta v$ is the assumed line widths \mbox{($\sim$2 km s$^{-1}$)}. Second, we used MADEX to create LTE models to simulate the emission of different rotational lines at different frequencies of the same species and to constrain
  their column densities. The column densities and abundances 3$\sigma$ upper limits for \mbox{$T_{\rm rot}$=20-30 K} are 
  listed in Table~\ref{Table_undetected_lines}.

Despite some controversy about their presence in PDRs \citep[see e.g.][]{Fortenberry_2013}, hydrocarbon 
anions are not detected at the sensitivity level of our line survey. We provide the following abundance ratio limits: 
\mbox{[C$_{2}$H$^{-}$]/[C$_{2}$H]<0.007$\%$} and \mbox{[C$_{4}$H$^{-}$]/[C$_{4}$H]<0.05$\%$}. They agree with previous
 unsuccessful anion searches in other interstellar environments \citep[e.g.][]{Agundez_2008b}.


\section{PDR models of the Orion Bar}

To investigate whether the inferred hydrocarbon column densities and spatial distribution agree with our current understanding of their gas-phase chemical formation, we have used an updated version of the Meudon code for photochemical studies \citep{LePetit_2006, LeBourlot_2012}. This 1D PDR model solves the FUV radiative transfer in an absorbing and diffusing medium of gas and dust \citep{Goicoechea_2007}. This allows the explicit computation of the FUV radiation field (continuum+lines) and, therefore, the explicit integration of self-consistent  photoionisation  and photodissociation rates as a function of cloud depth. The model also solves the thermal balance \citep[see][]{LePetit_2006} and thus the thermal profile through the PDR. Once the attenuation of the FUV radiation field and the temperature profile have been determined, steady-state chemical abundances are computed for a given chemical reaction network.

\begin{table}
\centering 
\caption{Upper limits for undetected hydrocarbons.}
\label{Table_undetected_lines}     
\begin{tabular}{c c c@{\vrule height 10pt depth 5pt width 0pt}}     
\hline\hline      
\bf Molecule  &                     \bf{N(X)} [cm$^{-2}$]       & \bf Abundance \tablefootmark{a} \\
\hline
$p$-C$_{2}$H$_{3}^{+}$            & (1.8-2.0)$\times$10$^{+12}$    &  (2.9-3.2)$\times$10$^{-11}$  \\ 
C$_{2}$D \,                        & (5.0-9.0)$\times$10$^{+11}$    &  (0.8-1.4)$\times$10$^{-11}$     \\ 
($o$+$p$)-H$_{2}$C$_{4}$         & (2.5-2.8)$\times$10$^{+11}$    &  (4.0-4.4)$\times$10$^{-12}$    \\ 
C$_{6}$H \,                        & (2.0-4.0)$\times$10$^{+11}$    &  (3.2-6.4)$\times$10$^{-12}$      \\
C$_{5}$H \,                        & (1.5-2.0)$\times$10$^{+11}$    &  (2.4-3.2)$\times$10$^{-12}$      \\  
C$_{6}$H$^{-}$                    & (0.7-1.5)$\times$10$^{+11}$    &  (1.1-2.4)$\times$10$^{-12}$   \\ 
C$_{2}$H$^{-}$                    & (2.0-3.0)$\times$10$^{+10}$    &  (3.2-4.8)$\times$10$^{-13}$   \\   
C$_{4}$H$^{-}$                    & (1.5-2.0)$\times$10$^{+10}$    &  (2.4-3.2)$\times$10$^{-13}$    \\   
 \hline
 \hline      
\end{tabular}
\tablefoot{
\tablefoottext{a}{The abundance of each species with respect to H nuclei is given by ${\frac{N(X)}{N_H}  =  \frac{N(X)}{N(H)+2N(H_{2})} }$, for \mbox{$N$(H$_{2}$)=3$\times$10$^{+22}$ cm$^{-2}$} (see Sect.~5.2) and \mbox{$N$(H)=3$\times$10$^{+21}$ cm$^{-2}$} \citep{vanderWerf_2013}.
} 
}
\end{table} 

\begin{figure}
\centering
\includegraphics[scale=0.195,angle=0]{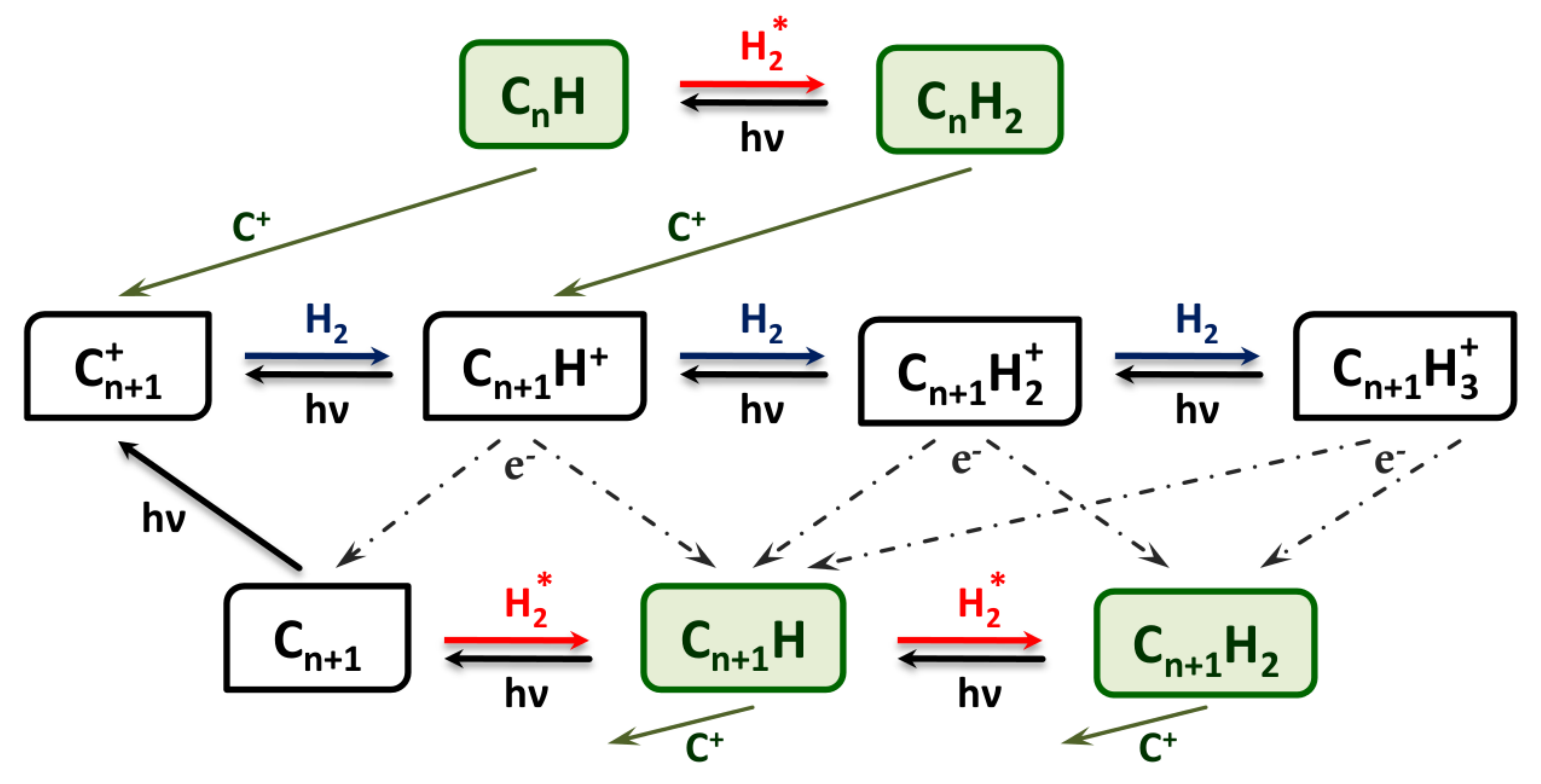} 
\caption{Scheme with the main gas-phase formation and destruction reactions of small hydrocarbons predicted by
our model of the Orion Bar PDR
at $A_{\rm V}\negmedspace\simeq$1.5 (close to their abundance peak).
 Reactions with activation energy barriers are indicated with red arrows. The blue arrows are barrierless 
 hydrogenation reactions.}
\label{fig:reactions}
\end{figure}

Our network contains $\thicksim$130 species and $\thicksim$2800 gas-phase reactions. It includes the formation of carbon
bearing molecules up to four carbon atoms. 
In our models we adopt a FUV radiation field $\chi\negmedspace=$2$\times$10$^{4}$ times the mean interstellar 
radiation field (ISRF) in Draine units \citep[e.g.][]{Marconi_1998}. 
When available, we used photodissociation rates given by \citet{vanDishoeck_1988} and \citet{vanDishoeck_2006}
(multiplied by $\chi$), which
are explicitly calculated for the Draine interstellar radiation field (ISRF). State-to-state reactions of vibrationally
excited H$_{2}$ with C$^{+}$, O, or OH are explicitly treated \citep[see][]{Agundez_2010}. We have further upgraded the 
carbon-bearing species network and used the most recent branching ratios for ion-molecule, neutral-neutral, dissociative 
recombination, and charge exchange reactions for carbon chains and hydrocarbon species described in \citet{Chabot_2013}.
Reactions of hydrocarbon radicals with H and H$_2$ are also included in the model
\citep[see e.g.][]{Cernicharo_2004, Agundez_2008a}. Figure~\ref{fig:reactions} 
shows the dominant gas-phase formation and destruction routes of hydrocarbons 
predicted by  our PDR models of the Orion Bar close to their abundance peak (at $A_{\rm V}\negthickspace\simeq$1.5, see below).

We also included the freeze-out of molecular and atomic species at different cloud depths. In our models ice mantles can desorb thermally and non-thermally (FUV-induced), but we do not contemplate grain surface chemistry. Given 
the high FUV radiation field and thus grain temperatures in the Orion Bar PDR \mbox{($T_{\rm d}$ \negthickspace$\approx$50-70 K;} \citealt{Arab_2012}), the less volatile species (e.g. H$_{2}$O) start to be affected by freeze-out beyond the PDR surface \citep[$A_{\rm V}\negthickspace>$5; e.g.][]{Hollenbach_2009}. 
The adopted elemental abundances are those of \citet{Goicoechea_2006}. 
We adopted a cosmic-ray ionisation-rate ($\zeta_{\rm CR}$) of 10$^{-16}$\,s$^{-1}$, but we note that the total ionisation rate in the Orion Bar
might be higher if one includes the X-ray ionisation effects produced by all X-ray stellar sources in the Trapezium cluster \citep[e.g.][]{Gupta_2010}.

\begin{figure*}
\centering
\includegraphics[scale=0.66,angle=0]{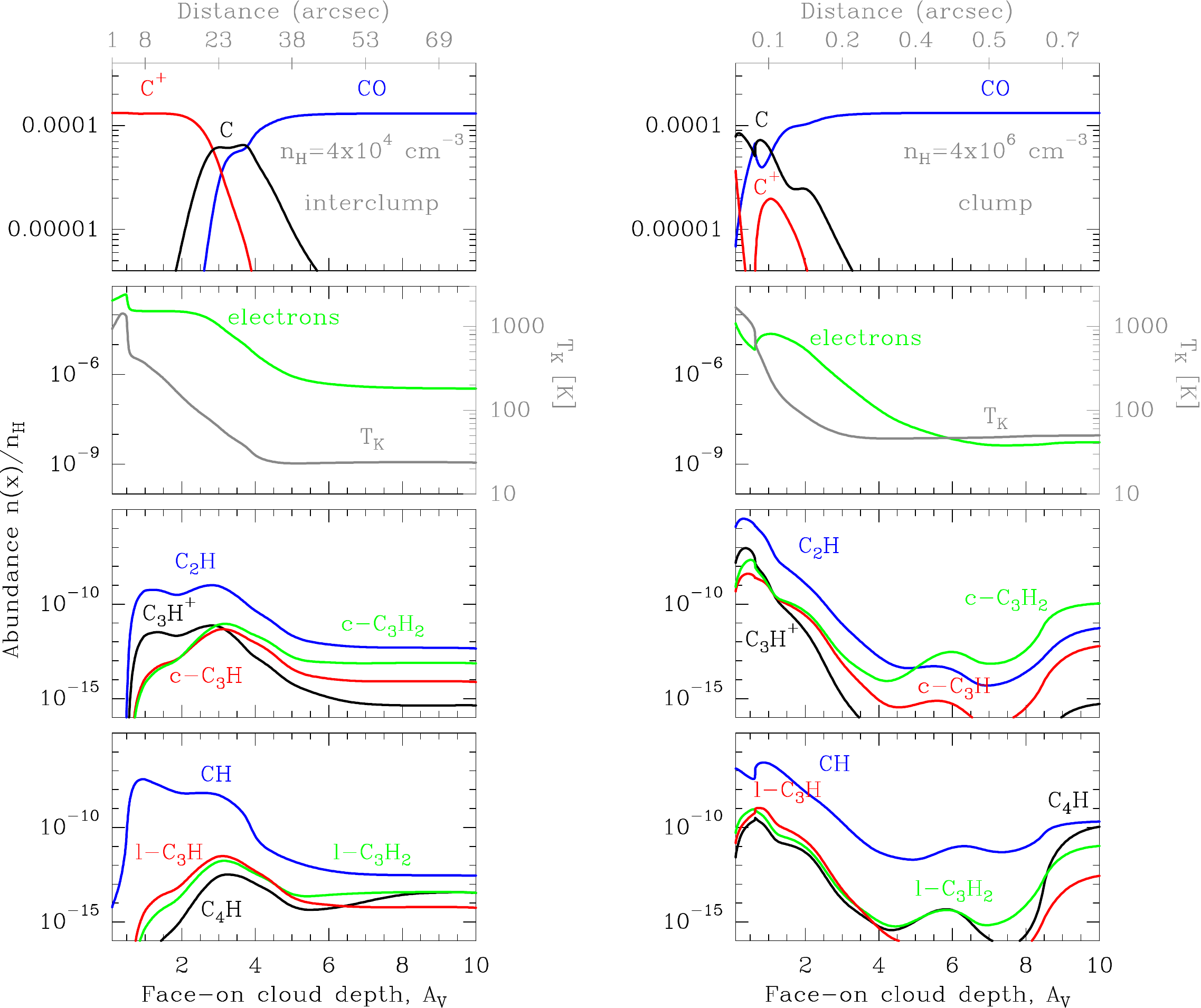} 
\caption{Constant density, photochemical models as a function of cloud visual extinction ($A_{\rm V}$) for $\chi$=2$\times$10$^4$.
\textit{Left}: $n_{_{\rm H}}$=4$\times$10$^{4}$\,cm$^{-3}$ characteristic of a moderate density interclump medium; 
\textit{Right}: $n_{_{\rm H}}$=4$\times$10$^{6}$\,cm$^{-3}$ characteristic of higher density clumps.
We note the different spatial scales, thermal profile, and molecular abundance and ionisation fraction gradients 
implied by the two models.}
\label{fig:oribar_interclump_clump}
\end{figure*}

In order to guide the interpretation of our observations, we ran two types of PDR models that have been proposed to reproduce 
the physical conditions in the Orion Bar: constant density and isobaric models.
We first ran models with constant hydrogen nuclei density in which the thermal pressure decreases from the illuminated 
cloud edge to the cloud interior:  an interclump medium with \mbox{$n_{_{\rm H}}$=4$\times$10$^{4}$\,cm$^{-3}$} and 
a denser clump component with \mbox{$n_{_{\rm H}}$=4$\times$10$^{6}$\,cm$^{-3}$} 
\citep[see][for more complicated descriptions]{Andree-Labsch_2014}. Figure~\ref{fig:oribar_interclump_clump} shows the physical and chemical structure of the two models. 
The spatial scales of the chemical stratification seen in the PDR (\mbox{C$^+$/C/CO/...}) is only compatible with the
presence of a moderate density interclump medium. The physical gradients in the 
denser clump model occur in much smaller  scales (that cannot be resolved with the IRAM 30m) but
produce enhanced columns of several species.  Hence, an ensemble of low filling factor embedded  clumps or dense gas 
structures  may be responsible for some chemical signatures that we see averaged in our observations.

Alternatively, isobaric models (with $P$$\simeq$10$^{8}$ K cm$^{-3}$), in which the gas density naturally
increases from a few \mbox{$\sim$10$^{4}$ cm$^{-3}$}  in the  cloud edge to a few \mbox{$\sim$10$^{6}$ cm$^{-3}$} in the interior, have been recently  invoked to explain the CH$^+$, OH$^+$ and high-$J$ CO lines detected by \textit{Herschel} (\citealt{Nagy_2013}; \citealt{van_der_Tak_2013}; Joblin et al. in prep.). 
A high thermal pressure model of this kind is shown in Fig.~\ref{fig:oribar_hcarbons_abundances_2014_isobaric}.
In the following we restrict our discussion and model predictions for the observed hydrocarbon molecules.

\subsection{Gas-phase formation of small hydrocarbons}

\begin{figure}[ht]
\centering
\includegraphics[scale=0.55,angle=0]{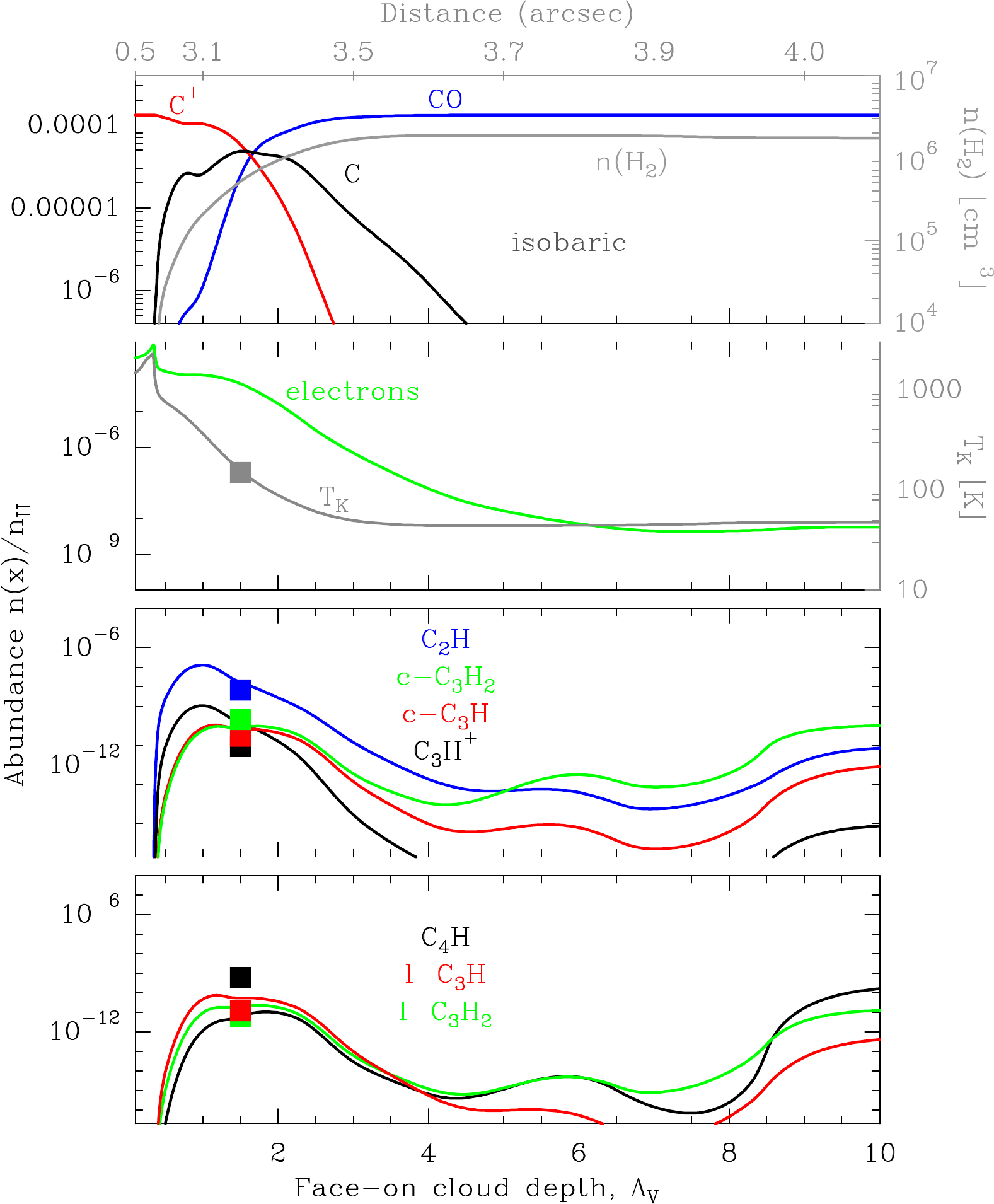} 
\caption{Reference isobaric photochemical model for the Orion Bar with $P$=10$^8$\,K\,cm$^{-3}$ and $\chi$=2$\times$10$^4$.
We note the increase of $n$(H$_2$) as the gas temperatures decreases. The grey square shows the beam-averaged gas temperature
estimated from our millimetre observations ($T_{\rm k}$$\simeq$150\,K), which corresponds to $A_{\rm V}$$\simeq$1.5 in the model.
The coloured squares show the beam-averaged abundances derived from the line survey.}
\label{fig:oribar_hcarbons_abundances_2014_isobaric}
\end{figure}

Figure~\ref{fig:oribar_hcarbons_abundances_2014_isobaric} shows model results for the  $P$$\simeq$10$^{8}$ K cm$^{-3}$ isobaric model that we take as our reference model. For such a highly UV-illuminated PDR, the first step of the gas-phase carbon chemistry is the \mbox{H$_{2}$ + C$^{+}$ $\rightarrow$ CH$^{+}$ + H} reaction, which becomes faster than the \mbox{H$_{2}$ + C$^{+}$ $\rightarrow$ CH$_{2}^{+}$ + photon} radiative association. The first reaction is endothermic by \mbox{E/k$\approx$4300 K}. Hot gas (a few hundred K) and/or the presence of (FUV pumped) vibrationally excited H$_{2}$ are needed to overcome the reaction endothermicity \citep[e.g.][]{Black_1987, Agundez_2010}. 
Such conditions apply to the edge of the Orion Bar, as demonstrated by the detection of rotationally excited CH$^{+}$ 
lines \citep[see e.g.][]{Habart_2010, Nagy_2013}. The presence of significant amounts of CH$^{+}$ produces enhanced 
abundances of CH$_{2}^{+}$ and CH$_{3}^{+}$ ions by subsequent reactions with H$_{2}$ (barrierless reactions). These 
simple hydrocarbon ions recombine and form CH and CH$_{2}$. In fact, CH$^{+}$ and CH show extended line emission in 
the entire Orion region (Goicoechea et al. in prep.) whereas CH$^{+}$ has not been detected in low FUV-flux
field PDRs like the Horsehead (Teyssier et al. in prep.). The coexistence of CH radicals and C$^{+}$ in the surface
of the PDR allows the formation of C$_{2}^{+}$ through the ion-neutral reaction \mbox{CH + C$^{+}$ $\rightarrow$ C$_{2}^{+}$ + H}. This starts the chemistry of species containing two carbon atoms. 

The C$_{2}^{+}$ ion reacts with H$_{2}$ to form C$_{2}$H$^{+}$, C$_{2}$H$_{2}^{+}$, and C$_{2}$H$_{3}^{+}$ by a series
of successive barrierless hydrogenation reactions. Recombination of these ions with electrons form the abundant C$_{2}$ and C$_{2}$H neutrals. Figure~\ref{fig:oribar_hcarbons_abundances_2014_isobaric} shows that most small hydrocarbons show a first abundance peak near the illuminated edge of the cloud where the predicted gas temperature sharply goes from \mbox{$\sim$1000 K} in the cloud surface, close to the ionisation front, to \mbox{$\sim$150 K} near the dissociation front at \mbox{$A_{\rm V}$ \negmedspace$\approx$1.5}. Such elevated temperatures contribute to enhancing the abundance of C$_{2}$H through the \mbox{C$_{2}$ + H$_{2}$ $\rightarrow$ C$_{2}$H + H} reaction, which has an activation energy barrier of \mbox{$E$/k$\approx$1500 K} \citep{Pitts_1982}. For the physical conditions prevailing in the edge of Orion Bar, this neutral-neutral reaction dominates the gas-phase formation of abundant C$_{2}$H. For this reason, the gas-phase production of C$_{2}$H may be more efficient in dense and hot PDRs than in cool PDRs
(we note the higher peak C$_2$H abundance in the clump model compared to the interclump model in
Fig.~\ref{fig:oribar_interclump_clump}). Like CH$^{+}$ or CH, the highest C$_{2}$H abundances are predicted at the illuminated edge of the cloud. Reaction of C$_{2}$H with H$_{2}$ then forms acetylene, C$_{2}$H$_{2}$. This reaction is also favoured at high temperatures \citep{Baulch_2005}. Since  C$^{+}$ is the most abundant carbon-bearing species in the PDR edge, 
further reactions of C$_{2}$H with C$^{+}$ produce C$_{3}^{+}$, that then reacts with H$_2$ to produce C$_{3}$H$^{+}$
(depending on the acetylene abundance, \mbox{C$_{2}$H$_{2}$ + C$^{+}$} can also contribute to C$_{3}$H$^{+}$ formation).
These are crucial intermediate precursors that form increasingly complex hydrocarbons. The chemistry of species containing three carbon atoms then proceeds. Reactions of C$_{3}$H$^{+}$ with H$_{2}$ produce the linear and cyclic forms of C$_{3}$H$_{2}^{+}$ and C$_{3}$H$_{3}^{+}$ isomers \citep[e.g.][]{Maluendes_1993, McEwan_1999}. 
Dissociative recombination of these ions then produces the 
\mbox{$c$-C$_{3}$H$_{2}$}, \mbox{$l$-H$_2$C$_{3}$}, and $l$- and $c$-C$_{3}$H isomers \citep[e.g.][]{Fosse_2001}.
The detection of $l$-C$_{3}$H$^{+}$ in the Horsehead \citep{Pety_2012} and Orion Bar, 
supports the above gas-phase routes for the synthesis of hydrocarbons containing several C atoms
in UV-illuminated gas (i.e. with available C$^{+}$).
We note that in the hydrocarbon abundance peak, their destruction is dominated by photodissociation
(e.g. for C$_2$H and $c$-C$_3$H$_2$ producing C$_2$, C$_3$, and $c$-C$_3$H, respectively, see~Fig.~\ref{fig:reactions}).  

Although not all chemical rates  and branching ratios involving hydrocarbons are known with precision, the Orion Bar is a good laboratory for testing this scheme (Fig.~\ref{fig:reactions}) because \emph{(i)} large column densities of C$^{+}$ exist, \emph{(ii)} the electron density is high enough to make recombination reactions efficient, and \emph{(iii)} the gas temperature is high enough to activate the barriers of reactions involving neutral 
carbon-bearing molecules with H and H$_{2}$.

\begin{table}
\centering 
\caption{Observed and best isobaric PDR model column densities.}
\label{Table_models}      	
\begin{tabular}{c c c @{\vrule height 10pt depth 5pt width 0pt}}   
\hline
\hline
	&	{\bf OBSERVED}	 &	{\bf MODEL}\tablefootmark{a}  \\  \cline{2-2} \cline{3-3}
      &  log$_{10}(N\,)$	&	log$_{10}(N\,)$	\\
	  &    [cm$^{-2}$]      & 	[cm$^{-2}$]	\\
\hline
C$_{2}$H			&	14.6		&	14.1\tablefootmark{b}-14.8\tablefootmark{c}	\\
$l$-C$_{3}$H$^{+}$	&	11.7		&	12.0\tablefootmark{b}-12.7\tablefootmark{c}	\\
$c$-C$_{3}$H$_{2}$	&	13.1		&	12.2\tablefootmark{b}-12.9\tablefootmark{c}	\\
$l$-H$_2$C$_{3}$	&	11.6		&	11.3\tablefootmark{b}-12.0\tablefootmark{c}	\\
$c$-C$_{3}$H			&	12.3		&	11.3\tablefootmark{b}-12.0\tablefootmark{c}	\\
$l$-C$_{3}$H			&	11.8		&	11.1\tablefootmark{b}-11.9\tablefootmark{c}	\\
C$_{4}$H			&	13.6		&	12.3\tablefootmark{b}-13.1\tablefootmark{c}	\\
\hline
\hline
\end{tabular}
\tablefoot{  \tablefoottext{a}{Isobaric model with $P$=10$^8$\,K\,cm$^{-3}$ and $\chi$=2$\times$10$^4$.}
Column densities integrated up to $N$(H$_2$)(face-on)$\backsimeq$10$^{22}$ cm$^{-2}$.
\tablefoottext{b}{Face-on configuration.}
\tablefoottext{c}{Column densities for an inclination angle of \mbox{$\alpha\negmedspace\backsimeq$11$^{\circ}$ with respect to an edge-on geometry.}}}
\end{table}

For the physical conditions prevailing in the Orion Bar, the gas temperature and ionisation fraction 
(the electron abundance or $x_e$) significantly drops at cloud depths larger than \mbox{$A_{\rm V}$ \negmedspace$\approx$4}. 
At these depths, most of the carbon becomes locked in CO and not in C$^{+}$ or C. This modifies the hydrocarbon chemistry.
Even deeper inside the cloud, molecules and atoms start to freeze-out so the
exact abundances of small hydrocarbons in cloud interiors are more uncertain. 
In particular, they critically depend on $x_e$, which determines the abundance of hydrocarbon ion precursors
\citep[see also][]{Fosse_2001}. The ionisation fraction in UV-shielded gas is set by the cosmic-ray ionisation-rate and by
 the poorly known
gas-phase abundance of low ionisation metals like Fe that carry most of the positive charge 
\citep[see][for the Horsehead]{Goicoechea_2009}.
The chemical time-scales become much longer than in the illuminated cloud edge and time-dependent effects 
are expected to be important \citep{Hollenbach_2009, Pilleri_2013}.
Our reference  isobaric model predicts that the hydrocarbon abundances peak again at
\mbox{$N_{\rm H}$(face-on)$\gtrsim$9$\times$10$^{21}$\,cm$^{-2}$}.
This is consistent with the detection of hydrocarbons in dark clouds 
 \citep[][]{Fosse_2001} and
with the spatial distribution of the low energy C$_2$H and \mbox{$c$-C$_3$H$_2$} lines in our maps (peaking
beyond the PDR, Fig.~\ref{fig:maps}). 

At least qualitatively, the abundances of small hydrocarbons beyond the cloud edge increase if $x_e$
decreases. 
The ionisation fraction in a cloud interior can be low
if the metallicity \mbox{(Fe, etc.)} is low or if the gas density is high (\mbox{$x_e \propto$ $\sqrt{\zeta_{\rm CR}/n_{\rm H}}$}; see the different $x_e$ gradients
and resulting hydrocarbon abundances at large cloud depths in the models shown in Fig.~\ref{fig:oribar_interclump_clump}).
Finally, $x_e$ can also be low if significant abundances ($\gtrsim$10$^{-7}$) of negatively charged species exist
(PAH, grains, or other large molecules to  which electrons could easily attach). 
This of course is far from being proven and  is still controversial.
In such a case, the abundances of hydrocarbons like $c$-C$_3$H$_2$ are expected to increase by large factors \citep[e.g.][]{Lepp_1988,Goicoechea_2009}.


\subsection{Comparison with observations}

Because the Orion Bar does not have a perfect \mbox{edge-on} orientation,  comparison of unidimensional PDR models and 
observations requires 
\emph{(i)} a knowledge of the inclination angle $\alpha$ with respect to a pure \mbox{edge-on} configuration and 
\emph{(ii)} knowledge of the equivalent \mbox{face-on} cloud depth (as in 1D models) of the observed \mbox{line of sight},
\mbox{$N_{\rm H}$(face-on) = $N_{\rm H}$(observed) $\times$ sin $\alpha$}. Different estimations have 
constrained the inclination angle to a maximum value of 15$^{\circ}$ (see discussion by \citealt{Melnick_2012} and 
references therein). This means a geometric column density enhancement of $\gtrsim$4 with respect to a 
pure \mbox{face-on} PDR model predictions. 

\begin{figure}
\centering
\includegraphics[scale=0.8,angle=0]{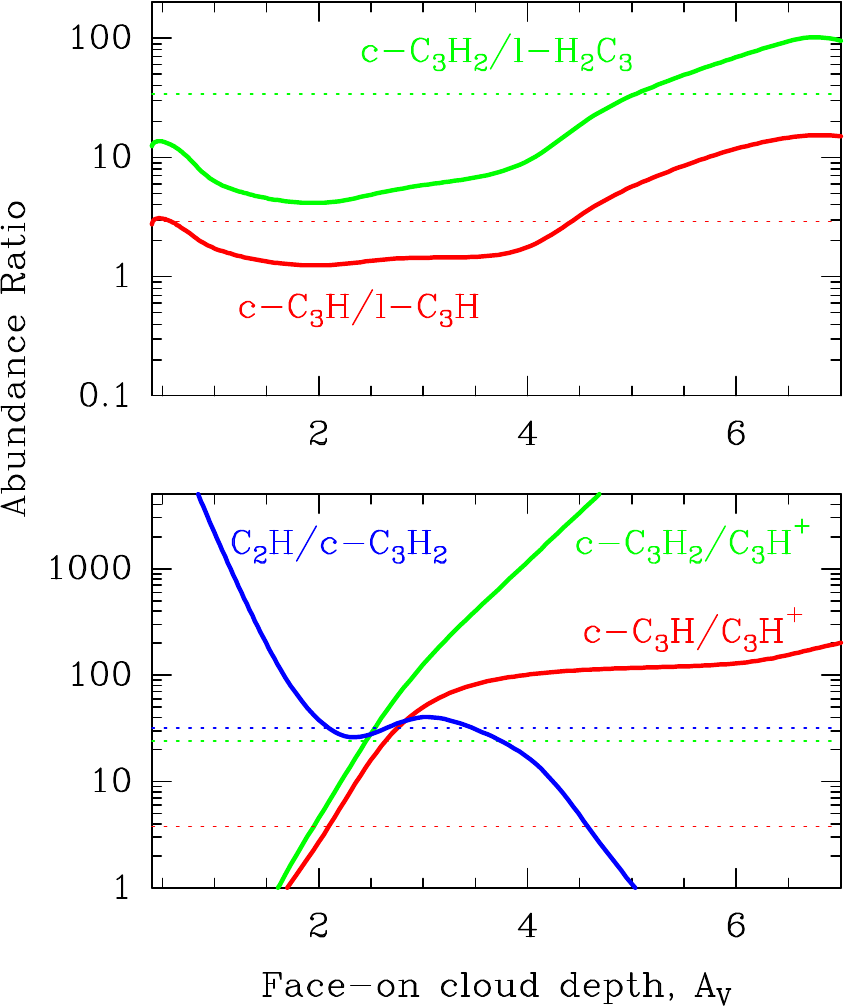} 
\caption{Reference isobaric photochemical model for the Orion Bar with $P$=10$^8$\,K\,cm$^{-3}$ and $\chi$=2$\times$10$^4$
showing selected hydrocarbon abundance ratios as a function of $A_{\rm V}$. The dashed horizontal lines show the 
ratios derived from the line survey.}
\label{fig:oribar_hcarbons_ratios_2014_isob_1e8}
\end{figure}

Table~\ref{Table_models} lists the column densities inferred from our observations towards the PDR survey position and the column densities predicted by the reference isobaric model ($P$=10$^8$\,K\,cm$^{-3}$) integrated from 
\mbox{$N$(H$_2$)(face-on)$\simeq$0} to \mbox{10$^{22}$ cm$^{-2}$} and for two different inclinations 
(face-on and \mbox{$\alpha\negmedspace\backsimeq$11$^{\circ}$}). We note that for the reference  model
with $\alpha\negmedspace\backsimeq$11$^{\circ}$, the face-on extinction $A_{\rm V}\negthickspace\simeq$10 is equivalent
to an angular scale $\sim$4$''$/sin\,11$^{\circ}\negmedspace\simeq$20$''$ at the distance of the Orion Bar PDR
 (see Fig.~\ref{fig:oribar_hcarbons_abundances_2014_isobaric}), 
roughly the average beam of our millimetre observations.

In this range of values, the match between observations and our reference models is reasonably good. 
In particular, the column densities of all small hydrocarbons can be reproduced within factors of $<$3.
The predicted variation of selected hydrocarbon abundance ratios with cloud depth (decreasing FUV field) is shown in 
Fig.~\ref{fig:oribar_hcarbons_ratios_2014_isob_1e8}.
Taking the column density of \mbox{$l$-C$_3$H} as a reference (the model matches the observed value and this 
molecule is expected to show enhanced abundances in high $x_e$ environments; see \citealt{Fosse_2001}) 
the PDR model also reproduces the observed \mbox{[C$_2$H]/[$l$-C$_3$H]}, \mbox{[$c$-C$_3$H$_2$]/[$l$-C$_3$H]}, and \mbox{[$l$-H$_2$C$_3$]/[$l$-C$_3$H]} column density ratios within a factor of 2. Nevertheless, the column density of other species, C$_4$H in particular, is underestimated by a factor of 3 in 
the reference model, showing that the agreement is clearly not perfect.
Still, given the complexity of the region and the geometrical simplicity of our models, this
 is much better than the \mbox{order-of-magnitude} differences 
reported between observations and models of low FUV-flux PDRs like the Horsehead \citep{Pety_2005, Pety_2012}. 
In this PDR, photo-destruction of PAHs or very small grains (VSGs) has been invoked to dominate the production of
 hydrocarbons like C$_2$H or $c$-C$_3$H$_2$.
In the Orion Bar, the FUV flux is much higher and our  model results suggest that photochemical models can explain the observed C$_2$H, C$_3$H, and C$_3$H$_2$ column densities without invoking a
major contribution of PAH photodestruction. This conclusion would agree with the 
different spatial distributions of the PAHs and the C$_2$H and $c$-C$_3$H$_2$ emission seen along the Orion Bar. An important difference
compared to the diffuse clouds or low UV-flux PDRs like the Horsehead is the much higher temperature attained by the
gas and the elevated abundances of vibrationally excited H$_{2}$ activating many neutral-neutral reactions
 that likely play a minor role in the Horsehead or in diffuse clouds (see also Sect.~7.4).

 For the more complex hydrocarbons (i.e. those with more than three carbon atoms), however, steady-state gas-phase models do not provide
 an entirely satisfactory fit. Time-dependent photochemical models, such as those applied to C-rich protoplanetary nebulae 
 \citep[e.g.][]{Cernicharo_2004} show that the steady-state abundances of several organic species are different from those obtained during the gas time evolution.
Hence, they may fit some specific hydrocarbons better despite the short time-scales in PDRs.


\section{Discussion}

\begin{figure}[t]
\centering
\includegraphics[scale=0.55,angle=0]{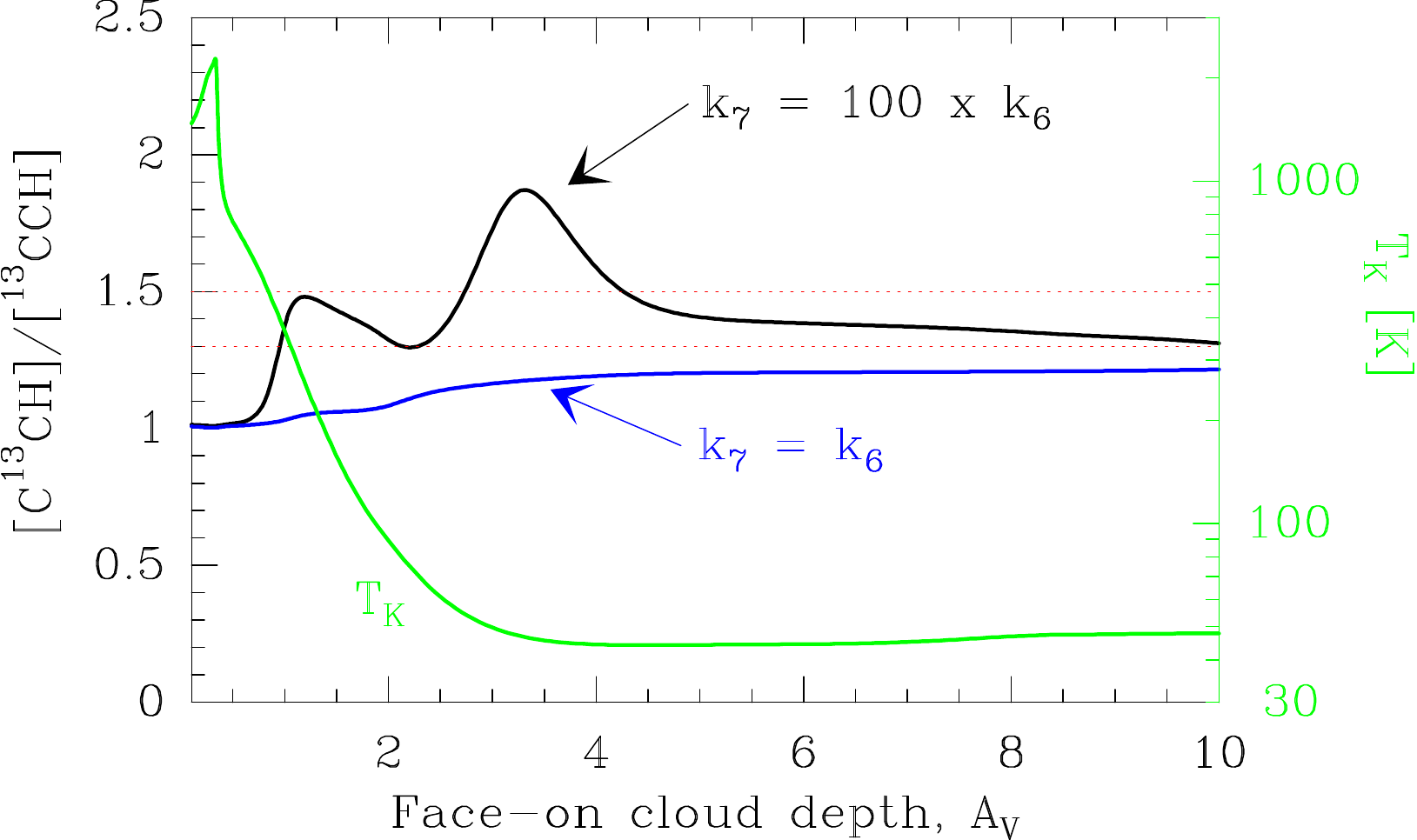} 
\caption{Reference isobaric photochemical model for the Orion Bar with $P$=10$^8$\,K\,cm$^{-3}$ and $\chi$=2$\times$10$^4$. $^{13}$C fractionation reactions have been included to explain the observed [C$^{13}$CH]/[$^{13}$CCH] ratio (shown in the region enclosed by the red dashed lines). The black curve show a model 
in which the rate of the $\rm{^{13}C^{+} + CCH}$ reaction is 100 times higher than
that of the  $\rm{^{13}CCH + H}$ reaction (see text).}
\label{fig:cch_frac}
\end{figure}

\subsection{Rotationally hot C$_{4}$H and $l$-C$_{3}$H$^{+}$}

The rotational population analysis presented in Sect.~5.2 shows that C$_{4}$H, and to a lesser extent $l$-C$_3$H$^+$, have unusually high rotational 
temperatures compared to other small hydrocarbon molecules detected in the millimetre domain. This is more clearly seen in C$_{4}$H, which has a similar dipole moment and rotational spectroscopy to C$_{2}$H but shows much higher rotational temperatures ($T_{\rm rot} \approx$ 77 K versus 26 K). The C$_{4}$H radical has a complicated vibronic spectroscopy due to the proximity of the degenerated $^{2}\Pi$ excited electronic state (with a much higher \mbox{4.3 Debye} dipole moment) only \mbox{$\sim$300 K} above the X$^{2}\Sigma^{+}$ ground state. The Renner-Teller effect, spin-orbit interactions, 
and other couplings complicate the low energy rovibronic structure of C$_{4}$H \citep[see e.g.][]{Senent_2010, Mazzotti_2011}.
We suspect that radiative pumping contributes to the excitation of  
the lowest lying bending modes of C$_4$H. Indeed, the derived rotational temperature for C$_4$H, $T_{\rm rot}\negmedspace\approx$77 K, is very similar to 
the dust temperature inferred towards the edge of the Orion Bar PDR \citep{Arab_2012}. 
Therefore, it is plausible that the absorption of IR continuum photons from warm grains heated by the 
strong FUV-radiation field  contributes to the C$_{4}$H excitation. In this context, the inferred $T_{\rm rot}$ would 
be more representative of the dust grain temperature in the PDR edge than of the gas temperature. 
The low energy modes of C$_4$H have been detected in the circumstellar envelope around the carbon-rich star IRC+10216 \citep{Guelin_1987}.
However, we have not detected  rovibrational lines from the lowest energy C$_{4}$H bending mode, $\nu$7 \citep{Yamamoto_1987b} probably due to the limited sensitivity of our line survey.  The upper limit C$_{4}$H [$\nu$7]/[$\nu$=0]<10$\%$ column density ratio we derive, however, is relatively high.

In Sect.~3.2.1 we concluded that $l$-C$_3$H$^+$ is a floppy molecule that very likely has low lying bending modes. Although we have not detected lines with their expected spectroscopic pattern, these levels can also be populated at relatively low gas temperatures through IR pumping. The $l$-C$_3$H$^+$ ion is a high dipole-moment molecule and, for the physical conditions in the Orion Bar PDR, its rotational levels are expected to be subthermally excited ($T_{\rm ex}\negmedspace\ll$$T_{\rm k}$). However, $T_{\rm rot}$($l$-C$_3$H$^+$)$\approx$46\,K 
is significantly higher\footnote{The $T_{\rm rot}$($l$-C$_3$H$^+$)=178(3)\,K value inferred by
\citet{McGuire_2014} from their 
observation of the J=9$\rightarrow$8, 10$\rightarrow$9, and 11$\rightarrow$10 lines is significantly higher than the value we obtain from our
 multi-line observations with the IRAM 30m telescope. The presence of significant $l$-C$_3$H$^+$ emission dilution in the larger CSO telescope beam
(\mbox{HPBW$_{\rm CSO}$[arcsec]$\approx$7000/Frequency[GHz],} \citealt[][]{Mangum_1993}) would lead to apparently larger values of $T_{\rm rot}$
if beam dilution is not corrected. 
In particular, we compute that a $l$-C$_3$H$^+$ emission source size of $\theta_{_{S}}\negthickspace\lesssim$17" (leading to a beam dilution of  
$\eta_{_{bf}}$(CSO)$\lesssim$0.23 at $\sim$225~GHz) reconciles both data sets.}
than the rotational temperatures inferred for all the other small hydrocarbons molecules (except C$_4$H). Again, this is an indication 
that IR pumping likely affects the $l$-C$_3$H$^+$ excitation.


\subsection{Fractionation of C$^{13}$CH and $^{13}$CCH isotopomers} 

The inferred [C$^{13}$CH]/[$^{13}$CCH]=\mbox{1.4$\pm$0.1 (3$\sigma$)} column density ratio towards the Orion
Bar implies differential $^{13}$C fractionation of CCH isotopologues. \citet{Sakai_2010} also observed both 
isotopomers towards the dark cloud TMC-1  and the star-forming core exhibiting warm carbon-chain chemistry L1527.
 They derived [C$^{13}$CH]/[$^{13}$CCH]=1.6$\pm$0.4 and 1.6$\pm$0.1, respectively.

\citet{Furuya_2011} suggested that the observed fractionation in cold and dense gas could be explained by
the isotopomer exchange reaction
\begin{equation}
\rm{^{13}CCH + H \rightleftarrows  C^{13}CH + H + \Delta \textit{E}},
\end{equation}
where  $\Delta E$$\simeq$8\,K is the difference between the zero point energy (ZPE) of
C$^{13}$CH and $^{13}$CCH \citep[see also][]{Tarroni_2003}.
Compared to a dark cloud, the Orion Bar shows different physical conditions driven by the presence
 of a strong UV radiation field: a large C$^+$ abundance, a higher fraction of H atoms with respect to H$_2$, 
 and much more elevated gas temperatures. In the warm PDR gas, and in the absence of an activation barrier \citep[suggested by][]{Furuya_2011}, 
 reaction (6) will not  enhance the C$^{13}$CH abundance  significantly above C$^{13}$CH (by only $\sim$5\% at $\sim$150\,K). 
Therefore, in addition to reaction (6), we suggest that reactions
\begin{equation}
\rm{ ^{13}C^{+} + CCH \rightleftarrows\, C^{13}CH + C^+ + \Delta \textit{E'}}
\end{equation}
and
\begin{equation}
\rm{ ^{13}C^{+} + CCH \rightleftarrows\,^{13}CCH + C^+ + \Delta \textit{E''},}
\end{equation}
can contribute to the  differential fractionation of C$^{13}$CH and $^{13}$CCH.
Both isotopomers  have relatively high ZPE differences with respect to CCH 
($\Delta E'\negthickspace\simeq$63\,K and $\Delta E''\negthickspace\simeq$55\,K, respectively)
and both reactions (7) and (8) are more endothermic in the backward direction than reaction~(6).
We note that in a high UV-flux PDR, the gas temperature is high enough to prevent significant $^{13}$C$^+$ depletion
through the reaction
\begin{equation}
\rm{ ^{13}C^{+} + CO \rightleftarrows\,C^+ +\, ^{13}CO + 34.8\,K}
\end{equation}
\citep[see][for low \mbox{[\CII]/[$^{13}$\CII]} line intensity ratios towards the Orion Bar]{Ossenkopf_2013}. 
Owing to the higher H atom abundance in  PDRs, reaction (7) needs to be faster (e.g. forming  C$^{13}$CH) 
than the backward  isotopomer exchange reaction (6) (e.g.~destroying  C$^{13}$CH).
Figure~\ref{fig:cch_frac} shows our reference isobaric model in which $^{13}$C fractionation reactions involving
$^{13}$C$^+$, $^{13}$CO, and H$^{13}$CO$^+$ have been included \citep[see][]{Langer_1984,Bourlot_1993}.
The blue curve shows the predicted depth-dependent [C$^{13}$CH]/[$^{13}$CCH] abundance ratio in a  model with the same rate 
for reactions (6), (7), and (8). In this model, the exchange reactions with H atoms  dominate and  C$^{13}$CH only 
fractionates at large cloud depths (by $\sim$20\% at $\sim$50\,K) where the gas temperature significantly decreases.
The black curve shows a model with \mbox{$k_7= k_8 =100 \times k_6$}. In this model, fractionation reactions 
with $^{13}$C$^{+}$ can dominate in the warm UV-illuminated gas and the  [C$^{13}$CH]/[$^{13}$CCH] ratio already increases at the cloud surface (becoming compatible with our observations). 
We note, however, that reaction (6) must always be present, otherwise the  [C$^{13}$CH]/[$^{13}$CCH] ratio will 
be much higher than the observed value as the gas temperature decreases. Quantum calculations
 and/or laboratory experiments  are needed to constrain the reaction rates and potential energy activation barriers
 of these processes in detail.


\subsection{Cyclic versus linear isomers} 

The linear and cyclic isomers of a given hydrocarbon species can have different behaviours with respect to neutral-neutral and ion-neutral reactions, and thus with respect to different physical conditions.  \citet{Fosse_2001} reported high \mbox{[$c$-C$_{3}$H$_{2}$]/[$l$-H$_2$C$_{3}$]} ($\sim$28) and \mbox{[$c$-C$_{3}$H]/[$l$-C$_{3}$H]} ($\sim$13) abundance ratios towards the cyanopolyyne peak in the cold dark cloud TMC-1. These ratios are higher than those observed in diffuse and translucent clouds:  \mbox{[$c$-C$_{3}$H$_{2}$]/[$l$-H$_2$C$_{3}$]$\simeq$(3-5)} \citep{Cernicharo_1999} and \mbox{[$c$-C$_{3}$H]/[$l$-C$_{3}$H]$\simeq$2} \citep{Turner_2000}. \citet{Teyssier_2005} showed that the cyclic-to-linear C$_{3}$H$_{2}$ column density ratio in the Horsehead nebula increases from the UV-illuminated layers to the shielded cloud interior. \citet{Fosse_2001} also explored the chemistry in TMC-1 and concluded that the cyclic-to-linear abundance ratio of  C$_3$H$_2$ increases with decreasing 
electron abundances. Our reference model for the Orion Bar also predicts that both the \mbox{[$c$-C$_{3}$H$_{2}$]/[$l$-H$_2$C$_{3}$]} and the \mbox{[$c$-C$_{3}$H]/[$l$-C$_{3}$H]} ratios increase with $A_{\rm V}$ as the FUV-radiation field is attenuated and $x_e$ decreases (see Fig.~\ref{fig:oribar_hcarbons_ratios_2014_isob_1e8}). This suggests that the
formation of the linear isomers (less stable energetically) is favoured in the warm UV-illuminated gas with high ionisation fractions. 
The low \mbox{[$c$-C$_{3}$H]/[$l$-C$_{3}$H]$\simeq$3} abundance ratio we infer towards the Orion Bar is the same as that found in the PDR around the protoplanetary nebula CRL 618 \citep{Pardo_2007} and slightly higher than that inferred towards the Horsehead PDR ($\sim$1.8) \citep{Teyssier_2004}. Such low abundance ratios ($\leq$3) therefore seem a  signature of the presence of FUV radiation and high ionisation fractions. 
On the other hand, the \mbox{[$c$-C$_{3}$H$_{2}$]/[$l$-H$_2$C$_{3}$]$\simeq$34} ratio we infer towards the Orion Bar is much higher than that observed in the Horsehead PDR ($\simeq$3.5) and in diffuse clouds. 
At this point, the reason for such a difference is not clear, but may suggest that in the Orion Bar, the $c$-C$_{3}$H$_{2}$ abundance is enhanced by formation routes not considered in our pure-gas phase models.


\subsection{C$_{2}$H versus c-C$_{3}$H$_{2}$ in other environments} 

\begin{figure}[b]
\includegraphics[scale=0.5,angle=0]{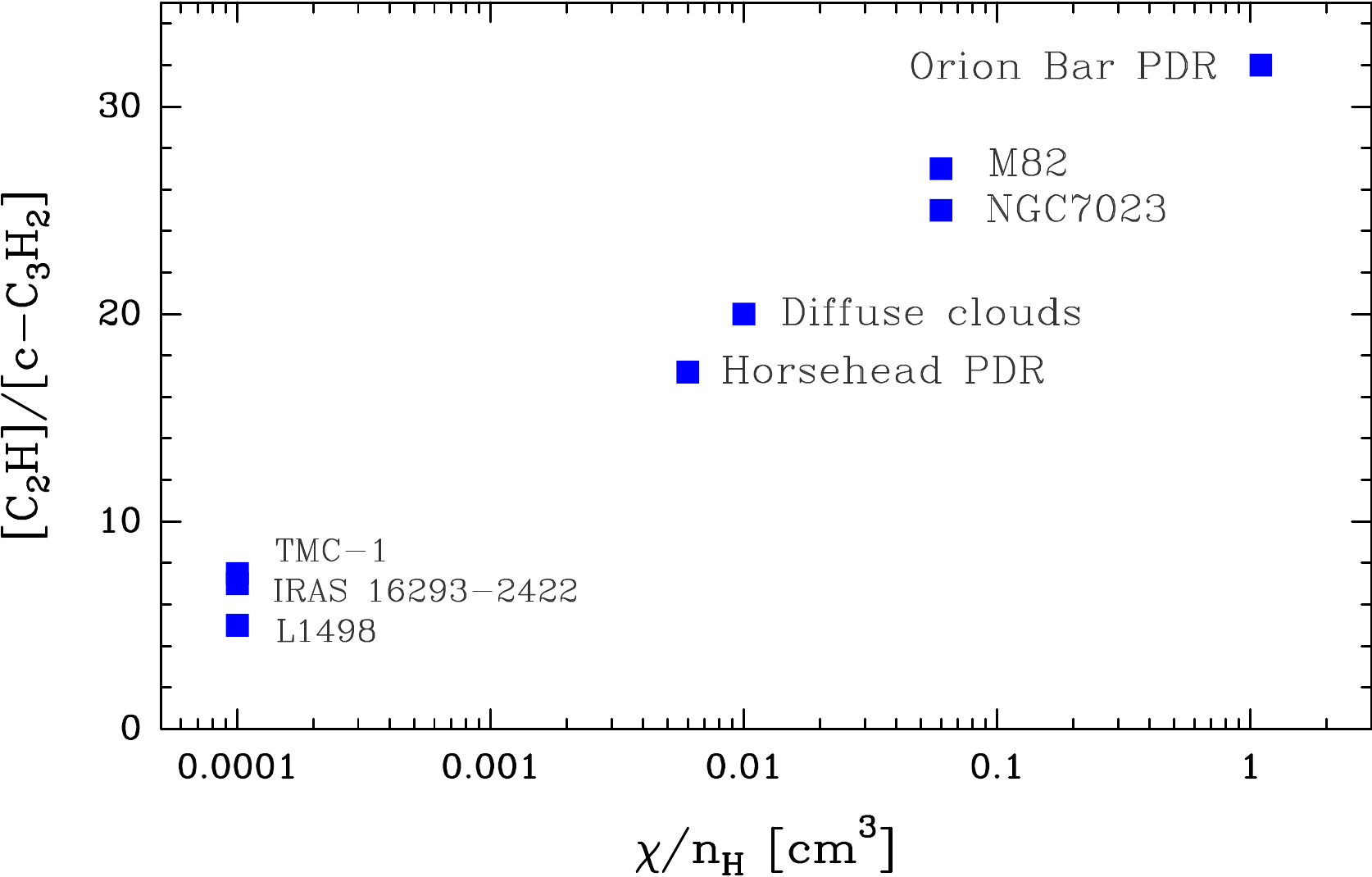} 
\caption{Observed [C$_{2}$H]/[$c$-C$_{3}$H$_{2}$] column density ratios as a function of \mbox{$\chi/n_{\rm H}$} dissociation parameter (see text) in
the Orion Bar PDR,
the nucleus of M82 \citep{Aladro_2011},
NGC7023 (NW PDR; \citealt{Fuente_1993,Fuente_2003}),
diffuse clouds (averaged abundances towards B0355, B0415, B2200, and B2251; \citealt{Liszt_2012}),
Horsehead PDR \citep{Pety_2012}, 
TMC-1 (towards the cyanopolyyne peak; \citealt{Ohishi_1992}),
IRAS 16293-2422 \citep{vanDishoeck_1995}, and
L1498 \citep{Tafalla_2006,Padovani_2009}.}\label{fig:ratios}
\end{figure}

In order to investigate the role of UV radiation and gas density in the formation of the small hydrocarbons, we compare the column 
density of several hydrocarbons in different environments. The considered sources are the Monoceros R2 (Mon R2) ultra-compact \HII\,region
 ($\chi\negmedspace>$10$^{5}$ in Draine units 
and \mbox{$\chi/n_{\rm H}\negmedspace\simeq$10$^{-1}$ cm$^{3}$} dissociation parameter, e.g. \citealt{Pilleri_2013}), 
the Orion Bar PDR ($\chi\negthickspace \approx$10$^{4}$ and \mbox{$\chi/n_{\rm H}\negmedspace\simeq$0.5 cm$^{3}$}, e.g. \citealt{Marconi_1998}),
the nucleus of the starburst galaxy M82 ($\chi\negthickspace \approx$6$\times$10$^{3}$ and \mbox{$\chi/n_{\rm H}\negmedspace\simeq$0.06 cm$^{3}$}, e.g. \citealt{Fuente_2008}),
the reflection nebula NGC7023 ($\chi\negthickspace \approx$10$^{3}$ and \mbox{$\chi/n_{\rm H}\negmedspace\simeq$0.06 cm$^{3}$}, e.g. \citealt{Joblin_2010}), 
diffuse clouds ($\chi\negthickspace \approx$1 and \mbox{$\chi/n_{\rm H}\negmedspace\simeq$10$^{-2}$ cm$^{3}$}, e.g. \citealt{Liszt_2012}), and
the Horsehead PDR ($\chi\negthickspace \approx$60 and \mbox{$\chi/n_{\rm H}\negmedspace\simeq$6$\times$10$^{-3}$ cm$^{3}$}, e.g. \citealt{Pety_2012}).
We also considered three sources shielded from external UV-illumination: 
the cold and dense cloud \mbox{TMC-1}, 
the low-mass protostar (hot corino) IRAS 16293-2422, 
and the dense core L1498 (the three with a \mbox{$\chi/n_{\rm H}\negmedspace\ll$10$^{-4}$ cm$^{3}$}). 
 
The variation of incident UV radiation flux affects the relative abundance of certain hydrocarbons. 
As expected for  widespread interstellar molecules, intense  C$_{2}$H and \mbox{$c$-C$_{3}$H$_{2}$} emission is detected in all the above sources.
 However, the observed \mbox{[C$_{2}$H]/[$c$-C$_{3}$H$_{2}$]} column density ratio varies from highly irradiated  sources like the Orion Bar
  ($\sim$32) to UV-shielded sources ($<$10) (see Fig.~\ref{fig:ratios}). In fact, the observed \mbox{[C$_{2}$H]/[$c$-C$_{3}$H$_{2}$]} ratios seem to scale with $\chi/n_{\rm H}$, the critical parameter determining most of the PDR properties. In the Orion Bar model this can be readily seen in the predicted decrease of the \mbox{[C$_{2}$H]/[$c$-C$_{3}$H$_{2}$]} abundance ratio with increasing $A_{\rm V}$ (Fig.~\ref{fig:oribar_hcarbons_ratios_2014_isob_1e8}). This trend can be explained by the effect of UV radiation on the chemical processes governing the hydrocarbon formation.
In particular, the chemistry in cold shielded gas is driven by the ionisation of H$_{2}$ by cosmic rays, and the hydrocarbons are mainly produced by 
ion-molecule barrierless reactions. Time-dependent effects and grain surface processes are also likely important deep inside clouds \citep{Pilleri_2013}. 
In strongly UV irradiated environments, the presence of C$^{+}$ and H$_{2}^{*}$ triggers the rapid formation of hydrocarbons ions like 
CH$^+$,  CH$_2$$^+$, and CH$_3$$^+$ (see Sect.~6.2). In the associated warm gas, neutral-neutral reactions (e.g. reactions of C$_n$ or neutral hydrocarbons with H and H$_2$) 
that do not play a role in the cold gas become efficient \citep[see e.g.][]{Cernicharo_2004}. They allow high abundances of hydrocarbon molecules to be maintained despite the large UV field. 
For the typical densities in PDRs ($\sim$10$^5$\,cm$^{-3}$) our gas-phase models predict higher column densities of C$_2$H 
in strongly irradiated PDRs ($\chi\negmedspace>$1000) than in low UV-flux PDRs. 
On the other hand, they predict decreasing column densities of \mbox{$c$-C$_3$H$_2$} for $\chi\negmedspace>$1000. 
Therefore, the \mbox{[C$_2$H]/[$c$-C$_3$H$_2$]} ratio is expected to increase with the strength of the UV radiation 
field and observations seem to confirm this. For high UV-fluxes, photodissociation is the main destruction mechanism of C$_2$H and
$c$-C$_3$H$_2$ up to a few $A_{\rm V}$. The $c$-C$_3$H$_2$ photodissociation rate is a factor of $\sim$4  higher than that of C$_2$H \citep[e.g.][]{vanDishoeck_2006}.
For the Orion Bar physical conditions this contributes to the fact that C$_2$H is predicted to peak slightly closer to the dissociation front,
and also to the general increase in $c$-C$_3$H$_2$ column density compared to C$_2$H when the UV field decreases.

The dense PDR around Mon~R2\,\HII\,region (not shown in Fig.~~\ref{fig:ratios}) displays the highest 
\mbox{[C$_{2}$H]/[$c$-C$_{3}$H$_{2}$]$\sim$125} ratio \citep{Pilleri_2013}. 
This is a more extreme environment where 
the gas is heated to very high temperatures, favouring
the production of simple hydrocarbons like CH$^+$ \citep{Pilleri_2014} or C$_2$H  \citep{Pilleri_2013} through the hot-gas PDR chemistry
described in Sect.~6.1.

Hydrocarbons are also detected in diffuse clouds \citep[e.g.][]{Liszt_2012, Liszt_2014}, where radiation fields ($\chi \negthickspace \approx$1) and densities \mbox{($n_{_{\rm H}} \negthickspace \approx$100 cm$^{-3}$)} are lower than in PDRs. 
However, owing to the similar $\chi/n_{_{\rm H}}$ dissociation parameter, diffuse clouds and low-FUV flux PDRs are expected to share common characteristics. The \mbox{[C$_{2}$H]/[$c$-C$_{3}$H$_{2}$]} ratio inferred in diffuse clouds ($\sim$20) is 
more similar to the one inferred in the Horsehead PDR ($\sim$17), a low UV-flux and low $\chi/n_{\rm H}$ PDR, and indeed is lower than the value observed in the Orion Bar ($\sim$32). 
 
In contrast to the low and high illumination PDRs, dark clouds and hot corinos show the lowest [C$_{2}$H]/[$c$-C$_{3}$H$_{2}$] ratios ($<$10). The exact ratios in these UV-shielded environments are probably very time-dependent, i.e. the molecular abundances depends on the evolutive stage of the clouds.

In a broader extragalactic context, the nucleus of M82, which is the most studied example of an extragalactic starburst, shares similar photochemical characteristics with high UV-flux galactic PDRs. The chemistry of its nucleus seems to be the result of an old starburst mainly affected by the influence of intense UV fields from massive stars, where star formation has almost exhausted the molecular gas reservoir \citep{Fuente_2008}. In the study of the chemical complexity of the NE lobe of the M82 galaxy carried out by \citet{Aladro_2011}, they found that C$_{2}$H is the brightest spectral feature in the 1.3 mm and 2 mm bands. The \mbox{[C$_{2}$H]/[$c$-C$_{3}$H$_{2}$]} ratio observed towards M82 ($\sim$27) lies in between that of the Orion Bar ($\sim$32) and of NGC7023 \citep[$\sim$18-32; see][]{Fuente_1993, Fuente_2003}. 

In summary, observations and models suggest that the \mbox{[C$_{2}$H]/[$c$-C$_{3}$H$_{2}$]} ratio is a good tracer of increasing 
$\chi/n_{_{\rm H}}$ values. Ratios above $\sim$10 suggest the presence of UV radiation ($\chi\negmedspace>$1), 
with \mbox{[C$_{2}$H]/[$c$-C$_{3}$H$_{2}$]}  ratios above $\sim$30 probing the presence of strong radiation fields 
($\chi\negmedspace>$10$^3$). In combination with the observation of low \mbox{[$c$-C$_3$H]/[$l$-C$_3$H]} isomeric ratios ($\leq$3), 
characteristic of high $x_e$ environments, these ratios are good diagnostics of the presence of an active photochemistry. 
On the other hand,  \mbox{[C$_{2}$H]/[$c$-C$_{3}$H$_{2}$]$\lesssim$10} and \mbox{[$c$-C$_3$H]/[$l$-C$_3$H]$\gtrsim$10} ratios are indicative 
of low $x_e$ gas shielded from external UV radiation.


\subsection{PAH/HAC photodestruction and grain surface chemistry}

Another difference between low and high UV-illumination environments is the spatial distribution of the
hydrocarbons emission. The C$_{2}$H and $c$-C$_{3}$H$_{2}$ emission spatially correlates very well in the Horsehead
PDR \citep{Pety_2005}, diffuse clouds \citep{Lucas_2000, Gerin_2011}, and in the Orion Bar PDR (Fig.~\ref{fig:maps}). Furthermore, a tight correlation between the PAHs and the small hydrocarbon emission was found in the Horsehead PDR from high angular resolution interferometric observations \citep{Pety_2005}. Following previous suggestions \citep[][]{Fosse_2000, Fuente_2003, Teyssier_2004}, \citet{Pety_2005} proposed that the
photo-fragmentation of PAHs likely increases the abundance of small hydrocarbons in the Horsehead. In the Orion Bar, 
the \mbox{8 $\mu$m} PAH emission and the C$_{2}$H and $c$-C$_{3}$H$_{2}$ emission clearly show a different spatial distribution (Fig.~\ref{fig:maps}).
Lacking  higher angular resolution observations and a complete model of the PAH/VSG photoerosion \citep{Pilleri_2012}, 
we can only conclude that in strongly irradiated PDRs like the Orion Bar, photodestruction of PAHs is not a necessary condition 
to explain the observed abundances of the simplest hydrocarbons.

Nevertheless, additional top-down formation routes for hydrocarbon molecules not included in our gas-phase models may take place in PDRs. In particular, recent ultra-high vacuum experiments with carbonaceous grains show that
 hydrogen atoms attached to the grain surface
can efficiently react and produce a large variety of organic molecules, from PAHs to acetylene \citep{Merino_2014}. 
We note that C$_2$H$_2$ photodissociation produces C$_2$H and the reaction of C$_2$H$_2$ with C$^+$ forms the observed hydrocarbon ion $l$-C$_3$H$^+$, an important gas-phase precursor of C$_3$H$_2$ and C$_3$H. 
In addition, laboratory experiments performed by \citet{Alata_2014} show that the photodestruction of 
hydrogenated amorphous carbon (HAC) grains, observed in the diffuse medium \citep{Duley_1983}, 
also leads to the production of small hydrocarbons (such as CH$_{4}$) that can trigger the gas-phase formation
of other hydrocarbons.

Since PAHs, carbonaceous grains, and H atoms are abundant in PDRs, 
both the photodestruction of PAHs/HACs and the chemistry that 
takes place at the surfaces of carbonaceous grains will need to be taken into account in future PDR models.


\section{Summary and conclusions}

We have investigated the spatial distribution and chemistry of small hydrocarbons in the strongly UV-irradiated Orion Bar PDR. 
We performed a complete millimetre line survey towards the Orion Bar dissociation front (the "CO$^+$ peak") covering a bandwidth of $\sim$220 GHz 
using the IRAM 30m telescope. These observations have been complemented with 
$\sim$2$'\times$2$'$ maps of the  C$_{2}$H and $c$-C$_{3}$H$_{2}$ emission. 
Approximately 40$\%$ of the detected lines have been assigned to hydrocarbons 
(C$_{2}$H, C$_{4}$H, $c$-C$_{3}$H$_{2}$, $c$-C$_{3}$H, C$^{13}$CH, $^{13}$CCH, $l$-C$_{3}$H, and $l$-H$_2$C$_{3}$ in decreasing order of abundance). 
We also present the detection of nine rotational lines of the newly discovered hydrocarbon ion $l$-C$_{3}$H$^{+}$, allowing us to improve its spectroscopic constants. 
No anions, lines from vibrationally excited states, or deuterated hydrocarbons were detected. 
A detailed analysis of the excitation conditions and chemistry was carried out. 
In particular, we obtained the following results:\\\\
$\bullet$ Although the Orion Bar is a harsh environment, the millimetre line survey shows a relatively rich molecular line spectra, with more than 200 lines arising from hydrocarbons.\\
$\bullet$ The inferred rotational temperatures range from 17 to 77~K (most species have \mbox{$T_{\rm rot}\negthickspace<$30 K}) and column densities ranging from 10$^{11}$ to 10$^{14}$ cm$^{-2}$. C$_{2}$H is the most abundant of the detected hydrocarbons ($\sim$10$^{-8}$ with respect to H nuclei).\\
$\bullet$ We obtain similar rotational temperatures for ortho and para forms of cyclic and linear C$_{3}$H$_{2}$. The inferred $c$-C$_{3}$H$_{2}$ ortho-to-para ratio is 2.8$\pm$0.6, consistent with the high temperature limit.\\
$\bullet$ The [C$^{13}$CH]/[$^{13}$CCH] ratio is 1.4$\pm$0.1 and shows that fractionation processes differently affect the two $^{13}$C isotopes of C$_{2}$H. We suggest that reactions of C$_{2}$H with $^{13}$C$^+$, as well as reactions of C$_{2}$H isotopologues with H atoms,
can explain the observed levels of  C$^{13}$CH fractionation in the Orion Bar.\\
$\bullet$ We constrain the beam-averaged physical conditions from non-LTE models of C$_{2}$H and
$c$-C$_{3}$H$_{2}$. 
The best fits for C$_{2}$H are obtained for $T_{\rm k}\negthickspace\gtrsim$150~K and $n$(H$_{2}$)$\gtrsim$10$^{5}$\,cm$^{-3}$. Slightly denser gas and lower temperatures are required to fit the 
 $c$-C$_{3}$H$_{2}$ lines.\\
$\bullet$ We provide accurate upper limit abundances for chemically related carbon bearing molecules that are not detected in the PDR: [C$_{2}$D]/[C$_{2}$H]<0.2$\%$, [C$_{2}$H$^{-}$]/[C$_{2}$H]<0.007$\%$, and [C$_{4}$H$^{-}$]/[C$_{4}$H]<0.05$\%$.\\ 
$\bullet$ Hydrocarbon molecules show moderate abundances towards the FUV-illuminated edge of the cloud, but they are also  abundant in the more shielded cloud interior. The observed decrease of the [C$_{2}$H]/[$c$-C$_{3}$H$_{2}$] column density ratio from the dissociation front to the molecular peak and observations towards different environments  suggest that the \mbox{[C$_{2}$H]/[$c$-C$_{3}$H$_{2}$]} abundance ratio increases with increasing $\chi$/$n_{\rm H}$ values. In addition, the observation of low \mbox{[$c$-C$_3$H]/[$l$-C$_3$H]} ratios ($\leq$3) in the Orion Bar PDR probes a high electron abundance environment.\\
$\bullet$ We compare the inferred column densities with updated photochemical models. Our models can reasonably match the observed column densities of most hydrocarbons (within factors of $<$3). The largest discrepancy is for C$_4$H: our model underestimates the C$_4$H column density by a factor of $\sim$3. Since the observed spatial distribution of the C$_{2}$H and $c$-C$_{3}$H$_{2}$ emission is similar but does not follow the PAH emission, we conclude that the photodestruction of PAHs is not a necessary requirement in high UV-flux PDRs to explain the observed abundances  
of the smallest hydrocarbons. Instead, endothermic reactions (or with barriers) between C$^{+}$, radicals, and H$_{2}$ can dominate their formation. Still, photoerosion of PAHs/HACs/VSGs and  surface chemistry on carbonaceous grains may be needed to explain the abundances of more complex hydrocarbons. \\
$\bullet$ The electron abundance influences the hydrocarbon chemistry beyond the cloud layers directly exposed to the
UV radiation field. Unfortunately, the ionisation fraction depends on the poorly known abundances of low ionisation metals, on the density profile, and on the controversial presence of negatively charged PAH, grains, or polyatomic anions in cloud interiors. Improving our knowledge of these aspects, and on the products and details of PAH/HAC/VSG photoerosion and grain surface chemistry processes, will help us to improve our knowledge of the interstellar carbon chemistry.

 
\begin{acknowledgements} 

This work has been partially funded by MINECO grants (CSD2009-00038, AYA2009-07304, and AYA2012-32032).
The authors acknowledge the valuable comments and suggestions of the anonymous referee.
We are grateful to the IRAM staff for their help during the observations.
We warmly thank G.B. Esplugues and M. Ag\'undez for helping us with some of the observations presented in this work and for useful discussions. 
S.C. acknowledges support from FPI-INTA grant. P.P. acknowledges financial support from the Centre National d'Etudes Spatiales (CNES).

\end{acknowledgements}

\bibliographystyle{aa}
\bibliography{references}

\appendix

\section{MADEX: a local, non-LTE LVG code}

The physical conditions in ISM clouds are such that molecular excitation is usually far from LTE. 
MADEX solves the \mbox{non-LTE} level excitation and line radiative transfer in a 1D isothermal homogeneous
medium assuming a large velocity gradient (LVG) and spherical geometry. 
In this approximation the statistical-equilibrium equations are solved assuming \textit{local} excitation 
conditions and a  geometrically averaged escape probability formalism for the emitted photons \cite[see details in][]{Sobolev_1960, Castor_1970}.
This description allows one to take into account radiative trapping and collisional excitation and deexcitation 
more easily and computationally faster than more sophisticated  \textit{non-local} codes in which the radiative coupling 
between different cloud positions is explicitly treated  (Montecarlo simulations, ALI methods, etc.).
As a small benchmark, and in order to place the conclusions of our work on a firm ground, here we compare MADEX results with those obtained 
with RADEX\footnote{http://www.sron.rug.nl/$\sim$vdtak/radex/radex.php}, 
a publicly available escape probability code (see \citealt{vanderTak_2007} for the basic formulae). 

We ran several models for CO (a low-dipole moment molecule with $\mu$=0.12~D) and for HCO$^+$ ($\mu$=3.90~D). These are typical examples of molecules with low and high critical densities, respectively
 (e.g. \mbox{$n_{cr}$(CO 2$\rightarrow$1)} of a few 10$^4$\,cm$^{-3}$ and  \mbox{$n_{cr}$(HCO$^+$ 2$\rightarrow$1)} of a few 10$^6$\,cm$^{-3}$).
We note that for optically thin emission lines and for densities $n$(H$_2$)$\gg$$n_{cr}$, collisions dominate over radiative excitations and level 
populations get closer to LTE ($T_{\rm ex}\negthickspace \rightarrow \negthickspace T_{\rm k}$) as the density increases.
For optically thick lines, line trapping effectively reduces $n_{cr}$ and lines can be thermalized at lower densities.
In the low density limit ($n$(H$_2$)$\ll$$n_{cr}$),  level populations are subthermally excited
and, as the density decreases, tend to thermalize to the  background radiation temperature (\mbox{$T_{\rm k} > T_{\rm ex}\rightarrow 2.7$~K} in the millimetre domain).

\begin{figure}[ht]
\centering
\includegraphics[scale=0.5,angle=0]{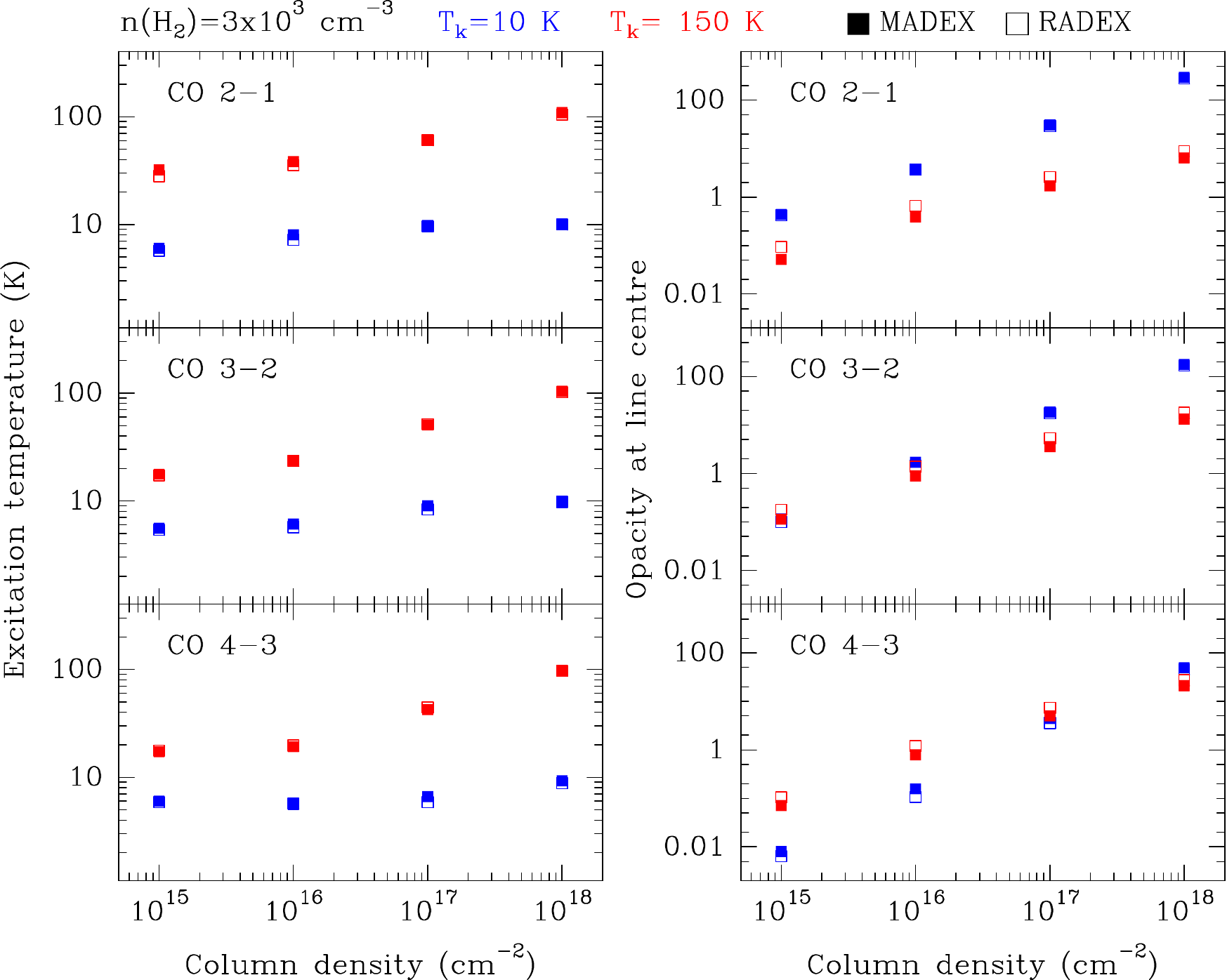} \\
\vspace{0.5cm}
\includegraphics[scale=0.5,angle=0]{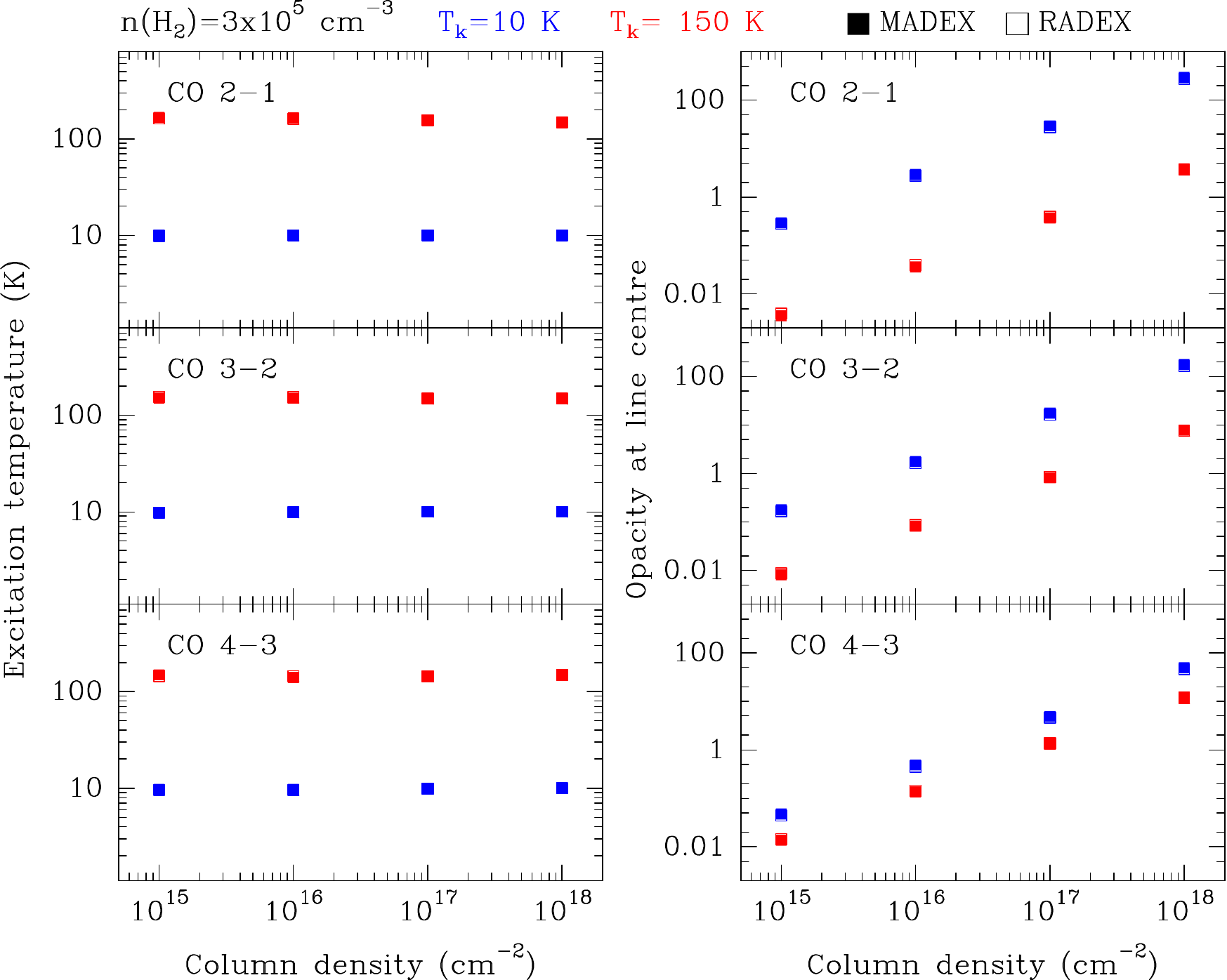} \\ 

\caption{CO isothermal models carried out with MADEX (filled squares) and RADEX (empty squares) non-LTE radiative transfer codes. 
Two gas densities are considered: $n$(H$_2$)=3$\times$10$^{3}$\,cm$^{-3}$ (\textit{upper panels}) and  $n$(H$_2$)=3$\times$10$^{5}$\,cm$^{-3}$
(\textit{lower panels}). Excitation temperatures and  line centre opacities (\textit{left} and \textit{right} panels, respectively)
are shown  for several rotational transitions in the millimetre domain as a function of CO column density.
Two gas temperatures are considered, 10~K (blue points) and 150~K (red points). }
\label{fig:bench1}
\end{figure}

Figure~\ref{fig:bench1} (\textit{upper panels}) shows  low density model results ($n$(H$_2$)=3$\times$10$^{3}$\,cm$^{-3}$) 
 for $N$(CO) from
10$^{15}$ to 10$^{18}$\,cm$^{-2}$ at  two different gas temperatures ($T_{\rm k}$=10 and 150\,K, blue and red points, respectively). 
The left and right figures show the computed excitation temperatures and line centre opacities, respectively, for the CO 2$\rightarrow$1, 3$\rightarrow$2, and 4$\rightarrow$3 
transitions. Figure~\ref{fig:bench1} (\textit{lower panels}) shows higher density models ($n$(H$_2$)=3$\times$10$^{5}$\,cm$^{-3}$) 
close to thermalization (\mbox{$T_{\rm ex}\simeq T_{\rm k}$}). We note that the selected range
of column densities represents a transition from optically thin to optically thick emission.
A line width of 1\,km\,s$^{-1}$ is adopted in all models.

\begin{figure}[ht]
\centering
\includegraphics[scale=0.5,angle=0]{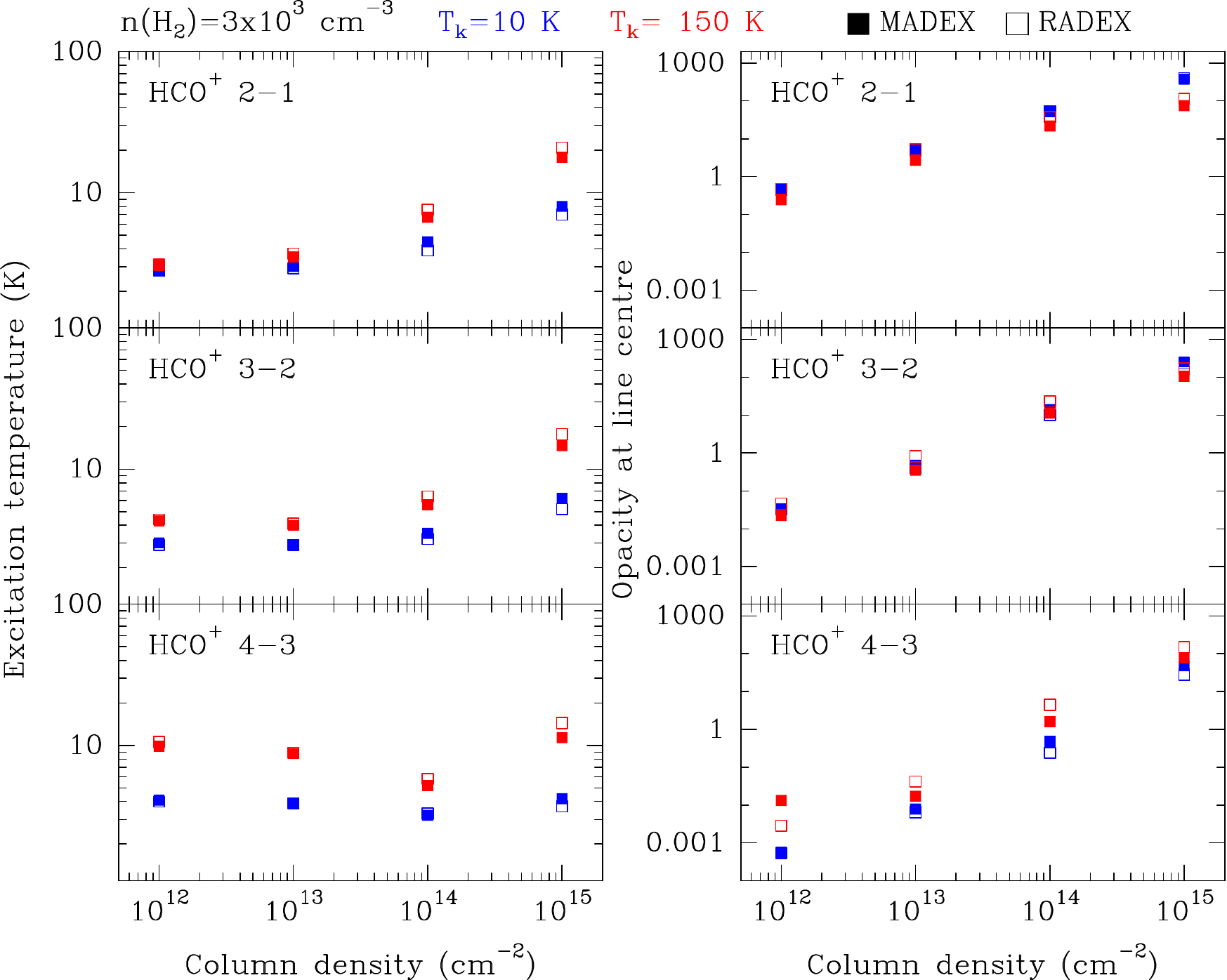}  \\
\vspace{0.5cm}
\includegraphics[scale=0.5,angle=0]{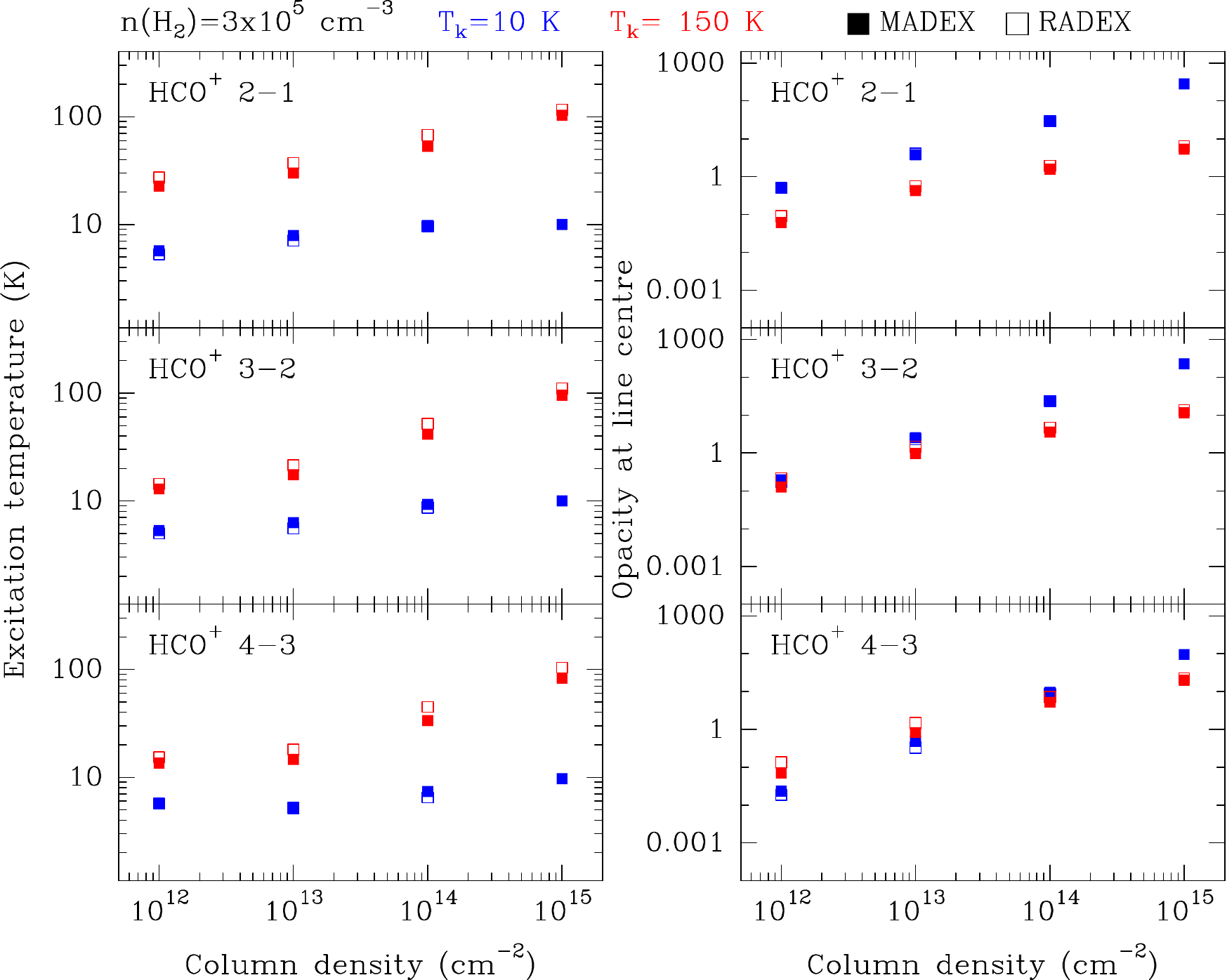}  \\
\caption{Same as Fig.~\ref{fig:bench1} but for HCO$^+$.}
\label{fig:bench2}
\end{figure}

Figure~\ref{fig:bench2} shows the same kind of models for HCO$^+$ (column densities from 10$^{12}$ to 10$^{15}$\,cm$^{-2}$).
Owing to the much higher critical densities of HCO$^+$ rotational transitions, their excitation is sub-thermal (\mbox{$T_{\rm ex}< T_{\rm k}$})
in most of the explored parameter space. In addition, their associated emission lines become optically thick for 
column densities smaller than those of CO.

The filled and empty square marks in Figs.~\ref{fig:bench1} and \ref{fig:bench2} represent computations performed with MADEX and RADEX 
codes, respectively.
We checked that for the considered models, the predicted excitation temperatures and line opacities agree within $\sim$20$\%$ and $\sim$40$\%$, respectively.
This translates into maximum brightness temperature differences of $\sim$50$\%$ in the most extreme cases.

\section{Identified hydrocarbon lines}

\begin{table*}[!h]
\begin{center}
\caption{Line parameters of C$_{2}$H.}  \label{Table_C2H} 
\begin{tabular}{c c c c c c c c c c@{\vrule height 10pt depth 5pt width 0pt}}     
\hline\hline      

\bf Transition & \bf Frequency & \bf E$_{_{\mathrm{\bf u}}}$ & \bf A$_{_{\mathrm{\bf ul}}}$ & {\bf S$_{_{\mathrm{\bf ij}}}$} & {\bf g$_{_{\mathrm{\bf u}}}$} & \bf $\displaystyle{\int}\bf T_{_{\mathrm{\bf MB}}}dv$ & \bf v$_{_{\mathrm{\bf LSR}}}$ & \bf $\Delta$v & \bf T$_{_{\mathrm{\bf MB}}}$ \rule[-0.3cm]{0cm}{0.8cm}\ \\ \cline{1-1}

${\mathrm{(N, J, F)_{u} \rightarrow (N, J, F)_{l}}}$ & [MHz] & [K] & [$\mathrm{s^{-1}}$] & & & [$\mathrm{K\, km\, s^{-1}}$] & [$\mathrm{km\, s^{-1}}$] & [$\mathrm{km\, s^{-1}}$]  & [$\mathrm{K}$]  \\

\hline 
${\mathrm{(1, 3/2, 1)\rightarrow(0, 1/2, 1)}}$ & 87284.105 $^{(F)}$ & 4.2 & 2.59$\times$10$^{-7}$ & 0.2 & 3 & 0.40(1)  & 10.52(2) & 3.03(6) & 0.12  \\
${\mathrm{(1, 3/2, 2)\rightarrow(0, 1/2, 1)}}$ & 87316.898 $^{(F)}$ & 4.2 & 1.53$\times$10$^{-6}$ & 1.7 & 5 & 3.82(1)  & 10.47(1) & 2.91(1) & 1.23  \\
${\mathrm{(1, 3/2, 1)\rightarrow(0, 1/2, 0)}}$ & 87328.585 $^{(F)}$ & 4.2 & 1.27$\times$10$^{-6}$ & 0.8 & 3 & 1.86(1)  & 10.47(1) & 2.89(2) & 0.61  \\
${\mathrm{(1, 1/2, 1)\rightarrow(0, 1/2, 1)}}$ & 87401.989 $^{(F)}$ & 4.2 & 1.27$\times$10$^{-6}$ & 0.8 & 3 & 1.89(1)  & 10.48(1) & 2.88(1) & 0.61  \\	 
${\mathrm{(1, 1/2, 0)\rightarrow(0, 1/2, 1)}}$ & 87407.165 $^{(F)}$ & 4.2 & 1.53$\times$10$^{-6}$ & 0.3 & 1 & 0.74(1)  & 10.49(1) & 2.84(4) & 0.25  \\
${\mathrm{(1, 1/2, 1)\rightarrow(0, 1/2, 0)}}$ & 87446.470 $^{(F)}$ & 4.2 & 2.61$\times$10$^{-7}$ & 0.2 & 3 & 0.39(1)  & 10.49(2) & 2.90(5) & 0.13 \\	 
 	                                                                                                         
\hline  
${\mathrm{(2, 5/2, 2)\rightarrow (1, 3/2, 2)}}$ & 174634.861 $^{(W)}$ & 12.6 & 1.00$\times$10$^{-6}$ & 0.1 & 5 & 1.32(19) &--- & ---  & ---  \\
${\mathrm{(2, 5/2, 3)\rightarrow (1, 3/2, 2)}}$ & 174663.199 $^{(W)}$ & 12.6 & 1.47$\times$10$^{-5}$ & 2.8 & 7 & \multirow{2}{*}[0cm]{$\Big \rangle$ 13.90(21)} & \multirow{2}{*}[-0.01cm]{---} &  \multirow{2}{*}[-0.01cm]{---} & \multirow{2}{*}[-0.01cm]{---}  \\
${\mathrm{(2, 5/2, 2)\rightarrow (1, 3/2, 1)}}$ & 174667.629 $^{(W)}$ & 12.6 & 1.36$\times$10$^{-5}$ & 1.9 & 5 & & &   &   \\
${\mathrm{(2, 3/2, 2)\rightarrow (1, 1/2, 1)}}$ & 174721.744 $^{(W)}$ & 12.6 & 1.16$\times$10$^{-5}$ & 1.6 & 5 & 2.84(17) & --- &  --- &  --- \\
${\mathrm{(2, 3/2, 1)\rightarrow (1, 1/2, 0)}}$ & 174728.071 $^{(W)}$ & 12.6 & 8.16$\times$10$^{-6}$ & 0.7 & 3 & \multirow{2}{*}[0cm]{$\Big \rangle$ 4.06(53)} & \multirow{2}{*}[-0.01cm]{---} & \multirow{2}{*}[-0.01cm]{---}  & \multirow{2}{*}[-0.01cm]{---}  \\
${\mathrm{(2, 3/2, 1)\rightarrow (1, 1/2, 1)}}$ & 174733.210 $^{(W)}$ & 12.6 & 5.08$\times$10$^{-6}$ & 0.4 & 3 & & &   &   \\
${\mathrm{(2, 3/2, 2)\rightarrow (1, 3/2, 2)}}$ & 174806.843 $^{(W)}$ & 12.6 & 2.67$\times$10$^{-6}$ & 0.4 & 5 & 1.52(26) &--- &  --- &  --- \\

\hline 
${\mathrm{(3, 7/2, 3)\rightarrow (2, 5/2, 3)}}$ & 261978.120 $^{(F)}$ & 25.1 & 1.95$\times$10$^{-6}$ & 0.1 & 7 & 0.33(1)  & 10.67(4) & 2.04(9)   & 0.15  \\
${\mathrm{(3, 7/2, 4)\rightarrow (2, 5/2, 3)}}$ & 262004.260 $^{(F)}$ & 25.2 & 5.31$\times$10$^{-5}$ & 3.9 & 9 & 9.19(5)  & 10.72(1) & 1.86(1)   & 4.65  \\
${\mathrm{(3, 7/2, 3)\rightarrow (2, 5/2, 2)}}$ & 262006.482 $^{(F)}$ & 25.1 & 5.10$\times$10$^{-5}$ & 2.9 & 7 & 10.10(6) & 10.72(1) & 2.39(1)   & 3.97  \\
${\mathrm{(3, 5/2, 3)\rightarrow (2, 3/2, 2)}}$ & 262064.986 $^{(F)}$ & 25.2 & 4.88$\times$10$^{-5}$ & 2.8 & 7 & 6.74(7)  & 10.78(1) & 1.90(2)   & 3.33  \\
${\mathrm{(3, 5/2, 2)\rightarrow (2, 3/2, 1)}}$ & 262067.469 $^{(F)}$ & 25.2 & 4.46$\times$10$^{-5}$ & 1.8 & 5 & 6.18(7)  & 10.83(1) & 2.39(3)   & 2.43  \\
${\mathrm{(3, 5/2, 2)\rightarrow (2, 3/2, 2)}}$ & 262078.934 $^{(F)}$ & 25.2 & 5.99$\times$10$^{-6}$ & 0.2 & 5 & 0.68(2)  & 10.75(3) & 2.09(8)   & 0.30  \\
${\mathrm{(3, 5/2, 3)\rightarrow (2, 5/2, 3)}}$ & 262208.614 $^{(F)}$ & 25.2 & 3.95$\times$10$^{-6}$ & 0.2 & 7 & 0.67(1)  & 10.77(2) & 2.29(6)   & 0.27  \\
${\mathrm{(3, 5/2, 3)\rightarrow (2, 5/2, 2)}}$ & 262236.957 $^{(F)}$ & 25.2 & 4.03$\times$10$^{-7}$ & 0.02 & 7 & 0.06(1)  & 10.71(9) & 1.35(22)  & 0.04  \\
${\mathrm{(3, 5/2, 2)\rightarrow (2, 5/2, 2)}}$ & 262250.928 $^{(F)}$ & 25.2 & 2.27$\times$10$^{-6}$ & 0.09 & 5 & 0.27(1)  & 10.67(5) & 2.17(15)  & 0.12  \\

\hline  
${\mathrm{(4, 9/2, 4)\rightarrow (3, 7/2, 4)}}$ & 349312.833 $^{(F)}$ & 41.9 & 2.98$\times$10$^{-6}$ & 0.09 & 9  & 0.22(1) & 10.72(22) & 2.32(44)  & 0.15$^{\star}$  \\
${\mathrm{(4, 9/2, 5)\rightarrow (3, 7/2, 4)}}$ & 349337.707 $^{(F)}$ & 41.9 & 1.30$\times$10$^{-4}$ & 4.9 & 11 & \multirow{2}{*}[0cm]{$\Big \rangle$ 18.09(5)} & \multirow{2}{*}[-0.01cm]{10.64(1)} & \multirow{2}{*}[-0.01cm]{2.93(1)}  & \multirow{2}{*}[-0.01cm]{4.65}  \\
${\mathrm{(4, 9/2, 4)\rightarrow (3, 7/2, 3)}}$ & 349338.989 $^{(F)}$ & 41.9 & 1.27$\times$10$^{-4}$ & 3.9 & 9  &          &    &   &   \\
${\mathrm{(4, 7/2, 4)\rightarrow (3, 5/2, 3)}}$ & 349399.274 $^{(F)}$ & 41.9 & 1.25$\times$10$^{-4}$ & 3.8 & 9  & \multirow{2}{*}[0cm]{$\Big \rangle$ 14.02(7)} & \multirow{2}{*}[-0.01cm]{10.69(1)} & \multirow{2}{*}[-0.01cm]{3.01(1)}  & \multirow{2}{*}[-0.01cm]{3.33}  \\
${\mathrm{(4, 7/2, 3)\rightarrow (3, 5/2, 2)}}$ & 349400.669 $^{(F)}$ & 41.9 & 1.19$\times$10$^{-4}$ & 2.9 & 7  &          & &   &   \\
${\mathrm{(4, 7/2, 3)\rightarrow (3, 5/2, 3)}}$ & 349414.640 $^{(F)}$ & 41.9 & 7.02$\times$10$^{-6}$ & 0.2 & 7  & 0.29(2)  & 10.80(11) & 1.89(28)  &  0.30$^{\star}$ \\
${\mathrm{(4, 7/2, 4)\rightarrow (3, 7/2, 4)}}$ & 349603.611 $^{(F)}$ & 41.9 & 5.18$\times$10$^{-6}$ & 0.2 & 9  & 0.34(1)  & 10.61(12) & 2.14(26)  & 0.27  \\
                                                                                                 
\hline 
\hline     
    
\end{tabular}
\end{center}

\vspace{0.2cm} {\bf General notes to Appendix B tables}: 

(i) Frequency, energy of the upper level of each transition (E$_{_{\mathrm{\bf u}}}$), Einstein coefficient for spontaneous emission (A$_{_{\mathrm{\bf ul}}}$), intrinsic line strength (S$_{_{\mathrm{\bf ij}}}$), and the level degeneracy (g$_{_{\mathrm{\bf u}}}$) from MADEX. The velocity-integrated intensity ($\mathrm{\int T_{_{MB}}dv}$), radial velocity (v$_{_{\mathrm{\bf LSR}}}$), and FWHM line width ($\Delta$v) obtained by Gaussian fit.

(ii) Parentheses indicate the uncertainty obtained by the Gaussian fitting programme. The fit uncertainty in units of the last significant digit is given in Tables \ref{Table_C2H}, \ref{Table_$c$-C3H2}, and \ref{Table_C4H}.

(iii) The weighted average of hyperfine and fine components is used for the analysis of C$_{2}$H, $l$-C$_{3}$H, $c$-C$_{3}$H, and C$_{4}$H molecules. 

(iv) Fully overlapping transitions are marked with connecting symbols.

(v) We note that the line intensities in Tables \ref{Table_13CCH}, \ref{Table_$l$-C3H}, \ref{Table_$c$-C3H}, and \ref{Table_$l$-C3H2} are given in units of mK.

(vi) Labels:  $^{(F)}$ Detected with FFTS backend. $^{(W)}$ The lines detected with  WILMA backend just give information about the integrated line intensity (see Sect.~2). $^{\star}$ Marginal detection.

\end{table*}

\begin{table*}[p]
\begin{center}
  
\caption{Line parameters of $^{13}$CCH and C$^{13}$CH.}  \label{Table_13CCH} 
\begin{tabular}{c c c c c c c c c c@{\vrule height 10pt depth 5pt width 0pt}}     
\hline\hline      

\bf Transition & \bf Frequency & \bf E$_{_{\mathrm{\bf u}}}$ & \bf A$_{_{\mathrm{\bf ul}}}$ & {\bf S$_{_{\mathrm{\bf ij}}}$} & {\bf g$_{_{\mathrm{\bf u}}}$} & \bf $\displaystyle{\int}\bf T_{_{\mathrm{\bf MB}}}dv$ & \bf v$_{_{\mathrm{\bf LSR}}}$ & \bf $\Delta$v & \bf T$_{_{\mathrm{\bf MB}}}$ \rule[-0.3cm]{0cm}{0.8cm}\ \\ \cline{1-1}

\ \ ${\mathrm{(N, J, F_{1}, F)_{u} \rightarrow (N, J, F_{1}, F)_{l}}}$ & [MHz] & [K] & [$\mathrm{s^{-1}}$] & & & [$\mathrm{mK\, km\, s^{-1}}$] & [$\mathrm{km\, s^{-1}}$] & [$\mathrm{km\, s^{-1}}$]  & [$\mathrm{mK}$]  \\

\hline 
{\bf $^{13}$CCH}  & & & & & & & & &  \\
${\mathrm{(1, 3/2, 2, 5/2)\rightarrow (0, 1/2, 1, 3/2)}}$ & \ \ 84119.329 $^{(F)}$ &  \ \ 4.0 & 1.37$\times10^{-6}$ & 2.0 & 6 &  \ \ \ 20.5(7.0)  & 11.0(0.3) & 1.4(0.4) & 14$^{\star}$  \\

${\mathrm{(3, 7/2, 4, 9/2)\rightarrow (2, 5/2 , 3, 7/2)}}$ & 252422.933 $^{(F)}$& 24.2 & 4.75$\times10^{-5}$ & 4.3 & 10 & \multirow{2}{*}[0cm]{$\Big \rangle$ 114.2(6.6)} & \multirow{2}{*}[-0.01cm]{10.7(0.1)} & \multirow{2}{*}[-0.01cm]{3.6(0.3)} & \multirow{2}{*}[-0.01cm]{34} \\
${\mathrm{(3, 7/2, 4, 7/2)\rightarrow (2, 5/2 , 3, 5/2)}}$ & 252424.122 $^{(F)}$& 24.2 & 4.63$\times10^{-5}$ & 3.3 & 8 &  & & &   \\
${\mathrm{(3, 7/2, 3, 5/2)\rightarrow (2, 5/2 , 2, 3/2)}}$ & 252447.991 $^{(F)}$& 24.2 & 4.42$\times10^{-5}$ & 2.4 & 6 &  \multirow{2}{*}[0cm]{$\Big \rangle$ 61.4(8.2)}  & \multirow{2}{*}[-0.01cm]{10.5(0.2)} & \multirow{2}{*}[-0.01cm]{2.5(0.3)} & \multirow{2}{*}[-0.01cm]{27}  \\
${\mathrm{(3, 7/2, 3, 7/2)\rightarrow (2, 5/2 , 2, 5/2)}}$ & 252449.265 $^{(F)}$& 24.2 & 4.74$\times10^{-5}$ & 3.4 & 8 &  & & &   \\
${\mathrm{(3, 5/2, 3, 7/2)\rightarrow (2, 3/2 , 2, 5/2)}}$ & 252457.865 $^{(F)}$& 24.2 & 4.16$\times10^{-5}$ & 3.0 & 8 &  \ \ \ 46.9(6.6) & 10.4(0.2) & 1.6(0.5) & 19 \\
${\mathrm{(3, 5/2, 3, 5/2)\rightarrow (2, 3/2 , 2, 3/2)}}$ & 252468.774 $^{(F)}$& 24.2 & 3.86$\times10^{-5}$ & 2.1 & 6 &  \ \ \ 18.4(8.2) & 10.3(0.3) & 1.4(0.7) & 16 \\
${\mathrm{(3, 5/2, 2, 5/2)\rightarrow (2, 3/2 , 1, 3/2)}}$ & 252480.925 $^{(F)}$& 24.2 & 3.17$\times10^{-5}$ & 1.7 & 6 &  \ \ \ 28.3(6.6) & 11.0(0.2) & 1.4(0.4) & 15 \\
${\mathrm{(3, 5/2, 2, 3/2)\rightarrow (2, 3/2 , 1, 1/2)}}$ & 252489.308 $^{(F)}$& 24.2 & 3.32$\times10^{-5}$ & 1.2 & 4 &  \ \ \ 16.0(3.3) & 10.6(0.1) & 1.4(0.4) & 12  \\

\hline 
{\bf C$^{13}$CH}  & & & & & & & & &  \\

${\mathrm{(1, 3/2, 2, 5/2)\rightarrow (0, 1/2, 1, 3/2)}}$ & \ \ 85229.326 $^{(F)}$ & \ \ 4.1 & 1.42$\times10^{-6}$ & 2.0 & 6 & 36.3(7.0) & 11.0(0.2) & 2.2(0.4) & 16$^{\star}$  \\
${\mathrm{(3, 7/2, 3, 5/2)\rightarrow (2, 5/2, 2, 5/2)}}$ & 255742.430 $^{(F)}$ & 24.5  & 3.36$\times10^{-6}$ & 0.2 & 6  & 12.0(6.7) & 10.4(0.1) & 0.6(0.3) & 21  \\
${\mathrm{(3, 7/2, 4, 9/2)\rightarrow (2, 5/2, 3, 7/2)}}$ & 255746.086 $^{(F)}$ & 24.5  & 4.94$\times10^{-5}$ & 4.3 & 10 & \multirow{2}{*}[0cm]{$\Big \rangle$ 151.7(15)} & \multirow{2}{*}[-0.01cm]{10.7(0.1)} & \multirow{2}{*}[-0.01cm]{3.0(0.4)} & \multirow{2}{*}[-0.01cm]{47}  \\
${\mathrm{(3, 7/2, 4, 7/2)\rightarrow (2, 5/2, 3, 5/2)}}$ & 255747.258 $^{(F)}$ & 24.5  & 4.83$\times10^{-5}$ & 3.4 & 8  &    &  &  &   \\
${\mathrm{(3, 7/2, 3, 5/2)\rightarrow (2, 5/2, 2, 3/2)}}$ & 255756.027 $^{(F)}$ & 24.5  & 4.52$\times10^{-5}$ & 2.4 & 6  & 51.3(16.6) & 10.7(0.4) & 2.4(0.9) & 19  \\
${\mathrm{(3, 7/2, 3, 7/2)\rightarrow (2, 5/2, 2, 5/2)}}$ & 255758.767 $^{(F)}$ & 24.5  & 4.81$\times10^{-5}$ & 3.3 & 8  & 57.1(18.3) & 10.5(0.2) & 1.9(0.7) & 31  \\
${\mathrm{(3, 5/2, 3, 7/2)\rightarrow (2, 3/2, 2, 5/2)}}$ & 255794.864 $^{(F)}$ & 24.5  & 4.34$\times10^{-5}$ & 3.0 & 8  & 46.1(11.6) & 10.6(0.2) & 1.6(0.5) & 28 \\
${\mathrm{(3, 5/2, 3, 5/2)\rightarrow (2, 3/2, 2, 3/2)}}$ & 255803.715 $^{(F)}$ & 24.5  & 4.34$\times10^{-5}$ & 2.3 & 6  & 38.9(15.0) & 10.7(0.2) & 1.1(0.4) & 35  \\
${\mathrm{(3, 5/2, 2, 5/2)\rightarrow (2, 3/2, 1, 3/2)}}$ & 255805.399 $^{(F)}$ & 24.5  & 3.62$\times10^{-5}$ & 1.9 & 6  & 34.5(20.0) & 10.8(0.4) & 1.9(0.8) & 17  \\
\hline
\hline    
    
\end{tabular}
\end{center} 
 
\end{table*}

\begin{table*}[p]
\centering
  \caption{Line parameters of $l$-C$_{3}$H$^{+}$.}  \label{Table_C3H+} 
\begin{tabular}{c c c c c c c c c c@{\vrule height 10pt depth 5pt width 0pt}}     
\hline\hline      

\bf Transition & \bf Frequency & \bf E$_{_{\mathrm{\bf u}}}$ & \bf A$_{_{\mathrm{\bf ul}}}$ & {\bf S$_{_{\mathrm{\bf ij}}}$} & {\bf g$_{_{\mathrm{\bf u}}}$} & \bf $\displaystyle{\int}\bf T_{_{\mathrm{\bf MB}}}dv$  & \bf $\Delta$v & \bf T$_{_{\mathrm{\bf MB}}}$ \rule[-0.3cm]{0cm}{0.8cm} & {\bf Notes } \\ \cline{1-1}

${\mathrm{J \rightarrow J-1}}$ & [MHz] & [K] & [$\mathrm{s^{-1}}$] & & & [$\mathrm{K\, km\, s^{-1}}$] & [$\mathrm{km\, s^{-1}}$]  & [$\mathrm{K}$] &  \\

\hline 

${\mathrm{4 \rightarrow 3}}$   	& \ \  89957.849(0.054) $^{(F)}$ & 10.8 & 3.389$\times10^{-5}$  & \ \ 4  &\ \  9 & 0.12(0.01) & 3.25(0.23) &  0.033   & \ \    \tablefootmark{a} \\
${\mathrm{5 \rightarrow 4}}$ 	   &  112445.713(0.047) $^{(F)}$ & 16.2 & 6.770$\times10^{-5}$  & \ \ 5  &   11    & 0.13(0.01) & 2.96(0.32) &  0.043 &  \ \     \tablefootmark{a}  \\
${\mathrm{6 \rightarrow 5}}$ 	   &  134932.733(0.010) $^{(W)}$ & 22.7 & 1.188$\times10^{-4}$  & \ \ 6  &   13    & 0.21(0.04) & ---      & ---   &  \ \   \tablefootmark{b}  \\
${\mathrm{7 \rightarrow 6}}$ 	   &  157418.719(0.016) $^{(W)}$ & 30.2 & 1.907$\times10^{-4}$  & \ \ 7  &   15    & 0.16(0.03) & ---      & ---   & \ \    \tablefootmark{b}  \\
${\mathrm{9 \rightarrow 8}}$       &  202386.776(0.065) $^{(F)}$ & 48.6 & 4.113$\times10^{-4}$  & \ \ 9  & 19    & 0.20(0.01) & 2.15(0.13) &  0.088 &  \ \    \tablefootmark{a}  \\
${\mathrm{10 \rightarrow 9 \ \ }}$ &  224868.307(0.114) $^{(F)}$ & 59.4 & 5.672$\times10^{-4}$  & 10     & 21    & 0.19(0.01) & 2.40(0.21) & 0.076  &  \ \     \tablefootmark{a}  \\
${\mathrm{11 \rightarrow 10}}$     &   247348.016(0.080) $^{(F)}$ & 71.2 & 7.582$\times10^{-4}$  & 11     & 23    & 0.14(0.01) & 1.68(0.18) & 0.080  &  \ \    \tablefootmark{a}  \\
${\mathrm{12 \rightarrow 11}}$ 	   &   269825.838(0.120) $^{(F)}$ & 84.2 & 9.878$\times10^{-4}$  & 12     & 25    & 0.17(0.02) & 2.00(0.28) & 0.079  &  \ \     \tablefootmark{a}  \\
${\mathrm{13 \rightarrow 12}}$ 	   &  292301.412(0.065) $^{(F)}$ & 98.2 & 1.260$\times10^{-3}$  & 13     &  27   & 0.19(0.02) & 1.92(0.17) &  0.095 &  \ \     \tablefootmark{a}  \\

\hline
\hline    
    
\end{tabular}
\tablefoot{
\tablefoottext{a}{Observed frequencies of the detected lines and their uncertainties obtained by fitting Gaussian at 200 kHz spectral resolution and measured in the local standard of rest frame (v$_{_{\mathrm{\bf LSR}}}$=10.7 $\mathrm{km\, s^{-1}}$ in the Orion Bar PDR).}
\tablefoottext{b}{Frequencies and uncertainties reported in \citet{Pety_2012}. }
}
\end{table*}


\begin{table*}[p]
\begin{center}
  
\caption{Line parameters of $l$-C$_{3}$H.}  \label{Table_$l$-C3H} 
\begin{tabular}{c c c c c c c c c c@{\vrule height 10pt depth 4pt width 0pt}}   
\hline\hline      

 \bf Transition & \bf Frequency & \bf E$_{_{\mathrm{\bf u}}}$ & \bf A$_{_{\mathrm{\bf ul}}}$ & {\bf S$_{_{\mathrm{\bf ij}}}$} & {\bf g$_{_{\mathrm{\bf u}}}$} & \bf $\displaystyle{\int}\bf T_{_{\mathrm{\bf MB}}}dv$ & \bf v$_{_{\mathrm{\bf LSR}}}$ & \bf $\Delta$v & \bf T$_{_{\mathrm{\bf MB}}}$ \rule[-0.3cm]{0cm}{0.8cm}\ \\ \cline{1-1}

 ${\mathrm{(J^{^{p}}, F)_{u} \rightarrow (J^{^{p}}, F)_{l}}}$   & [MHz] & [K] & [$\mathrm{s^{-1}}$] & & & [$\mathrm{mK\, km\, s^{-1}}$] & [$\mathrm{km\, s^{-1}}$] & [$\mathrm{km\, s^{-1}}$]  & [$\mathrm{mK}$]  \\

\hline

\bf $^{2}\Pi_{1/2}$ $\rightarrow$  $^{2}\Pi_{1/2}$ \ \ \ \ \ \  \ \ \  \ \ \ \ \ \  \ \ \ & & & &  &  &  &  &  &  \\  \cline{1-1}
\cline{1-1}
\cline{1-1} 
\cline{1-1}
   (9/2$^{+}$, 5 $\rightarrow$  7/2$^{-}$, 4)  &  97995.166 $^{(F)}$  &  12.5 & 6.12$\times$10$^{-5}$  & 4.9 &  11  &  35.1(3.5)       &  10.3(0.2)     & 1.4(0.2)  &  24     \\
   (9/2$^{+}$, 4 $\rightarrow$  7/2$^{-}$, 3)  &  97995.913 $^{(F)}$  &  12.5 & 5.95$\times$10$^{-5}$  & 3.9 &   9  &  30.1(3.0)       &  10.2(0.4)     & 1.4(0.2)  &  20     \\
   (9/2$^{-}$, 5 $\rightarrow$  7/2$^{+}$, 4)  &  98011.611 $^{(F)}$  &  12.5 & 6.13$\times$10$^{-5}$  & 4.9 &  11  &  28.0(2.8)       &  10.8(0.2)     & 1.3(0.2)  &  20     \\
   (9/2$^{-}$, 4 $\rightarrow$  7/2$^{+}$, 3)  &  98012.524 $^{(F)}$  &  12.5 & 5.96$\times$10$^{-5}$  & 3.9 &   9  &  32.6(3.3)       &  10.8(0.3)     & 1.3(0.2)  &  24     \\
  (13/2$^{-}$, 6 $\rightarrow$  11/2$^{+}$, 5)  &  149106.972 $^{(W)}$  &   46.1  & 2.11$\times$10$^{-4}$ &  5.6  &  13     &  \  \multirow{2}{*}[0cm]{$\Big \rangle$ 178.5(71.4)}       &   \multirow{2}{*}[-0.01cm]{---}     &   \multirow{2}{*}[-0.01cm]{---}     &  \multirow{2}{*}[-0.01cm]{---}    \\  
  (13/2$^{-}$, 7 $\rightarrow$ 11/2$^{+}$, 6)   &  149106.972 $^{(W)}$  &   46.1  & 2.14$\times$10$^{-4}$ &  6.6  &  15     &         &        &        &       \\
  (13/2$^{+}$, 7 $\rightarrow$ 11/2$^{-}$, 6)   &  149212.667 $^{(W)}$  &   46.2  & 2.14$\times$10$^{-4}$ &  6.6  &  15     &   \multirow{2}{*}[0cm]{$\Big \rangle$ 58.7(55.5)}       &    \multirow{2}{*}[-0.01cm]{---}   &   \multirow{2}{*}[-0.01cm]{---}     &     \multirow{2}{*}[-0.01cm]{---}  \\
  (13/2$^{+}$, 6 $\rightarrow$ 11/2$^{-}$, 5)   &  149212.667 $^{(W)}$  &   46.2  & 2.12$\times$10$^{-4}$ &  5.7  &  13     &         &        &        &       \\
  (15/2$^{+}$, 7 $\rightarrow$ 13/2$^{-}$, 6)   &  171958.650 $^{(W)}$  &   54.4  & 3.33$\times$10$^{-4}$ &  6.7  &  15     &   \multirow{2}{*}[0cm]{$\Big \rangle$ 87.0(8.7) \ }       &  \multirow{2}{*}[-0.01cm]{---}      &    \multirow{2}{*}[-0.01cm]{---}    &  \multirow{2}{*}[-0.01cm]{---}     \\
  (15/2$^{+}$, 8 $\rightarrow$ 13/2$^{-}$, 7)   &  171958.650 $^{(W)}$  &   54.4  & 3.36$\times$10$^{-4}$ &  7.7  &  17     &         &        &        &       \\
  (15/2$^{-}$, 7 $\rightarrow$ 13/2$^{+}$, 6)   &  172094.778 $^{(W)}$  &   54.4  & 3.34$\times$10$^{-4}$ &  6.7  &  15     &   \multirow{2}{*}[0cm]{$\Big \rangle$ 48.5(4.8) \ }       &   \multirow{2}{*}[-0.01cm]{---}     &   \multirow{2}{*}[-0.01cm]{---}     &   \multirow{2}{*}[-0.01cm]{---}    \\
  (15/2$^{-}$, 8 $\rightarrow$ 13/2$^{+}$, 7)   &  172094.778 $^{(W)}$  &   54.4  & 3.37$\times$10$^{-4}$ &  7.7  &  17     &         &        &        &       \\
  (19/2$^{+}$, 9  $\rightarrow$ 17/2$^{-}$, 8)  &  217571.404 $^{(F)}$  &   74.2  & 6.96$\times$10$^{-4}$ &  8.8  &  19     &  \multirow{2}{*}[0cm]{$\Big \rangle$ 57.6(13.4)}       &  \multirow{2}{*}[-0.01cm]{10.9(0.2)}    &  \multirow{2}{*}[-0.01cm]{1.4(0.3)}      &  \multirow{2}{*}[-0.01cm]{39}     \\
  (19/2$^{+}$, 10 $\rightarrow$ 17/2$^{-}$, 9)  \ \  &  217571.667 $^{(F)}$  &   74.2  & 7.00$\times$10$^{-4}$ &  9.7  &  21     &         &        &        &       \\
  (19/2$^{-}$, 9  $\rightarrow$ 17/2$^{+}$, 8)  &  217773.402 $^{(F)}$  &   74.2  & 6.98$\times$10$^{-4}$ &  8.8  &  19     &  \multirow{2}{*}[0cm]{$\Big \rangle$ 34.1(7.4) \ }       &   \multirow{2}{*}[-0.01cm]{10.6(0.2)}     &   \multirow{2}{*}[-0.01cm]{1.7(0.4)}     &  \multirow{2}{*}[-0.01cm]{17}     \\
  (19/2$^{-}$, 10 $\rightarrow$ 17/2$^{+}$, 9)  \ \  &  217773.513 $^{(F)}$  &   74.2  & 7.02$\times$10$^{-4}$ &  9.7  &  21     &         &        &        &       \\ 
  (27/2$^{-}$, 14  $\rightarrow$ 25/2$^{+}$, 13)   &  295172.315 $^{(F)}$  &  102.1  & 1.81$\times$10$^{-3}$ &  13.9  &  29    &  \multirow{2}{*}[0cm]{$\Big \rangle$ 30.0(7.7)}       &   \multirow{2}{*}[-0.01cm]{11.1(0.1)}     &   \multirow{2}{*}[-0.01cm]{0.6(0.2)}     &  \multirow{2}{*}[-0.01cm]{48}     \\
  (27/2$^{-}$, 13  $\rightarrow$ 25/2$^{+}$, 12)   &  295172.315 $^{(F)}$  &  102.1  & 1.81$\times$10$^{-3}$ &  12.9   &  27   &         &        &        &       \\ 
  (27/2$^{+}$, 14  $\rightarrow$ 25/2$^{-}$, 13)   &  295514.212 $^{(F)}$  &  102.2  & 1.82$\times$10$^{-3}$ &  13.9  &  29    &  \multirow{2}{*}[0cm]{$\Big \rangle$ 54.9(9.7)}       &   \multirow{2}{*}[-0.01cm]{10.2(0.1)}     &   \multirow{2}{*}[-0.01cm]{1.2(0.3)}     &  \multirow{2}{*}[-0.01cm]{41}     \\
  (27/2$^{+}$, 13  $\rightarrow$ 25/2$^{-}$, 12)   &  295514.212 $^{(F)}$  &  102.2  & 1.82$\times$10$^{-3}$ &  12.9  &  27   &         &        &        &       \\

\hline

\bf $^{2}\Pi_{3/2}$ $\rightarrow$  $^{2}\Pi_{3/2}$ \ \ \ \ \ \  \ \ \  \ \ \ \ \ \  \ \ \ & & & &  &  &  &  &  &  \\ \cline{1-1} 
\cline{1-1}
\cline{1-1} 
\cline{1-1}
   (9/2$^{-}$, 5 $\rightarrow$  7/2$^{+}$, 4) &  103319.276 $^{(F)}$  &  32.9 & 6.47$\times$10$^{-5}$  & 4.4  & 11  &  26.6(2.7)        & 10.9(0.2)       &  1.6(0.2)      & 16      \\
   (9/2$^{-}$, 4 $\rightarrow$  7/2$^{+}$, 3) &  103319.786 $^{(F)}$  &  32.9 & 6.30$\times$10$^{-5}$  & 3.5  &  9  &  20.2(2.0)        & 10.2(0.2)       &  1.4(0.2)      & 14      \\
   (9/2$^{+}$, 5 $\rightarrow$  7/2$^{-}$, 4) &  103372.483 $^{(F)}$  &  32.9 & 6.49$\times$10$^{-5}$  & 4.4  & 11  &  \multirow{2}{*}[0cm]{$\Big \rangle$ 25.1(2.5) \ }        & \multirow{2}{*}[-0.01cm]{10.0(0.4)}       &  \multirow{2}{*}[-0.01cm]{2.9(0.4)}      & \multirow{2}{*}[-0.01cm]{8}      \\
   (9/2$^{+}$, 4 $\rightarrow$  7/2$^{-}$, 3) &  103373.094 $^{(F)}$  &  32.9 & 6.31$\times$10$^{-5}$  & 3.5  &  9  &          &        &        &       \\
  (13/2$^{+}$, 7 $\rightarrow$ 11/2$^{-}$, 6)  &  141635.793 $^{(W)}$  &   25.1 & 1.92$\times$10$^{-4}$ & 6.9  & 15  &  \multirow{2}{*}[0cm]{$\Big \rangle$ 101.1(25.5)}       &    \multirow{2}{*}[-0.01cm]{---}    &   \multirow{2}{*}[-0.01cm]{---}     &   \multirow{2}{*}[-0.01cm]{---}    \\
  (13/2$^{+}$, 6 $\rightarrow$ 11/2$^{-}$, 5)  &  141636.431 $^{(W)}$  &   25.1 & 1.90$\times$10$^{-4}$ & 5.9  & 13  &         &        &        &       \\
  (13/2$^{-}$, 7 $\rightarrow$ 11/2$^{+}$, 6)  &  141708.728 $^{(W)}$  &   25.1 & 1.92$\times$10$^{-4}$ & 6.9  & 15  &  \multirow{2}{*}[0cm]{$\Big \rangle$ 120.7(31.9)}       &    \multirow{2}{*}[-0.01cm]{---}    &     \multirow{2}{*}[-0.01cm]{---}   &    \multirow{2}{*}[-0.01cm]{---}   \\
  (13/2$^{-}$, 6 $\rightarrow$ 11/2$^{+}$, 5)  &  141709.494 $^{(W)}$  &   25.1 & 1.90$\times$10$^{-4}$ & 5.9  & 13  &         &        &        &       \\
  (15/2$^{-}$, 8 $\rightarrow$ 13/2$^{+}$, 7)  &  163491.035 $^{(W)}$  &   32.9 & 2.90$\times$10$^{-4}$ & 7.9  & 17  &  \multirow{2}{*}[0cm]{$\Big \rangle$ 129.9(27.8)}       &   \multirow{2}{*}[-0.01cm]{---}     &    \multirow{2}{*}[-0.01cm]{---}    &   \multirow{2}{*}[-0.01cm]{---}    \\
  (15/2$^{-}$, 7 $\rightarrow$ 13/2$^{+}$, 6)  &  163491.557 $^{(W)}$  &   32.9 & 2.96$\times$10$^{-4}$ & 6.9  & 15  &         &        &        &       \\
  (15/2$^{+}$, 8 $\rightarrow$ 13/2$^{-}$, 7)  &  163597.232 $^{(W)}$  &   32.9 & 3.00$\times$10$^{-4}$ & 7.9  & 17  &  \multirow{2}{*}[0cm]{$\Big \rangle$ 144.6(19.9)}       &   \multirow{2}{*}[-0.01cm]{---}     &     \multirow{2}{*}[-0.01cm]{---}   &     \multirow{2}{*}[-0.01cm]{---}  \\
  (15/2$^{+}$, 7 $\rightarrow$ 13/2$^{-}$, 6)  &  163597.900 $^{(W)}$  &   32.9 & 2.96$\times$10$^{-4}$ & 6.9  & 15  &         &        &        &       \\
  (19/2$^{-}$, 10 $\rightarrow$ 17/2$^{+}$, 9) \ \ &  207279.369 $^{(F)}$  &   51.8 & 6.18$\times$10$^{-4}$ & 9.9  & 21  &  \multirow{2}{*}[0cm]{$\Big \rangle$ 79.3(7.9) \ \ }       &  \multirow{2}{*}[-0.01cm]{10.5(0.1)}      & \multirow{2}{*}[-0.01cm]{1.4(0.2)}       &   \multirow{2}{*}[-0.01cm]{53}    \\
  (19/2$^{-}$, 9 $\rightarrow$ 17/2$^{+}$, 8)  &  207279.779 $^{(F)}$  &   51.8 & 6.14$\times$10$^{-4}$ & 8.9  & 19  &         &        &        &       \\
  (19/2$^{+}$, 10 $\rightarrow$ 17/2$^{-}$, 9) \ \ &  207459.226 $^{(F)}$  &   51.8 & 6.19$\times$10$^{-4}$ & 9.9  & 21  &  \multirow{2}{*}[0cm]{$\Big \rangle$ 98.0(9.8) \ \ }       &  \multirow{2}{*}[-0.01cm]{10.5(0.2)}      &   \multirow{2}{*}[-0.01cm]{2.5(0.4)}     &  \multirow{2}{*}[-0.01cm]{37}     \\
  (19/2$^{+}$, 9 $\rightarrow$ 17/2$^{-}$, 8)  &  207459.800 $^{(F)}$  &   51.8 & 6.16$\times$10$^{-4}$ & 8.9  & 19  &         &        &        &       \\
  (21/2$^{+}$, 11 $\rightarrow$ 19/2$^{-}$, 10) & 229213.636 $^{(F)}$  &   62.8 & 8.40$\times$10$^{-4}$ & 10.9  & 23  &  \multirow{2}{*}[0cm]{$\Big \rangle$ 52.3(13.8)}       &  \multirow{2}{*}[-0.01cm]{10.4(0.2)}      &  \multirow{2}{*}[-0.01cm]{1.2(0.4)}      &   \multirow{2}{*}[-0.01cm]{40}    \\
  (21/2$^{+}$, 10 $\rightarrow$ 19/2$^{-}$, 9) \ \ & 229214.005 $^{(F)}$  &   62.8 & 8.36$\times$10$^{-4}$ & 9.9  & 21  &         &        &        &       \\
  (21/2$^{-}$, 11 $\rightarrow$ 19/2$^{+}$, 10) & 229432.781 $^{(F)}$  &   62.8 & 8.42$\times$10$^{-4}$ & 10.9  & 23  &  \multirow{2}{*}[0cm]{$\Big \rangle$ 49.8(13.8)}       &  \multirow{2}{*}[-0.01cm]{10.5(0.3)}      &  \multirow{2}{*}[-0.01cm]{1.2(0.2)}      &   \multirow{2}{*}[-0.01cm]{39}    \\
  (21/2$^{-}$, 10 $\rightarrow$ 19/2$^{+}$, 9) \ \ & 229433.316 $^{(F)}$  &   62.8 & 8.38$\times$10$^{-4}$ & 9.9  & 21  &         &        &        &       \\
  (23/2$^{-}$, 12 $\rightarrow$ 21/2$^{+}$, 11) & 251174.624 $^{(F)}$  &   74.8 & 1.11$\times$10$^{-3}$ & 11.9  & 25  &  \multirow{2}{*}[0cm]{$\Big \rangle$ 53.5(9.8)  \ }       &  \multirow{2}{*}[-0.01cm]{10.6(0.1)}      &   \multirow{2}{*}[-0.01cm]{1.6(0.3)}     &  \multirow{2}{*}[-0.01cm]{32}     \\
  (23/2$^{-}$, 11 $\rightarrow$ 21/2$^{+}$, 10) & 251174.624 $^{(F)}$  &   74.8 & 1.11$\times$10$^{-3}$ & 10.9  & 23  &         &        &        &       \\
  (23/2$^{+}$, 12 $\rightarrow$ 21/2$^{-}$, 11) & 251433.892 $^{(F)}$  &   74.9 & 1.11$\times$10$^{-3}$ & 11.9  & 25  &  \multirow{2}{*}[0cm]{$\Big \rangle$ 57.1(9.8) \ }       &  \multirow{2}{*}[-0.01cm]{10.8(0.2)}      &  \multirow{2}{*}[-0.01cm]{2.2(0.4)}      &  \multirow{2}{*}[-0.01cm]{25}    \\
  (23/2$^{+}$, 11 $\rightarrow$ 21/2$^{-}$, 10) & 251434.415 $^{(F)}$  &   74.9 & 1.11$\times$10$^{-3}$ & 10.9  & 23  &         &        &        &       \\

\hline
\hline    
    
\end{tabular}
\end{center} 
 
\end{table*}


\begin{table*}[p]
\begin{center}
  
\caption{Line parameters of $c$-C$_{3}$H.}  \label{Table_$c$-C3H} 
\begin{tabular}{c c c c c c c c c c@{\vrule height 10pt depth 5pt width 0pt}}   
\hline\hline      

 \bf Transition & \bf Frequency & \bf E$_{_{\mathrm{\bf u}}}$ & \bf A$_{_{\mathrm{\bf ul}}}$ & {\bf S$_{_{\mathrm{\bf ij}}}$} & {\bf g$_{_{\mathrm{\bf u}}}$} & \bf $\displaystyle{\int}\bf T_{_{\mathrm{\bf MB}}}dv$ & \bf v$_{_{\mathrm{\bf LSR}}}$ & \bf $\Delta$v & \bf T$_{_{\mathrm{\bf MB}}}$ \rule[-0.3cm]{0cm}{0.8cm}\ \\ \cline{1-1}
 
${\mathrm{(N_{K_{a}K_{c}}, J, F)_{u} \rightarrow (N_{K_{a}K_{c}}, J, F)_{l}}}$   & [MHz] & [K] & [$\mathrm{s^{-1}}$] & & & [$\mathrm{mK\, km\, s^{-1}}$] & [$\mathrm{km\, s^{-1}}$] & [$\mathrm{km\, s^{-1}}$]  & [$\mathrm{mK}$]  \\  
  
\hline

${\mathrm{(3_{1,2}, 5/2, 3)  \rightarrow (3_{1,3}, 5/2, 3)} }$  &  \ \ 85272.149 $^{(F)}$  &  14.9  &  3.40$\times$10$^{-6}$   &   0.6  &   7  & \multirow{2}{*}[0cm]{$\Big \rangle$ 21.6(5.9) \ \ }	    & \multirow{2}{*}[-0.01cm]{10.8(0.2)} & \multirow{2}{*}[-0.01cm]{1.2(0.4)}    &  \multirow{2}{*}[-0.01cm]{17}   \\
${\mathrm{(3_{1,2}, 5/2, 2)  \rightarrow (3_{1,3}, 5/2, 2)} }$  &   \ \  85272.522 $^{(F)}$  &  14.9  &  3.64$\times$10$^{-6}$   &   0.4  &   5  &  						            &                                       &                                         &                                   \\
${\mathrm{(3_{1,2}, 7/2, 4)  \rightarrow (3_{1,3}, 7/2, 4)} }$  &  \ \   85702.495 $^{(F)}$  &  14.9  &  3.89$\times$10$^{-6}$   &   0.8  &   9      &  7.5(4.7)                                             & 10.5(0.4)                           & 0.9(0.6)                              & \ \ 8                           \\
${\mathrm{(2_{1,2}, 5/2, 3)  \rightarrow (1_{1,1}, 3/2, 2)} }$  &  \ \   91494.349 $^{(F)}$  &  \ \  4.4  &  1.59$\times$10$^{-5}$   &   2.2  &   7  & 88.0(9.5)                                             & 10.5(0.2)                           & 3.4(0.5)                              & 29                             \\
${\mathrm{(2_{1,2}, 5/2, 2)  \rightarrow (1_{1,1}, 3/2, 1)} }$  &  \ \   91497.608 $^{(F)}$  &  \ \  4.4  &  1.38$\times$10$^{-5}$   &   1.3  &   5  & 52.4(9.5)                                             & 10.5(0.3)                           & 2.9(0.6)                              & 17                             \\
${\mathrm{(2_{1,2}, 3/2, 2)  \rightarrow (1_{1,1}, 1/2, 1)} }$  &  \ \   91699.471 $^{(F)}$  &   \ \ 4.4  &  1.37$\times$10$^{-5}$   &   1.3  &   5  & 52.3(5.9)                                             & 10.8(0.3)                           & 2.3(0.7)                              & 21                             \\
${\mathrm{(2_{1,2}, 3/2, 2)  \rightarrow (1_{1,1}, 3/2, 2)} }$  &   \ \  91780.518 $^{(F)}$  &  \ \  4.4  &  2.23$\times$10$^{-6}$   &   0.2  &   5  & 24.2(5.9)                                             & 10.5(0.2)                           & 1.7(0.4)                              & 14                             \\
${\mathrm{(3_{1,3}, 7/2, 4)  \rightarrow (2_{1,2}, 5/2, 3)} }$  &  132993.978 $^{(W)}$  &  10.8  &  6.06$\times$10$^{-5}$   &   3.5  &   9  & \multirow{2}{*}[0cm]{$\Big \rangle$ 227.3(17.6)}	    &  \multirow{2}{*}[-0.01cm]{---}        & \multirow{2}{*}[-0.01cm]{---}           & \multirow{2}{*}[-0.01cm]{---}     \\
${\mathrm{(3_{1,3}, 7/2, 3)  \rightarrow (2_{1,2}, 5/2, 2)} }$  &  132994.679 $^{(W)}$  &  10.8  &  5.75$\times$10$^{-5}$   &   2.6  &   7  &             					    &                                       &                                         &                                    \\
${\mathrm{(3_{1,3}, 5/2, 2)  \rightarrow (2_{1,2}, 3/2, 1)} }$  &  133186.451 $^{(W)}$  &  10.8  &  5.11$\times$10$^{-5}$   &   1.6  &   5  & \multirow{2}{*}[0cm]{$\Big \rangle$ 252.6(27.7)}      &  \multirow{2}{*}[-0.01cm]{---}        & \multirow{2}{*}[-0.01cm]{---}           & \multirow{2}{*}[-0.01cm]{---}      \\
${\mathrm{(3_{1,3}, 5/2, 3)  \rightarrow (2_{1,2}, 3/2, 2)} }$  &  133187.717 $^{(W)}$  &  10.8  &  5.70$\times$10$^{-5}$   &   2.5  &   7  &          					            &                                       &                                         &                                    \\                                                                                                     
${\mathrm{(4_{1,4}, 9/2, 5)  \rightarrow (3_{1,3}, 7/2, 4)} }$  &  172463.355 $^{(W)}$  &  19.1  &  1.43$\times$10$^{-4}$   &   4.6  &  11  & \multirow{2}{*}[0cm]{$\Big \rangle$ 32.3(3.2) \ \ }        &  \multirow{2}{*}[-0.01cm]{---}        & \multirow{2}{*}[-0.01cm]{---}           & \multirow{2}{*}[-0.01cm]{---}      \\
${\mathrm{(4_{1,4}, 9/2, 4)  \rightarrow (3_{1,3}, 7/2, 3)} }$  &  172463.718 $^{(W)}$  &  19.1  &  1.39$\times$10$^{-4}$   &   3.6  &   9  &            						    &                                       &                                         &                                    \\                                                                                                          
${\mathrm{(4_{1,4}, 7/2, 3)  \rightarrow (3_{1,3}, 5/2, 2)} }$  &  172660.964 $^{(W)}$  &  19.1  &  1.32$\times$10$^{-4}$   &   2.7  &   7  & \multirow{2}{*}[0cm]{$\Big \rangle$ 218.3(52.5)}      &  \multirow{2}{*}[-0.01cm]{---}        & \multirow{2}{*}[-0.01cm]{---}           & \multirow{2}{*}[-0.01cm]{---}      \\
${\mathrm{(4_{1,4}, 7/2, 4)  \rightarrow (3_{1,3}, 5/2, 3)} }$  &  172661.526 $^{(W)}$  &  19.1  &  1.38$\times$10$^{-4}$   &   3.6  &   9  &             					    &                                       &                                         &                                    \\
${\mathrm{(5_{1,5}, 11/2, 6) \rightarrow (4_{1,4}, 9/2, 5)} }$ \ \ &  211117.576 $^{(F)}$  &  29.2  &  2.74$\times$10$^{-4}$   &   5.6  &  13  & \multirow{2}{*}[0cm]{$\Big \rangle$ 182.2(14.6)}      & \multirow{2}{*}[-0.01cm]{10.7(0.1)} & \multirow{2}{*}[-0.01cm]{2.1(0.2)}   & \multirow{2}{*}[-0.01cm]{82}    \\
${\mathrm{(5_{1,5}, 11/2, 5) \rightarrow (4_{1,4}, 9/2, 4)} }$  \ \  &  211117.834 $^{(F)}$  &  29.2  &  2.68$\times$10$^{-4}$   &   4.7  &  11  &             					    &                                       &                                         &                                    \\
${\mathrm{(5_{1,5}, 9/2, 4)  \rightarrow (4_{1,4}, 7/2, 3)} }$  &  211318.450 $^{(F)}$  &  29.2  &  2.61$\times$10$^{-4}$   &   3.7  &   9  & \multirow{2}{*}[0cm]{$\Big \rangle$ 96.4(13.2) \ }       & \multirow{2}{*}[-0.01cm]{10.6(0.1)} & \multirow{2}{*}[-0.01cm]{1.7(0.3)}    & \multirow{2}{*}[-0.01cm]{54}    \\
${\mathrm{(5_{1,5}, 9/2, 5)  \rightarrow (4_{1,4}, 7/2, 4)} }$  &  211318.796 $^{(F)}$  &  29.2  &  2.68$\times$10$^{-4}$   &   4.7  &  11  &                                                       &                                       &                                         &                                    \\
${\mathrm{(4_{1,3}, 9/2, 5)  \rightarrow (3_{1,2}, 7/2, 4)} }$  &  216488.286 $^{(F)}$  &  25.3  &  2.51$\times$10$^{-4}$   &   4.1  &  11  & 71.7(8.9)                                             & 10.7(0.1)                           & 2.3(0.3)                              & 30                             \\
${\mathrm{(4_{1,3}, 9/2, 4)  \rightarrow (3_{1,2}, 7/2, 3)} }$  &  216492.634 $^{(F)}$  &  25.3  &  2.21$\times$10$^{-4}$   &   2.9  &   9  & 32.4(5.9)                                             & 10.9(0.1)                           & 1.3(0.3)                              & 22                             \\
${\mathrm{(4_{1,3}, 7/2, 4)  \rightarrow (3_{1,2}, 5/2, 3)} }$  &  216638.258 $^{(F)}$  &  25.3  &  2.26$\times$10$^{-4}$   &   3.0  &   9  & 42.8(14.8)                                            & 10.6(0.1)                           & 1.2(0.2)                              & 29                             \\
${\mathrm{(4_{1,3}, 7/2, 3)  \rightarrow (3_{1,2}, 5/2, 2)} }$  &  216641.130 $^{(F)}$  &  25.3  &  2.31$\times$10$^{-4}$   &   2.4  &   7  & 20.1(16.3)                                            & 11.0(0.2)                           & 0.9(0.1)                              & 21                             \\
${\mathrm{(6_{1,6}, 13/2, 7) \rightarrow (5_{1,5}, 11/2, 6)}}$  &  249544.145 $^{(F)}$  &  41.2  &  4.65$\times$10$^{-4}$   &   6.7  &  15  & \multirow{2}{*}[0cm]{$\Big \rangle$ 125.4(13.0)}      & \multirow{2}{*}[-0.01cm]{10.7(0.1)} & \multirow{2}{*}[-0.01cm]{1.8(0.2)}    & \multirow{2}{*}[-0.01cm]{66}    \\
${\mathrm{(6_{1,6}, 13/2, 6) \rightarrow (5_{1,5}, 11/2, 5)}}$  &  249544.343 $^{(F)}$  &  41.2  &  4.59$\times$10$^{-4}$   &   5.7  &  13  &                                                       &                                       &                                         &                                  \\
${\mathrm{(6_{1,6}, 11/2, 5) \rightarrow (5_{1,5}, 9/2, 4)} }$  \ \ &  249746.630 $^{(F)}$  &  41.2  &  4.50$\times$10$^{-4}$   &   4.7  &  11  & \multirow{2}{*}[0cm]{$\Big \rangle$ 155.5(17.9)}      & \multirow{2}{*}[-0.01cm]{10.5(0.2)} & \multirow{2}{*}[-0.01cm]{3.1(0.5)}    & \multirow{2}{*}[-0.01cm]{47}    \\
${\mathrm{(6_{1,6}, 11/2, 6) \rightarrow (5_{1,5}, 9/2, 5)} }$  \ \ &  249746.873 $^{(F)}$  &  41.2  &  4.59$\times$10$^{-4}$   &   5.7  &  13  &                                                       &                                       &                                         &                                 \\
${\mathrm{(5_{1,4}, 11/2, 6) \rightarrow (4_{1,3}, 9/2, 5)} }$  \ \ &  252697.373 $^{(F)}$  &  37.4  &  4.09$\times$10$^{-4}$   &   4.9  &  13  & \multirow{2}{*}[0cm]{$\Big \rangle$ 123.6(8.2)}       & \multirow{2}{*}[-0.01cm]{10.5(0.1)} & \multirow{2}{*}[-0.01cm]{2.5(0.2)}    & \multirow{2}{*}[-0.01cm]{49}    \\
${\mathrm{(5_{1,4}, 11/2, 5) \rightarrow (4_{1,3}, 9/2, 4)} }$ \ \  &  252698.198 $^{(F)}$  &  37.4  &  4.01$\times$10$^{-4}$   &   4.1  &  11  &                                                       &                                       &                                         &                                 \\
${\mathrm{(5_{1,4}, 9/2, 4)  \rightarrow (4_{1,3}, 7/2, 3)} }$  &  252881.049 $^{(F)}$  &  37.4  &  3.89$\times$10$^{-4}$   &   3.2  &   9  & \multirow{2}{*}[0cm]{$\Big \rangle$ 72.5(9.9) \ }        & \multirow{2}{*}[-0.01cm]{10.6(0.1)} & \multirow{2}{*}[-0.01cm]{1.6(0.3)}    & \multirow{2}{*}[-0.01cm]{44}    \\
${\mathrm{(5_{1,4}, 9/2, 5)  \rightarrow (4_{1,3}, 7/2, 4)} }$  &  252881.590 $^{(F)}$  &  37.4  &  4.02$\times$10$^{-4}$   &   4.1  &  11  &             					    &                                       &                                         &                                 \\
${\mathrm{(7_{1,7}, 15/2, 8) \rightarrow (6_{1,6}, 13/2, 7)}}$  &  287920.669 $^{(F)}$  &  55.0  &  7.28$\times$10$^{-4}$   &   7.7  &  17  & \multirow{2}{*}[0cm]{$\Big \rangle$ 148.0(16.8)}       & \multirow{2}{*}[-0.01cm]{10.5(0.1)}         & \multirow{2}{*}[-0.01cm]{1.7(0.2)}           & \multirow{2}{*}[-0.01cm]{80}     \\
${\mathrm{(7_{1,7}, 15/2, 7) \rightarrow (6_{1,6}, 13/2, 6)}}$  &  287920.669 $^{(F)}$  &  55.0  &  7.21$\times$10$^{-4}$   &   6.8  &  15  &            						    &                                       &                                         &                                 \\
${\mathrm{(7_{1,7}, 13/2, 6) \rightarrow (6_{1,6}, 11/2, 5)}}$  &  288124.063 $^{(F)}$  &  55.0  &  7.12$\times$10$^{-4}$   &   5.8  &  13  & \multirow{2}{*}[0cm]{$\Big \rangle$ 94.5(16.8)}       & \multirow{2}{*}[-0.01cm]{10.6(0.2)}         & \multirow{2}{*}[-0.01cm]{2.0(0.4)}           & \multirow{2}{*}[-0.01cm]{43}     \\
${\mathrm{(7_{1,7}, 13/2, 7) \rightarrow (6_{1,6}, 11/2, 6)}}$  &  288124.063 $^{(F)}$  &  55.0  &  7.22$\times$10$^{-4}$   &   6.8  &  15  &             					    &                                       &                                         &                                 \\
${\mathrm{(6_{1,5}, 13/2, 7) \rightarrow (5_{1,4}, 11/2, 6)}}$  &  289270.928 $^{(W)}$  &  51.3  &  6.41$\times$10$^{-4}$   &   5.9  &  15  & \multirow{2}{*}[0cm]{$\Big \rangle$ 78.7(20.6)}       & \multirow{2}{*}[-0.01cm]{---}         & \multirow{2}{*}[-0.01cm]{---}           & \multirow{2}{*}[-0.01cm]{---}     \\
${\mathrm{(6_{1,5}, 13/2, 6) \rightarrow (5_{1,4}, 11/2, 5)}}$  &  289271.481 $^{(W)}$  &  51.3  &  6.32$\times$10$^{-4}$   &   5.1  &  13  &             					    &                                       &                                         &                                 \\
${\mathrm{(6_{1,5}, 11/2, 5) \rightarrow (5_{1,4}, 9/2, 4)} }$  \ \ &  289461.153 $^{(W)}$  &  51.3  &  6.21$\times$10$^{-4}$   &   4.2  &  11  & \multirow{2}{*}[0cm]{$\Big \rangle$ 39.4(13.1)}       & \multirow{2}{*}[-0.01cm]{---}         & \multirow{2}{*}[-0.01cm]{---}           & \multirow{2}{*}[-0.01cm]{ ---}     \\
${\mathrm{(6_{1,5}, 11/2, 6) \rightarrow (5_{1,4}, 9/2, 5)} }$  \ \ &  289461.805 $^{(W)}$  &  51.3  &  6.33$\times$10$^{-4}$   &   5.1  &  13  &             					    &                                       &                                         &                                 \\

\hline
\hline    
    
\end{tabular}
\end{center} 
 
\end{table*}


\begin{table*}[p]
\begin{center}
  
\caption{Line parameters of $c$-C$_{3}$H$_{2}$.}  \label{Table_$c$-C3H2} 
\begin{tabular}{c c c c c c c c c c c@{\vrule height 10pt depth 4.6pt width 0pt}}     
\hline\hline      

\bf Transition & {\bf Sym} & \bf Frequency & \bf E$_{_{\mathrm{\bf u}}}$ & \bf A$_{_{\mathrm{\bf ul}}}$ & {\bf S$_{_{\mathrm{\bf ij}}}$} & {\bf g$_{_{\mathrm{\bf u}}}$} & \bf $\displaystyle{\int}\bf T_{_{\mathrm{\bf MB}}}dv$ & \bf v$_{_{\mathrm{\bf LSR}}}$ & \bf $\Delta$v & \bf T$_{_{\mathrm{\bf MB}}}$ \rule[-0.3cm]{0cm}{0.8cm}\ \\ \cline{1-1}

${\mathrm{(J_{K_{a}K_{c}})_{u} \rightarrow (J_{K_{a}K_{c}})_{l}}}$ &  & [MHz] & [K] & [$\mathrm{s^{-1}}$] & & & [$\mathrm{K\, km\, s^{-1}}$] & [$\mathrm{km\, s^{-1}}$] & [$\mathrm{km\, s^{-1}}$]  & [$\mathrm{K}$]  \\

\hline 

${\mathrm{4_{2,2} \rightarrow 4_{1,3}}}$  & \emph{para} & \ \ 80723.179 $^{(F)}$ & 28.8 & 1.46$\times$10$^{-5}$ & 1.8 & 9 & 0.10(1) & 10.56(14)  & 2.60(41) & 0.03   \\
${\mathrm{2_{0,2} \rightarrow 1_{1,1}}}$  & \emph{para} & \ \ 82093.548 $^{(F)}$ & \ \ 6.4  & 2.07$\times$10$^{-5}$ & 1.4 & 5 &   0.55(1) & 10.64(2) & 2.75(6) & 0.19 \\
${\mathrm{6_{6,1} \rightarrow 6_{5,2}}}$  & \emph{ortho} & \ \ 82583.441 $^{(F)}$ &  66.3  & 1.31$\times$10$^{-5}$  & 2.2 & 13 &   0.01(1) & 10.82(32) & 0.86(44) & 0.01$^{\star}$  \\
${\mathrm{3_{1,2} \rightarrow 3_{0,3}}}$ & \emph{ortho} & \ \ 82966.196 $^{(F)}$ & 13.7 & 1.09$\times$10$^{-5}$ & 1.0 & 7 & 0.48(1) & 10.71(2) & 2.61(6) & 0.17 \\
${\mathrm{3_{2,2} \rightarrow 3_{1,3}}}$  & \emph{para} & \ \ 84727.687 $^{(F)}$ & 16.1 & 1.15$\times$10$^{-5}$ & 1.0 & 7 & 0.14(1) & 10.79(5) & 2.15(11) & 0.06 \\
${\mathrm{2_{1,2} \rightarrow 1_{0,1}}}$ & \emph{ortho} & \ \ 85338.900 $^{(F)}$ & \ \ 4.1  & 2.55$\times$10$^{-5}$ & 1.5 & 5 &   2.03(2) & 10.72(1) & 2.72(2) & 0.70 \\
${\mathrm{4_{3,2} \rightarrow 4_{2,3}}}$ & \emph{ortho} & \ \ 85656.415 $^{(F)}$ & 26.7 & 1.67$\times$10$^{-5}$ & 1.8 & 9 & 0.24(1) & 10.75(3) & 2.20(8) & 0.10 \\
${\mathrm{5_{4,2} \rightarrow 5_{3,3}}}$  & \emph{para} & \ \ 87435.317 $^{(F)}$ & 45.3 & 2.04$\times$10$^{-5}$ & 2.5 & 11 & 0.04(1) & 10.72(16) & 2.01(73) & 0.02 \\
${\mathrm{6_{5,2} \rightarrow 6_{4,3}}}$  & \emph{ortho} & \ \ 90344.081 $^{(F)}$ & 62.4 & 2.38$\times$10$^{-5}$ & 3.1 & 13 & 0.04(1) & 10.72(21) & 1.92(54) & 0.02 \\
${\mathrm{7_{4,3} \rightarrow 7_{3,4}}}$ & \emph{ortho} & 112490.768 $^{(F)}$ & 80.3 & 4.50$\times$10$^{-5}$ & 3.5 & 15 & 0.05(1) & 10.69(14) & 0.98(41) & 0.02 \\
${\mathrm{3_{0,3} \rightarrow 2_{1,2}}}$ & \emph{ortho} & 117151.183 $^{(W)}$ & \ \ 9.7 & 7.67$\times$10$^{-5}$ & 2.4 & 7 &  2.51(15) & --- & --- & --- \\
${\mathrm{3_{1,3} \rightarrow 2_{0,2}}}$ & \emph{para} & 117546.231 $^{(W)}$ & 12.1 & 7.77$\times$10$^{-5}$ & 2.5 & 7 &  0.74(6) & --- & --- & --- \\
${\mathrm{3_{1,2} \rightarrow 2_{2,1}}}$ & \emph{ortho} & 145089.611 $^{(W)}$ & 13.7 & 7.44$\times$10$^{-5}$ & 1.3 & 7 & 1.30(11) &  --- & --- & ---   \\
${\mathrm{2_{2,0} \rightarrow 1_{1,1}}}$  & \emph{para} & 150436.558 $^{(W)}$ & \ \ 9.7 & 5.89$\times$10$^{-5}$ & 0.6 & 5 & 0.32(6)  & ---  & --- & ---   \\
${\mathrm{4_{0,4} \rightarrow 3_{1,3}}}$  & \emph{para} & 150820.665 $^{(W)}$ & 19.3 & 1.80$\times$10$^{-4}$ & 3.5 & 9 & 1.00(4)  & ---  & --- & ---   \\
${\mathrm{4_{1,4} \rightarrow 3_{0,3}}}$ & \emph{ortho} & 150851.898 $^{(W)}$ & 17.0 & 1.80$\times$10$^{-4}$ & 3.5 & 9 & 3.00(19) & ---  & --- &  ---  \\
${\mathrm{6_{2,4} \rightarrow 6_{1,5}}}$  & \emph{para} & 150954.689 $^{(W)}$ & 54.7 & 6.87$\times$10$^{-5}$ & 1.9 & 13 & 0.06(2)  & ---  & --- & ---  \\
${\mathrm{6_{3,4} \rightarrow 6_{2,5}}}$ & \emph{ortho} & 151039.149 $^{(W)}$ & 52.4 & 6.88$\times$10$^{-5}$ & 1.9 & 13 & 0.20(3) &  --- & --- & ---   \\
${\mathrm{5_{1,4} \rightarrow 5_{0,5}}}$ & \emph{ortho} & 151343.875 $^{(W)}$ & 33.1 & 4.35$\times$10$^{-5}$ & 1.0 & 11 & 0.34(4) &  --- & --- & ---   \\
${\mathrm{5_{2,4} \rightarrow 5_{1,5}}}$  & \emph{para} & 151361.102 $^{(W)}$ & 35.4 & 4.35$\times$10$^{-5}$ & 1.0 & 11 &  0.09(2) & ---  & --- &  ---  \\
${\mathrm{3_{2,2} \rightarrow 2_{1,1}}}$  & \emph{para} & 155518.308 $^{(W)}$ & 16.1 & 1.23$\times$10$^{-4}$ & 1.7 & 7 & 0.68(4)  & ---  & --- & ---   \\
${\mathrm{4_{2,2} \rightarrow 3_{2,1}}}$  & \emph{para} & 204788.929 $^{(F)}$ & 28.8 & 1.37$\times$10$^{-4}$ & 1.0 & 9 & 0.20(1) & 10.83(5)  & 2.03(14) & 0.09  \\
${\mathrm{3_{3,0} \rightarrow 2_{2,1}}}$ & \emph{ortho} & 216278.738 $^{(F)}$ & 17.1 & 2.81$\times$10$^{-4}$ & 1.4 & 7 & 1.29(1)  & 10.74(9) & 2.29(2) & 0.53 \\
${\mathrm{6_{0,6} \rightarrow 5_{1,5}}}$  & \emph{para} & 217822.057 $^{(F)}$ & 38.6 & 5.93$\times$10$^{-4}$ & 5.5 & 13 & \multirow{2}{*}[0cm]{$\Big \rangle$ 3.14(1)} & \multirow{2}{*}[-0.01cm]{10.75(1)} & \multirow{2}{*}[-0.01cm]{2.18(1)} & \multirow{2}{*}[-0.01cm]{1.35} \\
${\mathrm{6_{1,6} \rightarrow 5_{0,5}}}$ & \emph{ortho} & 217822.180 $^{(F)}$ & 36.3 & 5.93$\times$10$^{-4}$ & 5.5 & 13 &   &   &  &    \\
${\mathrm{5_{1,4} \rightarrow 4_{2,3}}}$ & \emph{ortho} & 217940.045 $^{(F)}$ & 33.1 & 4.43$\times$10$^{-4}$ & 3.4 & 11 & 1.60(1) &  10.78(1) & 2.13(1) & 0.70 \\
${\mathrm{5_{2,4} \rightarrow 4_{1,3}}}$  & \emph{para} & 218160.462 $^{(F)}$ & 35.4 & 4.44$\times$10$^{-4}$ & 3.4 & 11 & 0.56(1) & 10.79(2) & 2.20(4) & 0.24 \\
${\mathrm{8_{2,6} \rightarrow 8_{1,7}}}$  & \emph{para} & 218448.823 $^{(F)}$ & 86.9 & 1.64$\times$10$^{-4}$ & 2.0 & 17 & \multirow{2}{*}[0cm]{$\Big \rangle$ 0.10(1)} & \multirow{2}{*}[-0.01cm]{10.75(10)} & \multirow{2}{*}[-0.01cm]{2.27(25)} & \multirow{2}{*}[-0.01cm]{0.04} \\
${\mathrm{8_{3,6} \rightarrow 8_{2,7}}}$ & \emph{ortho} & 218449.400 $^{(F)}$ & 84.6 & 1.64$\times$10$^{-4}$ & 2.0 & 17 &   &   &  &    \\
${\mathrm{7_{1,6} \rightarrow 7_{0,7}}}$  & \emph{ortho} & 218732.683 $^{(F)}$ & 58.8 & 9.82$\times$10$^{-5}$ & 1.0 & 15 &  \multirow{2}{*}[0cm]{$\Big \rangle$ 0.16(1)} & \multirow{2}{*}[-0.01cm]{10.75(5)} & \multirow{2}{*}[-0.01cm]{2.07(11)} & \multirow{2}{*}[-0.01cm]{0.07} \\
${\mathrm{7_{2,6} \rightarrow 7_{1,7}}}$  & \emph{para} & 218732.767 $^{(F)}$ & 61.2 & 9.82$\times$10$^{-5}$ & 1.0 & 15 &   &   &  &    \\
${\mathrm{4_{3,2} \rightarrow 3_{2,1}}}$  & \emph{ortho} & 227169.143 $^{(F)}$ & 26.7 & 3.43$\times$10$^{-4}$ & 1.9 & 9 & 1.10(2) & 10.74(2) & 2.11(5) & 0.49 \\
${\mathrm{3_{2,1} \rightarrow 2_{1,2}}}$  & \emph{ortho} & 244222.155 $^{(F)}$ & 15.8 & 6.49$\times$10$^{-5}$ & 0.2 & 7 & 0.38(2) & 10.76(5) & 2.23(13) & 0.16 \\
${\mathrm{5_{2,3} \rightarrow 4_{3,2}}}$  & \emph{ortho} & 249054.415 $^{(F)}$ & 38.7 & 4.57$\times$10$^{-4}$ & 2.4 & 11 & 0.79(1) & 10.80(1) & 1.93(4) & 0.39 \\
${\mathrm{7_{0,7} \rightarrow 6_{1,6}}}$  & \emph{ortho} & 251314.362 $^{(F)}$ & 48.3 & 9.35$\times$10$^{-4}$ & 6.5 & 15 &  \multirow{2}{*}[0cm]{$\Big \rangle$ 2.22(1)} & \multirow{2}{*}[-0.01cm]{10.78(1)} & \multirow{2}{*}[-0.01cm]{1.98(1)} & \multirow{2}{*}[-0.01cm]{1.05} \\
${\mathrm{7_{1,7} \rightarrow 6_{0,6}}}$  & \emph{para} & 251314.369 $^{(F)}$ & 50.7 & 9.35$\times$10$^{-4}$ & 6.5 & 15 &   &   &  &    \\
${\mathrm{6_{1,5} \rightarrow 5_{2,4}}}$  & \emph{para} & 251508.713 $^{(F)}$ & 47.5 & 7.42$\times$10$^{-4}$ & 4.4 & 13 & 0.37(1) & 10.76(2) & 1.82(6) & 0.19 \\
${\mathrm{6_{2,5} \rightarrow 5_{1,4}}}$  & \emph{ortho} & 251527.325 $^{(F)}$ & 45.1 & 7.42$\times$10$^{-4}$ & 4.4 & 13 & 1.16(1) & 10.80(1) & 2.00(2) & 0.55 \\
${\mathrm{10_{3,7} \rightarrow 10_{2,8}}}$  & \emph{para} & 251773.193 $^{(F)}$ & 137.6 \ \ & 2.98$\times$10$^{-4}$ & 2.9 & 21 &    \multirow{2}{*}[0cm]{$\Big \rangle$ 0.02(1)} & \multirow{2}{*}[-0.01cm]{10.75(13)} & \multirow{2}{*}[-0.01cm]{0.93(23)} & \multirow{2}{*}[-0.01cm]{0.02$^{\star}$} \\
${\mathrm{10_{4,7} \rightarrow 10_{3,8}}}$  & \emph{ortho} & 251773.421 $^{(F)}$ & 135.3 \ \ & 2.98$\times$10$^{-4}$ & 2.9 & 21 &   &   &  &    \\
${\mathrm{8_{1,7} \rightarrow 8_{0,8}}}$  & \emph{para} & 252409.829 $^{(F)}$ & 76.5 & 1.34$\times$10$^{-4}$ & 1.0 & 17 &  \multirow{2}{*}[0cm]{$\Big \rangle$ 0.10(1)} & \multirow{2}{*}[-0.01cm]{10.76(1)} & \multirow{2}{*}[-0.01cm]{2.01(26)} & \multirow{2}{*}[-0.01cm]{0.05} \\
${\mathrm{8_{2,7} \rightarrow 8_{1,8}}}$  & \emph{ortho} & 252409.834 $^{(F)}$ & 74.1 & 1.34$\times$10$^{-4}$ & 1.0 & 17 &   &   &  &    \\
\multicolumn{2}{l}{\bf{\sc {Continued On Next Page.}}}  &  & &  &  &  &   &  & &  \\

\hline
\hline    
    
\end{tabular}
\end{center} 

\end{table*}

\begin{table*}[p]
\begin{center}
  
\begin{tabular}{c c c c c c c c c c c@{\vrule height 10pt depth 5pt width 0pt}}     
\hline\hline      

\bf Transition & {\bf Sym} & \bf Frequency & \bf E$_{_{\mathrm{\bf u}}}$ & \bf A$_{_{\mathrm{\bf ul}}}$ & {\bf S$_{_{\mathrm{\bf ij}}}$} & {\bf g$_{_{\mathrm{\bf u}}}$} & \bf $\displaystyle{\int}\bf T_{_{\mathrm{\bf MB}}}dv$ & \bf v$_{_{\mathrm{\bf LSR}}}$ & \bf $\Delta$v & \bf T$_{_{\mathrm{\bf MB}}}$ \rule[-0.3cm]{0cm}{0.8cm}\ \\ \cline{1-1}

${\mathrm{(J_{K_{a}K_{c}})_{u} \rightarrow (J_{K_{a}K_{c}})_{l}}}$ &  & [MHz] & [K] & [$\mathrm{s^{-1}}$] & & & [$\mathrm{K\, km\, s^{-1}}$] & [$\mathrm{km\, s^{-1}}$] & [$\mathrm{km\, s^{-1}}$]  & [$\mathrm{K}$]  \\

\hline 
\multicolumn{2}{l}{\bf{\sc {Continued From Table B.6.}}}  &  & &  &  &  &   &  & &  \\

${\mathrm{5_{3,3} \rightarrow 4_{2,2}}}$  & \emph{para} & 254987.657 $^{(F)}$ & 41.1 & 5.17$\times$10$^{-4}$ & 2.5 & 11 &  0.31(1) & 10.72(4) & 2.18(10) & 0.13 \\
${\mathrm{5_{3,2} \rightarrow 4_{4,1}}}$  & \emph{ortho} & 260479.764 $^{(W)}$ & 42.4 & 1.77$\times$10$^{-4}$ & 0.8 & 11 & 0.25(8)  & ---  & --- & ---   \\
${\mathrm{4_{4,1} \rightarrow 3_{3,0}}}$  & \emph{ortho} & 265759.483 $^{(F)}$ & 29.9 & 7.99$\times$10$^{-4}$ & 2.8 & 9 & 0.94(1) & 10.79(1) & 1.94(3) & 0.46   \\
  ${\mathrm{4_{4,0} \rightarrow   3_{3,1}}}$  & \emph{para} &    282381.108 $^{(F)}$ &   32.5  &  8.11$\times$10$^{-4}$ &   2.37 &   9   & 0.42(2)  &  10.80(4)   &  2.09 (10) & 0.19   \\
  ${\mathrm{8_{0,8} \rightarrow   7_{1,7}}}$  & \emph{para} &    284805.229 $^{(F)}$ &   64.3  &  1.39$\times$10$^{-3}$ &   7.45 &  17   & \multirow{2}{*}[0cm]{$\Big \rangle$ 2.05(2)}  &   \multirow{2}{*}[-0.01cm]{10.64(1)}   &   \multirow{2}{*}[-0.01cm]{2.21(3)}   & \multirow{2}{*}[-0.01cm]{0.87}   \\ 
  ${\mathrm{8_{1,8} \rightarrow   7_{0,7}}}$  & \emph{ortho} &   284805.230 $^{(F)}$ &   62.0  &  1.39$\times$10$^{-3}$ &   7.45 &  17   &          &             &            &        \\
  ${\mathrm{6_{2,4} \rightarrow   5_{3,3}}}$  & \emph{para} &    284913.028 $^{(F)}$ &   54.7  &  8.38$\times$10$^{-4}$ &   3.44 &  13   & 0.26(2)  &  10.65(8)   &  2.19(21)  & 0.11	   \\
  ${\mathrm{7_{1,6} \rightarrow   6_{2,5}}}$  & \emph{ortho} &   284998.025 $^{(F)}$ &   58.8  &  1.15$\times$10$^{-3}$ &   5.42 &  15   & \multirow{2}{*}[0cm]{$\Big \rangle$ 1.50(2)}  &   \multirow{2}{*}[-0.01cm]{10.38(2)}   &   \multirow{2}{*}[-0.01cm]{2.82(5)}   & \multirow{2}{*}[-0.01cm]{0.50}    \\
  ${\mathrm{7_{2,6} \rightarrow   6_{1,5}}}$  & \emph{para} &    284999.377 $^{(F)}$ &   61.2  &  1.15$\times$10$^{-3}$ &   5.42 &  15   &          &             &            &         \\
  ${\mathrm{6_{3,4} \rightarrow   5_{2,3}}}$  & \emph{ortho} &   285795.689 $^{(F)}$ &   52.4  &  8.49$\times$10$^{-4}$ &   3.45 &  13   & 0.79(2)  &  10.65(3)   &   2.17(7)  & 0.34    \\
  ${\mathrm{5_{4,2} \rightarrow   4_{3,1}}}$  & \emph{para} &    300191.723 $^{(F)}$ &   45.3  &  7.61$\times$10$^{-4}$ &   2.26 &  11   & 0.28(3)  &  11.13(17)  &   3.01(41) & 0.09     \\
  ${\mathrm{5_{5,0} \rightarrow   4_{4,1}}}$  & \emph{ortho} &   349264.002 $^{(F)}$ &   46.6  &  1.81$\times$10$^{-3}$ &   3.41 &  11   & 0.63(5)  &  10.60(12)  &   3.11(38) & 0.19     \\
  ${\mathrm{7_{3,4} \rightarrow   6_{4,3}}}$  & \emph{ortho} &   351523.296 $^{(F)}$ &   74.9  &  1.36$\times$10$^{-3}$ &   3.43 &  15   & 0.26(5)  &  10.86(19)  &   1.91(45) & 0.13     \\
 ${\mathrm{10_{1,10} \rightarrow  9_{0,9}}}$ \ \ \ & \emph{ortho} &   351781.573 $^{(F)}$ &   94.1  &  2.68$\times$10$^{-3}$ &   9.45 &  21   & \multirow{2}{*}[0cm]{$\Big \rangle$ 0.79(6)}  &   \multirow{2}{*}[-0.01cm]{10.61(7)}   &    \multirow{2}{*}[-0.01cm]{1.87(15)} & \multirow{2}{*}[-0.01cm]{0.40}     \\
 ${\mathrm{10_{0,10} \rightarrow  9_{1,9}}}$ \ \  \  & \emph{para} &    351781.573 $^{(F)}$ &   96.5  &  2.68$\times$10$^{-3}$ &   9.45 &  21   &          &             &            &          \\
  ${\mathrm{9_{1,8} \rightarrow   8_{2,7}}}$  & \emph{ortho} &   351965.963 $^{(F)}$ &   91.0  &  2.33$\times$10$^{-3}$ &   7.41 &  19   & \multirow{2}{*}[0cm]{$\Big \rangle$ 0.49(5)}  &   \multirow{2}{*}[-0.01cm]{10.54(7)}   &    \multirow{2}{*}[-0.01cm]{1.50(15)} & \multirow{2}{*}[-0.01cm]{0.31}     \\
  ${\mathrm{9_{2,8} \rightarrow   8_{1,7}}}$  & \emph{para} &    351965.969 $^{(F)}$ &   93.3  &  2.33$\times$10$^{-3}$ &   7.41 &  19   &          &             &            &          \\
  ${\mathrm{8_{3,6} \rightarrow   7_{2,5}}}$  & \emph{ortho} &   352193.664 $^{(F)}$ &   84.6  &  1.91$\times$10$^{-3}$ &   5.42 &  17   & 0.35(5)  &  10.74(13)  &   1.60(16) & 0.21     \\

\hline
\hline    
    
\end{tabular}
\end{center} 

\end{table*}


\begin{table*}[p]
\begin{center}

\caption{Line parameters of $l$-H$_{2}$C$_{3}$.}  \label{Table_$l$-C3H2} 
\begin{tabular}{c c c c c c c c c c c@{\vrule height 10pt depth 5pt width 0pt}}     
\hline\hline      

\bf Transition & {\bf Sym} & \bf Frequency & \bf E$_{_{\mathrm{\bf u}}}$ & \bf A$_{_{\mathrm{\bf ul}}}$ & {\bf S$_{_{\mathrm{\bf ij}}}$} & {\bf g$_{_{\mathrm{\bf u}}}$} & \bf $\displaystyle{\int}\bf T_{_{\mathrm{\bf MB}}}dv$ & \bf v$_{_{\mathrm{\bf LSR}}}$ & \bf $\Delta$v & \bf T$_{_{\mathrm{\bf MB}}}$ \rule[-0.3cm]{0cm}{0.8cm}\ \\ \cline{1-1}

${\mathrm{(J_{K_{a}K_{c}})_{u} \rightarrow (J_{K_{a}K_{c}})_{l}}}$ &  & [MHz] & [K] & [$\mathrm{s^{-1}}$] & & & [$\mathrm{mK\, km\, s^{-1}}$] & [$\mathrm{km\, s^{-1}}$] & [$\mathrm{km\, s^{-1}}$]  & [$\mathrm{mK}$]  \\

\hline

${\mathrm{4 _{1,4} \rightarrow  3_{1,3}}}$    & \emph{ortho}  &  \ \  82395.089 $^{(F)}$  &  \ \ 8.9   & 4.56$\times$10$^{-5}$   & 3.8  & 9 &  65.5(9.3) & 10.7(0.1) & 2.1(0.4) & 29  \\   
${\mathrm{4 _{0,4} \rightarrow  3_{0,3}}}$    & \emph{para}  &  \ \  83165.345 $^{(F)}$  &  10.0   & 5.00$\times$10$^{-5}$   &  4.0 & 9 &  22.8(9.3) & 10.5(0.6) & 2.8(1.1) &  8$^{\star}$   \\   
${\mathrm{4_{1,3} \rightarrow 3_{1,2 }}}$     &  \emph{ortho}&  \ \  83933.699 $^{(F)}$  &  \ \ 9.1   & 4.82$\times$10$^{-5}$   & 3.8 & 9  & 75.2(8.2)  &  10.7(0.2) & 2.0(0.4) & 29   \\ 
${\mathrm{5 _{1,5} \rightarrow 4_{1,4 }}}$    &  \emph{ortho} & 102992.379 $^{(F)}$  & 13.8   & 9.33$\times$10$^{-5}$   & 4.8   & 11 & 67.5(8.4)  &  10.5(0.2) & 2.7(0.4) & 26   \\
${\mathrm{5_{0,5} \rightarrow 4_{0,4}}}$   &  \emph{para} & 103952.926 $^{(F)}$   & 15.0  & 9.99$\times$10$^{-5}$ &  5.0   &   11 &  28.5(7.2) & 10.8(0.3) & 2.5(0.8) & 11   \\
${\mathrm{5_{1,4} \rightarrow 4_{1,3}}}$     &  \emph{ortho} & 104915.583 $^{(F)}$  &  14.1   & 9.86$\times$10$^{-5}$   & 4.8   &  11 & 91.3(8.4)  & 10.5(0.2)  & 3.7(0.4) & 23  \\  
${\mathrm{7 _{1,7} \rightarrow 6_{1,6}}}$     &  \emph{ortho} & 144183.804 $^{(W)}$  &  26.7   & 2.68$\times$10$^{-4}$   & 6.9    & 15 & 96.9(30.7)  & ---  & --- & ---   \\ 
${\mathrm{7_{1,6} \rightarrow 6_{1,5}}}$     & \emph{ortho}  & 146876.061 $^{(W)}$  & 27.2  &  2.83$\times$10$^{-4}$   & 6.9    & 15 & 105.6(33.4)  & --- & --- & ---   \\
${\mathrm{8 _{1,8} \rightarrow 7_{1,7}}}$     & \emph{ortho}  & 164777.547 $^{(W)}$ &  34.6   & 4.06$\times$10$^{-4}$   & 7.9    & 17 & 78.9(25.2)  &  --- &  --- &  ---  \\
${\mathrm{8_{1,7} \rightarrow 7 _{1,6}}}$     &  \emph{ortho} & 167854.234 $^{(W)}$ &  35.3   & 4.29$\times$10$^{-4}$   & 7.9   & 17 &  110.2(21.3) & ---  &  --- &  ---  \\
${\mathrm{10_{1,10} \rightarrow 9_{1,9}}}$ \ \ \    & \emph{ortho}  & 205960.125 $^{(F)}$ &  53.4   & 8.06$\times$10$^{-4}$   & 9.9 &  21 &  65.0(13.0) & 10.7(0.2) & 1.7(0.5) & 36  \\
${\mathrm{10_{0,10} \rightarrow 9_{0,9}}}$ \ \ \  &  \emph{para} &  207843.289 $^{(F)}$ & 54.9  & 8.37$\times$10$^{-4}$  & 10.0 \ \  & 21 & 27.8(8.7)  &  10.7(0.2) & 1.5(0.4) & 17$^{\star}$   \\
${\mathrm{10_{1,9 } \rightarrow 9_{1,8}}}$ \ \   & \emph{ortho}  & 209805.427 $^{(F)}$ & 54.4   & 8.52$\times$10$^{-4}$   & 9.9   & 21 & 80.4(17.5)  & 10.7(0.2)  & 2.2(0.5) & 35  \\
${\mathrm{11_{1,11} \rightarrow 10_{1,10 }}}$    &  \emph{ortho} &  226548.575 $^{(F)}$  &  64.3   & 1.08$\times$10$^{-3}$   & 10.9   \ \  &   23    & 65.4(18.3) & 10.6(0.2)  &  1.6(0.4) &  50  \\
${\mathrm{12_{0,12} \rightarrow 11_{0,11}}}$    &  \emph{para} & 249367.939 $^{(F)}$  & 77.8  &  1.46$\times$10$^{-3}$  & 12.0 \ \ &  25  &  95.8(14.7) & 10.9(0.2)  & 2.5(0.5) & 36$^{\star}$   \\
${\mathrm{12_{1,11} \rightarrow 11_{1,10}}}$     &  \emph{ortho} & 251748.328 $^{(F)}$  & 77.6  & 1.49$\times$10$^{-3}$   & 11.9  \ \    & 25   & 37.2(9.9)  & 10.7(0.1)  & 1.0(0.3) & 36   \\
${\mathrm{13_{0,13} \rightarrow 12_{0,12}}}$   &  \emph{para} &   270121.513 $^{(W)}$ & 90.8  & 1.86$\times$10$^{-3}$  & 13.0 \ \ &  27  & 14.5(2.9)$^{\star}$  & --- & --- & ---   \\
${\mathrm{13_{1,12} \rightarrow 12_{1,11}}}$   &  \emph{ortho} &   272716.158 $^{(F)}$ & 90.6  & 1.90$\times$10$^{-3}$  & 12.9 \ \ &  27  & 35.9(8.8)  & 10.9(0.1) & 1.0(0.2) & 34$^{\star}$   \\

\hline
\hline    
    
\end{tabular}
\end{center} 

\end{table*}


\begin{table*}[p]
\begin{center}
  
\caption{Line parameters of C$_{4}$H.}  \label{Table_C4H} 
\begin{tabular}{c c c c c c c c c c@{\vrule height 10pt depth 5pt width 0pt}}     
\hline\hline      

\bf Transition & \bf Frequency & \bf E$_{_{\mathrm{\bf u}}}$ & \bf A$_{_{\mathrm{\bf ul}}}$ & {\bf S$_{_{\mathrm{\bf ij}}}$} & {\bf g$_{_{\mathrm{\bf u}}}$} & \bf $\displaystyle{\int}\bf T_{_{\mathrm{\bf MB}}}dv$ & \bf v$_{_{\mathrm{\bf LSR}}}$ & \bf $\Delta$v & \bf T$_{_{\mathrm{\bf MB}}}$ \rule[-0.3cm]{0cm}{0.8cm}\ \\ \cline{1-1}

${\mathrm{(N, J)_{u} \rightarrow (N, J)_{l}}}$ & [MHz] & [K] & [$\mathrm{s^{-1}}$] & & & [$\mathrm{K\, km\, s^{-1}}$] & [$\mathrm{km\, s^{-1}}$] & [$\mathrm{km\, s^{-1}}$]  & [$\mathrm{K}$]  \\

\hline

${\mathrm{(9, 19/2) \rightarrow (8, 17/2)}}$      & \ \ 85634.023 $^{(F)}$ 	& 20.5 & 2.63$\times$10$^{-6}$  & \ \ 9.5 & 20 & 0.21(1)  &  10.55(5) & 2.66(13) & 0.074   \\
${\mathrm{(9, 17/2) \rightarrow (8, 15/2)}}$      & \ \ 85672.563 $^{(F)}$ 	& 20.6 & 2.62$\times$10$^{-6}$  & \ \ 8.5 & 18 & 0.19(1)  &   10.46(5) & 2.61(12) & 0.067   \\
${\mathrm{(10, 21/2) \rightarrow (9, 19/2)}}\ \ $ & \ \ 95150.402 $^{(F)}$ 	& 25.1 & 3.62$\times$10$^{-6}$  & 10.5 & 22 & 0.20(1) & 10.55(5) & 2.61(13) & 0.071   \\
${\mathrm{(10, 19/2) \rightarrow (9, 17/2)}}\ \ $ & \ \ 95188.935 $^{(F)}$ 	& 25.1 & 3.61$\times$10$^{-6}$  & \ \ 9.5 & 20 & 0.19(1) &   10.31(7) & 2.90(17) & 0.061   \\
${\mathrm{(11, 23/2) \rightarrow (10, 21/2)}}$    & 104666.575 $^{(F)}$	& 30.1 & 4.84$\times$10$^{-6}$  & 11.5 & 24 & 0.21(1) & 10.61(4) & 2.60(10) & 0.078   \\
${\mathrm{(11, 21/2) \rightarrow (10, 19/2)}}$    & 104705.099 $^{(F)}$ 	& 30.2 & 4.83$\times$10$^{-6}$  & 10.5 & 22 & 0.20(1) & 10.59(6) & 2.84(16) & 0.066   \\
${\mathrm{(12, 25/2) \rightarrow (11, 23/2)}}$    & 114182.520 $^{(F)}$ 	& 35.6 & 6.31$\times$10$^{-6}$  & 12.5 & 26 & 0.25(2) & 10.70(8) & 2.51(21)  & 0.093   \\
${\mathrm{(12, 23/2) \rightarrow (11, 21/2)}}$    & 114221.035 $^{(F)}$ 	& 35.6 & 6.29$\times$10$^{-6}$  & 11.5 & 24 & 0.23(1) & 10.72(6) & 2.42(14)  & 0.100   \\

${\mathrm{(14, 29/2) \rightarrow (13, 27/2)}}$    & 133213.644 $^{(W)}$ & 47.9 & 1.01$\times$10$^{-5}$  & 14.5 & 30 & 0.27(5) & --- & ---  & ---   \\
${\mathrm{(14, 27/2) \rightarrow (13, 25/2)}}$    & 133252.140 $^{(W)}$ & 48.0 & 1.01$\times$10$^{-5}$  & 13.5 & 28 & 0.26(6) & --- & ---  & ---   \\
${\mathrm{(15, 31/2) \rightarrow (14, 29/2)}}$    & 142728.783 $^{(W)}$ & 54.8 & 1.24$\times$10$^{-5}$  & 15.5 & 32 & 0.23(5) & --- & ---  & ---   \\
${\mathrm{(15, 29/2) \rightarrow (14, 27/2)}}$    & 142767.267 $^{(W)}$ & 54.8 & 1.24$\times$10$^{-5}$  & 14.5 & 30 & 0.28(4) & --- & ---  & ---   \\
${\mathrm{(16, 33/2) \rightarrow (15, 31/2)}}$    & 152243.611 $^{(W)}$ & 62.1 & 1.51$\times$10$^{-5}$  & 16.5 & 34 & 0.21(3) & --- & ---  & ---   \\
${\mathrm{(16, 31/2) \rightarrow (15, 29/2)}}$    & 152282.083 $^{(W)}$ & 62.1 & 1.51$\times$10$^{-5}$  & 15.5 & 32 & 0.30(4) & --- & ---  & ---   \\
${\mathrm{(17, 35/2) \rightarrow (16, 33/2)}}$    & 161758.109 $^{(W)}$ & 69.9 & 1.81$\times$10$^{-5}$  & 17.5 & 36 & 0.26(4) & --- & ---  & ---   \\
${\mathrm{(17, 33/2) \rightarrow (16, 31/2)}}$    & 161796.568 $^{(W)}$ & 69.9 & 1.81$\times$10$^{-5}$  & 16.5 & 34 & 0.24(3) & --- & ---  & ---   \\
${\mathrm{(18, 37/2) \rightarrow (17, 35/2)}}$    & 171272.255 $^{(W)}$ & 78.1 & 2.16$\times$10$^{-5}$  & 18.5 & 38 & 0.16(3)  & --- &  --- &  ---  \\
${\mathrm{(18, 35/2) \rightarrow (17, 33/2)}}$    & 171310.702 $^{(W)}$ & 78.1 & 2.15$\times$10$^{-5}$  & 17.5 & 36 & $^{(s)}$ & --- &  --- & ---   \\

${\mathrm{(22, 45/2) \rightarrow (21, 43/2)}}$    & 209324.922 $^{(F)}$ & 115.5 & 3.95$\times$10$^{-5}$  & 22.5 & 46 & 0.25(1)  & 10.78(4) & 2.04(10) & 0.117   \\
${\mathrm{(22, 43/2) \rightarrow (21, 41/2)}}$    & 209363.306 $^{(F)}$ & 115.6 & 3.95$\times$10$^{-5}$  & 21.5 & 44 & 0.26(1)  & 10.68(4) & 2.35(12) & 0.104   \\

${\mathrm{(23, 47/2) \rightarrow (22, 45/2)}}$    & 218837.007 $^{(F)}$ & 126.0 & 4.52$\times$10$^{-5}$  & 23.5 & 48 & 0.23(1)  & 10.82(4) & 2.15(10) & 0.101   \\
${\mathrm{(23, 45/2) \rightarrow (22, 43/2)}}$    & 218875.374 $^{(F)}$ & 126.1 & 4.52$\times$10$^{-5}$  & 22.5 & 46 & 0.21(1)  & 10.72(4) & 1.84(10) & 0.106   \\

${\mathrm{(24, 49/2) \rightarrow (23, 47/2)}}$    & 228348.618 $^{(F)}$ & 137.0 & 5.14$\times$10$^{-5}$  & 24.5 & 50 & 0.20(2)  & 10.76(9) & 2.08(24) & 0.092   \\
${\mathrm{(24, 47/2) \rightarrow (23, 45/2)}}$    & 228386.966 $^{(F)}$ & 137.0 & 5.14$\times$10$^{-5}$  & 23.5 & 48 & 0.17(2)  & 10.69(13) & 2.35(35) & 0.068  \\

${\mathrm{(25, 51/2) \rightarrow (24, 49/2)}}$    & 237859.735 $^{(F)}$ & 148.4 & 5.81$\times$10$^{-5}$  & 25.5 & 52 & 0.16(2)  & 10.84(9) & 1.62(28) & 0.090   \\
${\mathrm{(25, 49/2) \rightarrow (24, 47/2)}}$    & 237898.065 $^{(F)}$ & 148.5 & 5.81$\times$10$^{-5}$  & 24.5 & 50 & 0.21(2)  & 10.65(11) & 2.27(27) & 0.087   \\

${\mathrm{(26, 53/2) \rightarrow (25, 51/2)}}$    & 247370.338 $^{(F)}$ & 160.3 & 6.54$\times$10$^{-5}$  & 26.5 & 54 & 0.20(1)  & 10.89(6) & 1.96(15)  & 0.094   \\
${\mathrm{(26, 51/2) \rightarrow (25, 49/2)}}$    & 247408.648 $^{(F)}$ & 160.3 & 6.54$\times$10$^{-5}$  & 25.5 & 52 & 0.20(1)  & 10.71(6) & 2.19(15)  & 0.084   \\

${\mathrm{(27, 55/2) \rightarrow (26, 53/2)}}$    & 256880.407 $^{(F)}$ & 172.6 & 7.33$\times$10$^{-5}$  & 27.5 & 56 & 0.19(2)  & 10.83(7) & 1.70(20)  & 0.104   \\
${\mathrm{(27, 53/2) \rightarrow (26, 51/2)}}$    & 256918.696 $^{(F)}$ & 172.7 & 7.33$\times$10$^{-5}$  & 26.5 & 54 & 0.18(1)  & 10.75(7) & 2.08(17)  & 0.082   \\

${\mathrm{(28, 57/2) \rightarrow (27, 55/2)}}$    & 266389.921 $^{(F)}$ & 185.4 & 8.18$\times$10$^{-5}$  & 28.5 & 58 & 0.14(1)  &  10.85(8) & 1.96(18) & 0.067  \\
${\mathrm{(28, 55/2) \rightarrow (27, 53/2)}}$    & 266428.190 $^{(F)}$ & 185.5 & 8.18$\times$10$^{-5}$  & 27.5 & 56 & 0.12(1)  & 10.60(8)  & 1.90(20) & 0.057  \\

${\mathrm{(29, 59/2) \rightarrow (28, 57/2)}}$ &  275898.861 $^{(F)}$ & 198.6 &  9.10$\times$10$^{-5}$ &   29.5 &   60    & 0.17(2)   & 10.73(8)  & 1.60(22)    &  0.102  \\
${\mathrm{(29, 57/2) \rightarrow (28, 55/2)}}$ &  275937.107 $^{(F)}$ & 198.7 &  9.10$\times$10$^{-5}$ &   28.5 &   58    & 0.15(2)   & 10.67(10) & 1.76(27)    &  0.081  \\
${\mathrm{(30, 61/2) \rightarrow (29, 59/2)}}$ &  285407.207 $^{(F)}$ & 212.3 &  1.01$\times$10$^{-4}$ &   30.5 &   62    & 0.15(2)   & 10.53(15) & 2.30(32)    &  0.062  \\
${\mathrm{(30, 59/2) \rightarrow (29, 57/2)}}$ &  285445.430 $^{(F)}$ & 212.4 &  1.01$\times$10$^{-4}$ &   29.5 &   60    & 0.14(3)   & 10.52(34) & 2.50(59)    &  0.052  \\
${\mathrm{(31, 63/2) \rightarrow (30, 61/2)}}$ &  294914.937 $^{(F)}$ & 226.5 &  1.11$\times$10$^{-4}$ &   31.5 &   64    & 0.11(1)   & 10.59(7)  & 1.89(23)    &  0.105  \\
${\mathrm{(31, 61/2) \rightarrow (30, 59/2)}}$ &  294953.137 $^{(F)}$ & 226.6 &  1.11$\times$10$^{-4}$ &   30.5 &   62    & 0.06(1)   & 10.74(10) & 1.45(24)    &  0.074  \\
${\mathrm{(32, 65/2) \rightarrow (31, 63/2)}}$ &  304422.034 $^{(F)}$ & 241.1 &  1.22$\times$10$^{-4}$ &   32.5 &   66    & 0.16(2)   & 10.38(18) & 2.17(50)    & \ \ 0.068$^{\star}$   \\
${\mathrm{(32, 63/2) \rightarrow (31, 61/2)}}$ &  304460.209 $^{(F)}$ & 241.2 &  1.22$\times$10$^{-4}$ &   31.5 &   64    & 0.11(4)   & 10.95(37) & 1.90(32)    &  \ \ 0.056$^{\star}$   \\

\hline
\hline    
    
\end{tabular}
\end{center} 

\end{table*}


\end{document}